\documentclass{aa} 
\usepackage{hyperref}
\usepackage{natbib}
\bibpunct{(}{)}{;}{a}{}{,} 
\usepackage{graphicx}
\usepackage{txfonts}
\usepackage{adjustbox,lipsum}

\begin{document}

   \title{The survey of Planetary Nebulae in Andromeda (M31): I. Imaging the disk and halo with MegaCam@CFHT} 
    \titlerunning{The survey of PNe in M31 I - Imaging}
   \author{Souradeep Bhattacharya
          \inst{1}\fnmsep\thanks{Based on observations obtained with MegaCam, a joint project of CFHT and CEA/DAPNIA, at the Canada-France-Hawaii Telescope (CFHT) which is operated by the National Research Council (NRC) of Canada, the Institut National des Science de l'Univers of the Centre National de la Recherche Scientifique (CNRS) of France, and the University of Hawaii. The observations at the CFHT were performed with care and respect from the summit of Maunakea which is a significant cultural and historic site.} \and
          Magda Arnaboldi\inst{1} \and 
          Johanna Hartke\inst{1,2} \and
          Ortwin Gerhard\inst{3} \and
          Valentin Comte\inst{1,4} \and
          Alan McConnachie\inst{5} \and
          Nelson Caldwell\inst{6}
          }
   \institute{European Southern Observatory, Karl-Schwarzschild-Str. 2, 85748 Garching, Germany \\ 
   \email{sbhattac@eso.org} \and
   European Southern Observatory, Alonso de Cordova 3107, Vitacura, Casilla 19001, Santiago de Chile, Chile\and
   Max-Planck-Institut für Extraterrestrische Physik, Giessenbachstrasse, 85748 Garching, Germany \and
   Aix Marseille Universite, CNRS, LAM – Laboratoire d’Astrophysique de Marseille, 38 rue F. Joliot-Curie, 13388 Marseille, France \and
   NRC Herzberg Institute of Astrophysics, 5071 West Saanich Road, Victoria, BC V9E 2E7, Canada \and
   Harvard-Smithsonian Center for Astrophysics, 60 Garden Street, Cambridge, MA 02138
             }

   \date{Submitted: 5$^{th}$ November, 2018; Accepted: 6$^{th}$ March, 2019}

 
  \abstract
    { The Andromeda (M31) galaxy subtends nearly a 100 sq. deg. on the sky. Any study of its halo must therefore account for the severe contamination from the Milky Way halo stars whose surface density displays a steep gradient across the entire M31 field-of-view.} 
   {Our goal is to identify a population of stars firmly associated with the M31 galaxy. 
   Planetary Nebulae (PNe) are one such population that are excellent tracers of light, chemistry and motion in galaxies.
   We present a 16 sq. deg. survey of the disk and inner halo of M31 with the MegaCam wide-field imager at the CFHT to identify PNe, characterize their luminosity-specific PN number and  luminosity function in M31.}
   {PNe were identified based on their bright [\ion{O}{iii}] 5007 $\AA$ emission and absence of a continuum through automated detection techniques. Subsamples of the faint PNe were independently confirmed by matching with resolved \textit{Hubble Space Telescope} sources from the Panchromatic Hubble Andromeda Treasury and spectroscopic follow-up observations with HectoSpec at the MMT.}
   {The current survey reaches 2 magnitudes fainter than the previous most-sensitive survey. We thus identify 4289 PNe, of which only 1099 were previously known. By comparing the PN number density with the surface brightness profile of M31 out to $\sim 30$ kpc along the minor-axis, we find that the stellar population in the inner halo has a 7 times larger luminosity-specific PN number value than that of the disk.
   We measure the luminosity function of the PN population and find a bright cut-off and a slope consistent with the previous determination by \citet{ciardullo89}. Interestingly, it shows a significant rise at the faint end, present in all radial bins covered by the survey. Such a rise in the M31 PN luminosity function is much steeper than that observed for the Magellanic clouds and Milky Way bulge.}
   {The significant radial variation of the PN specific frequency value indicates that the stellar population at deprojected minor-axis radii larger than $\sim 10$ kpc is different from that in the disk of M31.
   The rise at the faint-end of the PN luminosity function is a property of the late phases of the stellar population. M31 shows two major episodes of star formation and the rise in the faint end of the PNLF is possibly associated with the older stellar population. It may also be a result of varying opacity of the PNe.}


   \maketitle
%

\section{Introduction}

The Andromeda galaxy, M31, is the closest giant spiral disk to our Milky Way (MW). 
It lies at a distance of $\sim$ 780 kpc with a high inclination to the line-of-sight (i$\sim 77\deg$) making it ideally suited for studies of its halo regions. Since galaxies are believed to be formed by hierarchical mass assembly, their outskirts with long dynamical timescales are expected to have coherent debris from past accretion events for the greatest longevity \citep{fm16}. Indeed, through the Pan-Andromeda Archaeological Survey \citep[PAndAS;][]{mcc09} map of the resolved stellar population number counts, we now know about the substructures present in the M31 halo 
\citep[Giant Stellar Stream, G1 and NE clump, NE and W shelves;][]{mcc18} and that the stellar halo extends out to 165 kpc. 


Because of the faint surface brightness of the M31 halo, $\mu_v > 25$ outside the main disk ($\sim 15$ kpc major axis distance), the use of discrete stellar tracers is superior to integrated absorption-line spectroscopy in providing a global mapping of the halo kinematics. Globular clusters (GCs) have been shown to efficiently trace the outer halo (outside 50 kpc) substructures of M31 \citep{mackey10, vel14} bolstering the idea that these clusters and the substructures they trace have been accreted in various merger events. However the inner halo substructures (within 50 kpc) are not well-traced by the GCs. Some of these substructures, especially the Giant Stream, may have resulted from a single merger event $\sim 2$ Gyr ago \citep{bernard15, ham18} perhaps between M31 and the large (M$_{M32P}\sim2.5\times10^{10}$ M$_\sun$) progenitor of M32 \citep{dsouza18}.

We can gain information on the motions of the stars in the low surface brightness regions of M31 by studying Planetary Nebulae (PNe) that act as discrete tracers of stars in the halo. PNe are the glowing shells of gas and dust observed around stars that have recently left the asymptotic giant branch (AGB) and are evolving towards the white dwarf stage.  They are traditionally considered the late phases of stars with masses between $\sim 0.7$ and 8 M$_\sun$, but have been shown to exhibit a wide variety of striking morphologies pointing towards a binary evolution in many systems \citep{jb17}. Since the timescales between the AGB and PN phases are short, the distribution and kinematics of PNe are expected to be identical to their parent population, having the same angular momentum distribution as the stellar population \citep[e.g.][]{hui95, arnaboldi96, arnaboldi98, men01}. Studying PNe as a population provides insight into galactic structure and evolution. Because of their relatively strong [\ion{O}{iii}] 5007$\AA$ emission, PNe can be readily identified. They have been shown to be efficient tracers of stellar light in different galaxies like M87 \citep{longobardi13}, M49 \citep{hartke17}, and many other early-type galaxies \citep{coc09,cor13,pul18}. The luminosity-specific PN number ($\alpha$-parameter) varies slightly with B-V colour of galaxies with higher values for late-type galaxies and lower values for early-type galaxies \citep{buz06}. Different $\alpha$-parameter values point to differences in stellar populations even within the same galaxy \citep[e.g. M49;][]{hartke17}. 

The characteristic [\ion{O}{iii}] 5007$\AA$ PN luminosity function (PNLF) has proven itself as a reliable secondary distance indicator for determining galactic distances out to $\sim$20 Mpc, by virtue of its invariant absolute bright cut-off, M*. The faint end of the PNLF was shown by \citet{jacoby80} to follow an exponential function expected from slowly evolving central stars embedded in rapidly expanding, optically thin nebulae \citep{hw63}. The PNLF was first described by \citet{ciardullo89} estimated empirically from the brightest PN they found in the center of M31 as:
\begin{equation}
    $$N(M) \propto e^{0.307M}(1-e^{3(M^{*}-M)})$$
\end{equation}
The bright end exponential cut-off, supported with accurate measures of foreground extinction, is currently measured at $M^{*}= -4.54 \pm 0.05$ \citep{ciardullo13}. Apart from M* reducing in low metallicity populations \citep[e.g.][]{ciardullo92,ciardullo02,hmp09}, the PNLF cutoff has proved to be largely invariant with metallicity and age of the parent stellar population, and, galaxy type. 

The faint-end of the PNLF has been shown to vary considerably, depending on the details of the stellar population. It is seen to be correlated with the star formation history of the parent stellar population, with steeper slopes associated with older stellar populations and conversely flatter slopes with younger populations \citep{ciardullo04, ciardullo10,longobardi13, hartke17}. The changes in the PNLF slope can result from superposition of multiple stellar populations which can then be disentangled using the PNe kinematics \citep{longobardi15, hartke18}. The PNLF has also been shown to display a dip for some galaxies. The dip is seen $\sim$ 3.5 mag below the bright cut-off in the Small Magellanic Cloud \citep[SMC;][]{jd02}, $\sim$ 2.5 mag below the bright cut-off for NGC 6822  \citep{hmp09} and also slightly in the Large Magellanic Clouds \citep{rp10}, and $\sim$ 1 mag below the bright cut-off in M87 \citep{longobardi15}. This dip is suggested to be related with the opacity of the PN and may be characterized by accounting for circumstellar extinction in the PNe. Indeed, circumstellar extinction correction does modify the PNLF \citep{rp10,davis18} but is difficult to estimate for the faint end of the PNLF (beyond $\sim$ 2.5 mag below the bright cut-off) to test for changes in opacity of the PN. Additionally PNe mimics like \ion{H}{ii} regions and symbiotic stars may be misidentified as PNe thereby affecting the PNLF.

Since the first empirical study of the PNLF by \citet{ciardullo89} using 104 objects in the M31 bulge, the number of PNe known in M31 has increased by leaps and bounds. Most notably, \citet[][hereafter M06]{merrett06} utilized the custom-built Planetary Nebula Spectrograph \citep[PNS;][]{dou02} at the William Herschel Telescope (WHT) to identify 2615 PNe in the disk and bulge of M31 and simultaneously obtain their [\ion{O}{iii}] 5007$\AA$ magnitude and line-of-sight velocity (LOSV). M06 corroborated the PNLF found by \citet{ciardullo89} and increased the photometric depth to $\sim$ 3.5 - 4 mag below the bright cut-off along with the increased uniform coverage. Since then, \citet{martin18} and \citet{Li18} have identified more PNe in the central regions and circumnuclear region of M31 respectively adding to the large number of PNe already known in M31. 

While some of the PNe identified by M06 have been shown to be \ion{H}{ii} regions \citep{san12,vey14}, the M06 PNe sample remains the largest uniform sample of PNe in any galaxy. The LOSV of the M06 PNe predicted links between the NE Shelf and the Giant Stream substructures of M31 which have been further explored in deep spectroscopic chemical tagging studies by \citet{fang15, fang18} to further establish these links. However while some identified PNe have been associated with the halo, a uniform survey of PNe in much of the metal-poor halo of M31 and the inner-halo substructures is necessary not only to unambiguously trace them but also to probe the variation in the PNLF further out from the disk to corroborate the invariant nature of its bright cut-off and observe the evolutionary effects on its faint end. 

In this paper, we survey the inner 16 sq. degree of M31 (corresponding to 20-30 kpc from the center), covering the disk, parts of the inner halo and some of the inner halo substructures. We detect PNe using the on-off-band technique to a depth further than M06. The observations and data reduction are described in Section 2. Identification of PNe is described in Section 3. Section 4 describes the identification of the PNe counterparts in a subsample to \textit{Hubble Space Telescope} (HST) data. We obtain the $\alpha$-parameter in section 5 and analyze the PNLF in Section 6. The discussions of the results are presented in Section 7 and we summarize and conclude in Section 8.


\begin{figure}[t]
	\centering
	\includegraphics[height=1.1\columnwidth,angle=0]{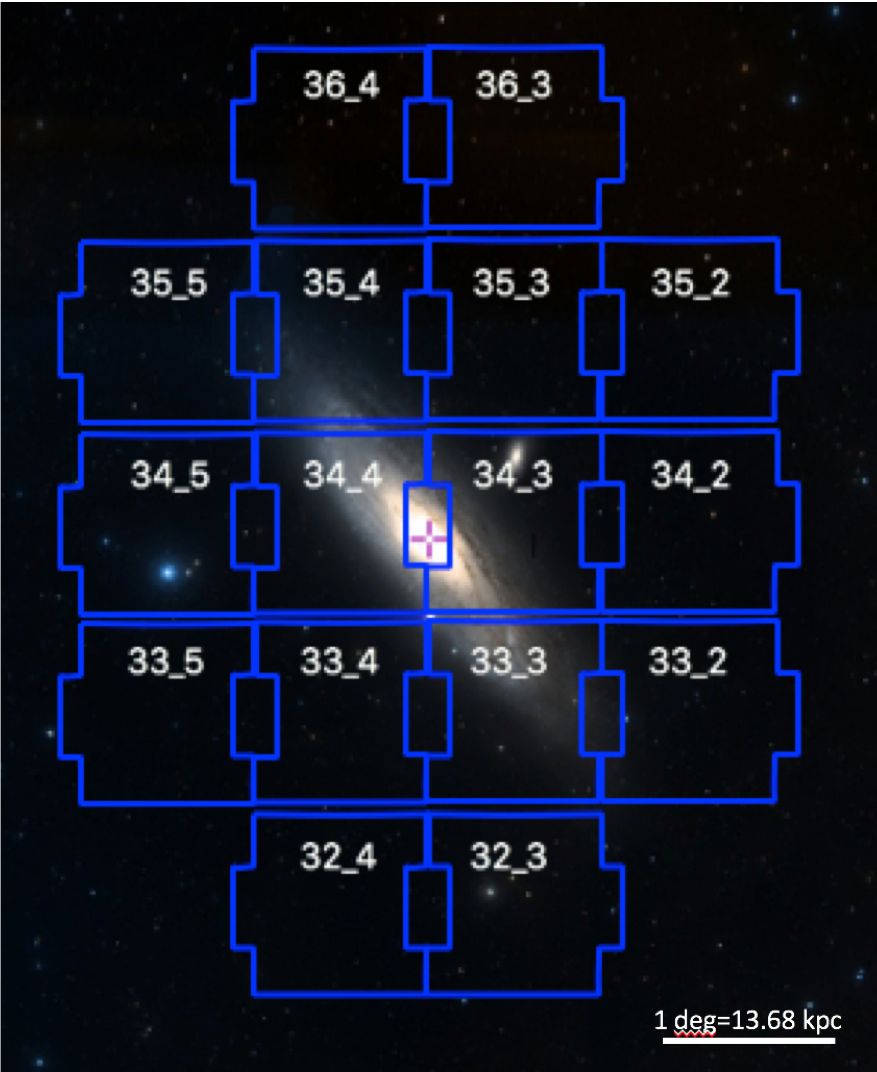}
	\caption{The fields observed with CFHT MegaCam shown in blue, labeled with their observation ID. North is up, east is left. The background image is from SDSS obtained using the Aladin Sky Atlas (\cite{bon00}).}
	\label{fig:pointings}
\end{figure}

\section{CFHT MegaCam M31 PNe survey}

\subsection{Imaging and observations}

The observations were carried out with the MegaCam wide-field imager \citep{boulade03} mounted on the 3.6-meter Canada France Hawaii Telescope (CFHT), located near the summit of the Mauna Kea mountain on Hawaii’s Big Island at an altitude of 4204 meters. MegaCam is comprised of a mosaic of 40 individual 2048 $\times$ 4612 CCDs, resulting in a contiguous field of view of 0.96 $\times$ 0.94 degrees with a pixel scale of 0.187$''$/pixel. The inner 20-30 kpc radius of the M31 halo were observed through 16 pointings of the MegaCam imager. The fields are shown in Figure~\ref{fig:pointings}. The observations were carried out, with photometric conditions, during two runs: October 9-11 and November 6-7, 2016. Over the course of the observations, seeing varied between 0.5$''$ and 1.1$''$ while the airmass varied between 1.03 and 1.48.

M31 is observed through a narrow-band [\ion{O}{iii}] filter ($\lambda\rm_c = 5007~\AA, ~\Delta\lambda = 102~\AA$, on-band) and a broad-band \textit{g}-filter ($\lambda\rm_c = 4750~\AA, ~\Delta\lambda = 1540~\AA$, off-band). The photometry is calibrated with observations of spectrophotometric standard stars. Each on-band image generally consists of 4 dithered exposures with a total exposure time of 1044 s and each off-band image of 3 dithered exposures with a total exposure time of 300 s. For some fields there are more dithered images for both on-band and off-band, leading to higher exposure times. A summary of the field positions and exposure times for the on-band and off-band exposures is presented in Table~\ref{table : obs}. The exposure time was chosen such that PNe with an apparent narrow-band magnitude of 6 mag from the bright cut-off m$\rm_{5007}$ = 20.2 mag, in the m$\rm_{5007}$ system described by \citet{jacoby89}, can still be detected. 

\subsection{Data reduction}

The data are pre-processed using the Elixir\footnote{http://www.cfht.hawaii.edu/Instruments/Elixir/home.html} pipeline \citep{magnier04}, which accomplishes the bias, flat, and fringe corrections and also determines the photometric zero point of the observations. The zero points for the [\ion{O}{iii}] and \textit{g}-band frames in \textit{AB} magnitudes, normalised to a 1 s exposure, are Z$_{\rm[\ion{O}{iii}]}$ = 23.434 and Z$_g$ = 26.5. For every observed field, a weight-map is computed corresponding to each exposure using WeightWatcher\footnote{http://www.astromatic.net/software/weightwatcher} \citep{marmo08}. This is used to assign higher weights to pixels which are more reliable compared to their local backgrounds. For each field, the exposures corresponding to the on-band image and those corresponding to the off-band image are then combined using SWARP\footnote{http://www.astromatic.net/software/swarp} \citep{bertin02} to produce the respective on-band and off-band images. The exposures are combined using median-type combination in conjunction with their respective weight maps, while a background subtraction is carried out with a background mesh size of 25 pixels. The images thus obtained are used for source extraction. 

\begin{table}[t]
\caption{Summary of the field positions, exposure times, and seeing for the narrow-band (on-band) and broad-band (off-band) images. {Limiting magnitudes (m$\rm_{5007, lim}$; described in Sect~\ref{sect:lim}) for each field are also provided.} }
\centering
\adjustbox{max width=\columnwidth}{
\begin{tabular}{ccccccc}
\hline
Field & $\alpha$ (J2000) & $\delta$ (J2000) & Exp$_{\rm{[\ion{O}{iii}]}}$ & Exp$_{\rm{g}}$ & S$_{\rm{FWHM}}$ & m$_{\rm{5007, lim}}$\\
  & (h:m:s) & (${\circ}:':''$) & (s) & (s) & ($''$) & (mag)\\
\hline
36\_4 & 00:45:51.9 & 43:24:20.5 & $4\times261$ & $3\times100$ & 0.97 & 26.08 \\
36\_3 & 00:40:14.7 & 43:24:23.7 & $4\times261$ & $3\times100$ & 0.95 & 26.26 \\
35\_5 & 00:51:18.0 & 42:21:48.2 & $4\times261$ & $4\times100$ & 0.85 & 26.30 \\
35\_4 & 00:45:48.0 & 43:23:02.4 & $8\times261$ & $5\times100$ & 0.64 & 26.40 \\
35\_3 & 00:40:17.2 & 42:23:05.6 & $4\times261$ & $6\times100$ & 0.66 & 26.17 \\
35\_2 & 00:34:45.9 & 42:21:57.6 & $4\times261$ & $3\times100$ & 0.68 & 26.12 \\
34\_5 & 00:51:09.5 & 41:20:18.6 & $4\times261$ & $3\times100$ & 0.90 & 26.26 \\
34\_4 & 00:45:44.1 & 41:21:18.7 & $5\times261$ & $3\times100$ & 0.85 & 26.16 \\
34\_3 & 00:40:18.7 & 41:21:21.6 & $4\times261$ & $3\times100$ & 0.92 & 25.93 \\
34\_2 & 00:34:53.3 & 41:20:27.6 & $6\times261$ & $9\times100$ & 0.47 & 26.02 \\
33\_5 & 00:51:02.7 & 40:18:41.6 & $4\times261$ & $3\times100$ & 0.77 & 25.94 \\
33\_4 & 00:45:41.3 & 40:19:48.0 & $4\times261$ & $3\times100$ & 0.76 & 25.64 \\
33\_3 & 00:40:20.9 & 40:19:50.8 & $4\times261$ & $3\times100$ & 0.84 & 25.89 \\
33\_2 & 00:35:00.6 & 40:18:57.6 & $4\times261$ & $3\times100$ & 0.54 & 25.90 \\
32\_4 & 00:45:39.0 & 39:18:16.9 & $4\times261$ & $3\times100$ & 0.60 & 26.13 \\
32\_3 & 00:40:25.1 & 39:18:26.8 & $4\times261$ & $3\times100$ & 0.59 & 26.10 \\
\hline
\end{tabular}
\label{table : obs}
}
\end{table}

\section{Selection of PNe candidates and catalog extraction}
\label{catex}

Having a bright [\ion{O}{iii}] 5007 $\AA$ and no continuum emission, extragalactic PNe can be identified as objects detected in the on-band [\ion{O}{iii}] image which are not detected in the off-band continuum images, or have an excess [\ion{O}{iii}] - g colour. Additionally, PNe are typically unresolved at extragalactic distances from ground-based observations and so we only considered point-like objects for analysis. We used the automatic selection procedure developed and validated in \citet{arnaboldi02,arnaboldi03}, which has been optimized for large imaging surveys by \citet{longobardi13} and \citet{hartke17}.

\begin{figure}[t]
	\centering
	\includegraphics[width=\columnwidth,angle=0]{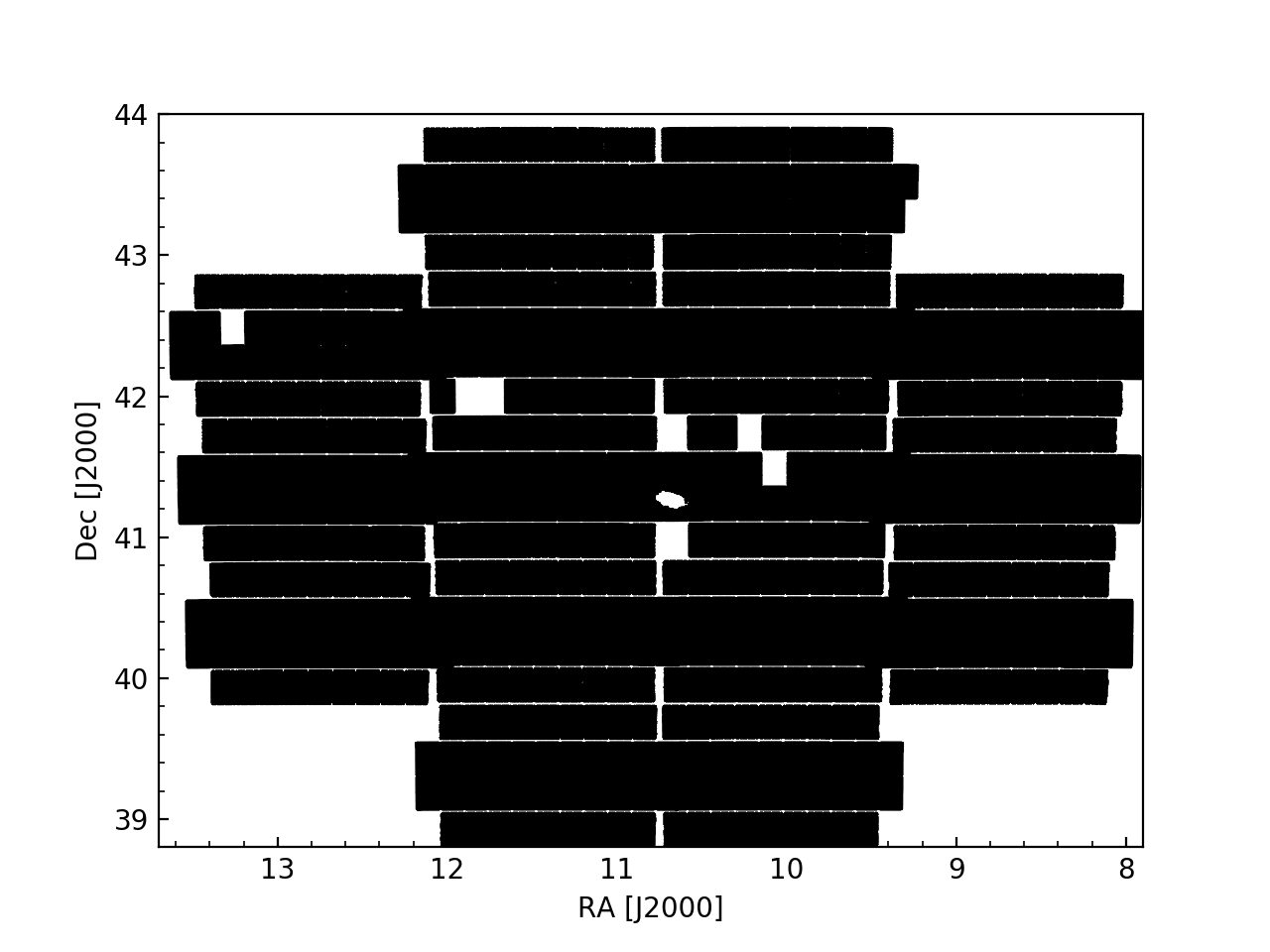}
	\caption{{The effective survey area is shown in black. CCD edges and noisy or saturated regions have been masked.}}
	\label{fig:survey_area}
\end{figure}

\subsection{Source extraction}
\label{sext}
We used SExtractor \citep{bertin96}, a source detection algorithm that detects and measures flux from point-like and extended sources, to detect and carry out photometry of the sources on the images. For each field, we measure the narrow-band and broad-band magnitudes, m$\rm_n$ and m$\rm_b$, in dual-image mode for sources detected on the narrow-band image. The broad-band magnitudes were extracted in the same apertures as on the narrow-band image. Sources were detected in the narrow-band image requiring that 25 adjacent pixels or more have flux values 1.2 $\times$ $\sigma$ rms above the background. Local backgrounds were calculated for the detected sources on apertures having a width of 25 pixels. and magnitudes were measured with different apertures having widths of 15, 17, 19 and 21 pixels. Magnitudes were also measured with an aperture having a width of 5 pixels corresponding to the core of the sources m$\rm_{core}$. Sources for which a broad-band magnitude could not be detected at the position of the [\ion{O}{iii}] detection were assigned a m$\rm_b$ corresponding to 1 $\times$ $\sigma$ rms above the background in the broad-band image, similar to \citet{arnaboldi02}. The narrow-band magnitude, m$\rm_n$, in the AB system is converted to the m$\rm_{5007}$ system as: $m_{5007}=m\rm_n+ 2.27$.  Refer to Sect~\ref{flux} for details. The central part of our survey covering part of the bulge of M31 are mostly saturated. In some observed fields, a few CCDs were noisy. Such noisy regions and CCD edges, affected by dithering and saturation, were masked as detailed in Sect~\ref{mask}. {The effective survey area is shown in Figure~\ref{fig:survey_area}.} 

\begin{figure}[t]
	\centering
	\includegraphics[width=\columnwidth,angle=0]{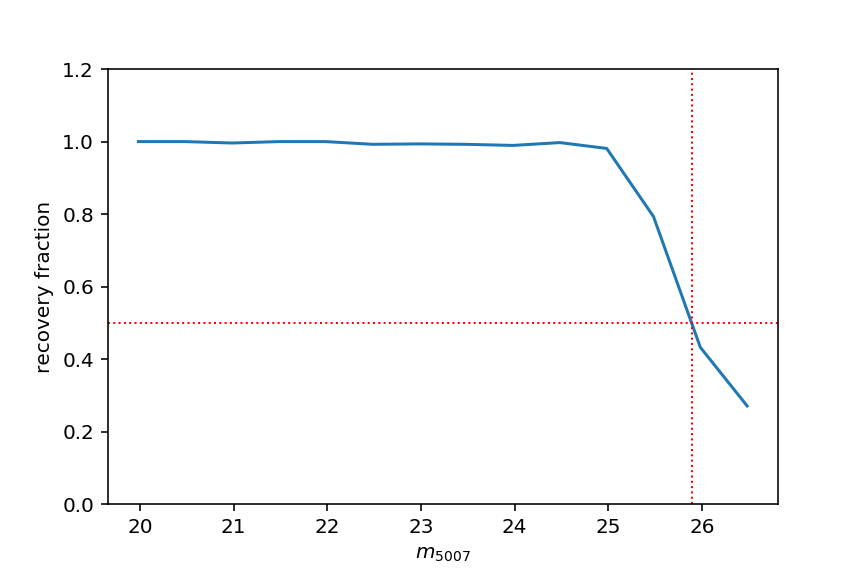}
	\caption{The recovery fraction of simulated sources for a single field to illustrate the limiting magnitude at the 50\% completeness limit.}
	\label{fig:lim_mag}
\end{figure}

\subsection{Limiting magnitude}
\label{sect:lim}
In order to determine the limiting magnitude of our sample for each field, we simulated a synthetic point-like population onto the on-band image {(with its corresponding exposure time and zero-point)} using the Image Reduction and Analysis Facility (IRAF\footnote{http://iraf.noao.edu/}) task {\it mkobjects}. The synthetic population follows a PNLF as detailed in \citet{ciardullo89} and the sources have a Moffat PSF profile as detailed in Sect~\ref{psf}. The sources are then extracted. The magnitude aperture most suited to recovering the simulated sources is found to be 15 pixels as detailed in Sect~\ref{aper}.  The limiting magnitude is defined as the magnitude at which the recovery fraction of the simulated sources drops below 50\% (Figure~\ref{fig:lim_mag}). {This limiting magnitude varies between m$\rm_{5007} = 25.64$, for the shallowest observed field, Field\# 33\_4, to m$\rm_{5007} = 26.4$, for the deepest observed field, Field\# 35\_4. The limiting magnitude is provided for each field in Table~\ref{table : obs}}.

\begin{figure}[t]
	\centering
	\includegraphics[width=\columnwidth,angle=0]{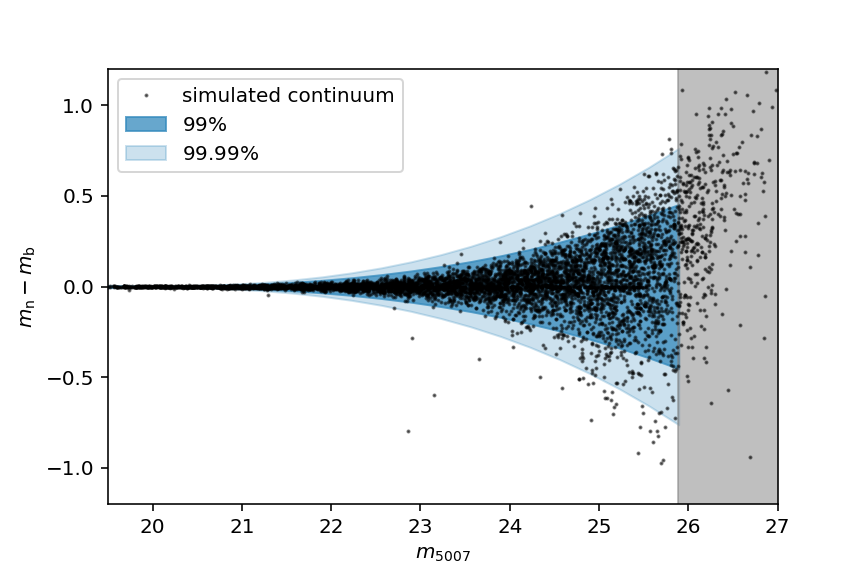}
	\caption{The CMD showing the synthetic continuum sources and the 99\% and 99.99\% limits on their positions for a single field. The region beyond the limiting magnitude of this field is shown in grey.}
	\label{fig:continuum}
\end{figure}

\subsection{Colour selection}
\label{csel}
We selected PNe candidates based on their position on the m$\rm_{5007}$ versus m$\rm_n - m\rm_b$ CMD. These are sources with a colour excess m$\rm_n - m\rm_b$ < $-1$ that are brighter than the limiting magnitude. The colour excess corresponds to an equivalent width EW$_{obs}$ = 110 $\AA$ \citep{teplitz00} and was chosen in order to limit contamination from background galaxies.

Sources which do not have an excess in m$\rm_n - m\rm_b$ are classified as continuum sources. However, some colour excess maybe seen for some of these sources during the source extraction from the images, especially for those with fainter magnitudes. We thus simulated the same synthetic point-like population onto the off-band image and checked their position on the extracted m$\rm_{5007}$ versus m$\rm_n - m\rm_b$ CMD (Figure~\ref{fig:continuum}). 
We calculate the 99\% and 99.99\% limits on their positions in the CMD, below which the probability of detecting continuum sources was reduced to the 1\% and 0.01\% level respectively. Sources with colour-excess within the 99.99\% limit are also discarded as possible PNe. 

\begin{figure}[t]
	\centering
	\includegraphics[width=\columnwidth,angle=0]{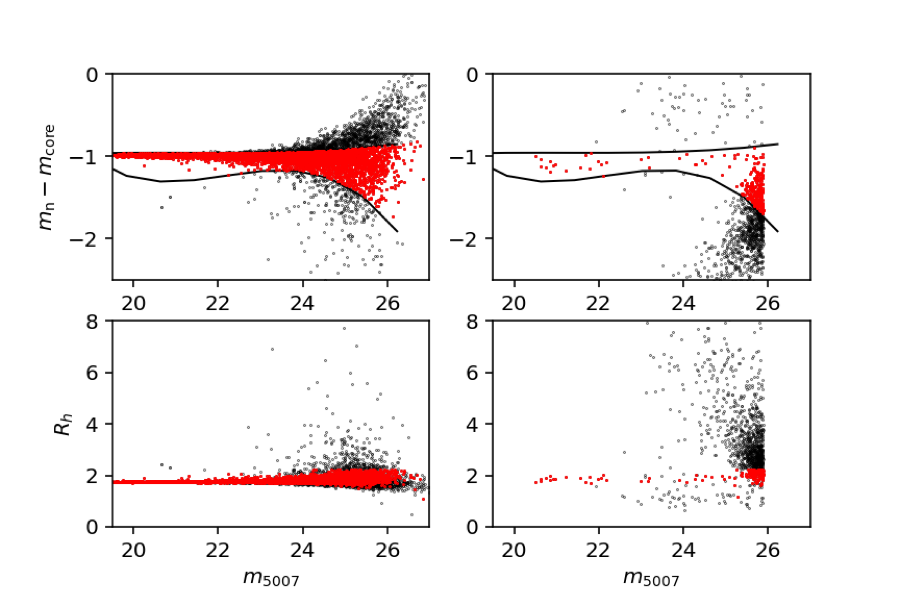}
	\caption{[Top] The difference between m$\rm_n$ and m$\rm_{core}$ for all sources in black and those within the 95\%-limits of the simulated population in red for the simulated population (left) and as applied to the real sources (right) for a single field. [Bottom] Same as the top panel but for half-light radius, R$_h$.}
	\label{fig:pt}
\end{figure}

\begin{figure}[t]
	\centering
	\includegraphics[width=\columnwidth,angle=0]{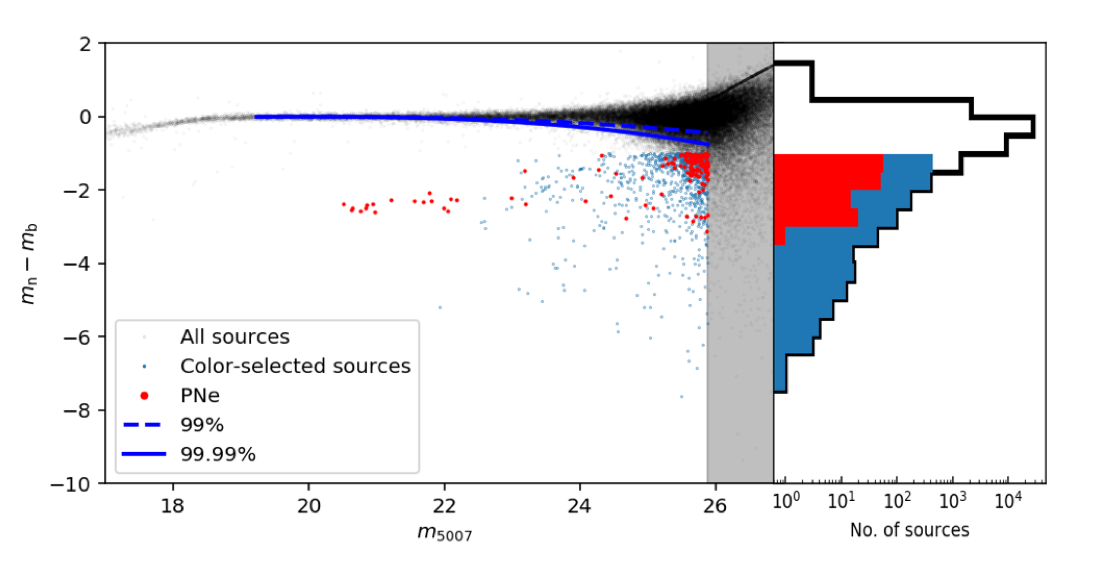}
	\caption{{The CMD for a single field showing all the detected sources (brighter than the limiting magnitude) in black, the colour-selected sources in blue and identified PNe in red. The 99\% and 99.99\% limits for the continuum sources are shown in blue while the region beyond the 50\% completeness limit is shown in grey. The histogram (in logarithmic scale) shows clearly the number of PNe recovered as a function of colour with all the detected sources (brighter than the limiting magnitude) in black, the colour-selected sources in blue and identified PNe in red.}}
	\label{fig:cmd}
\end{figure}

\begin{figure*}[t]
	\centering
	\includegraphics[height=\textwidth,angle=0]{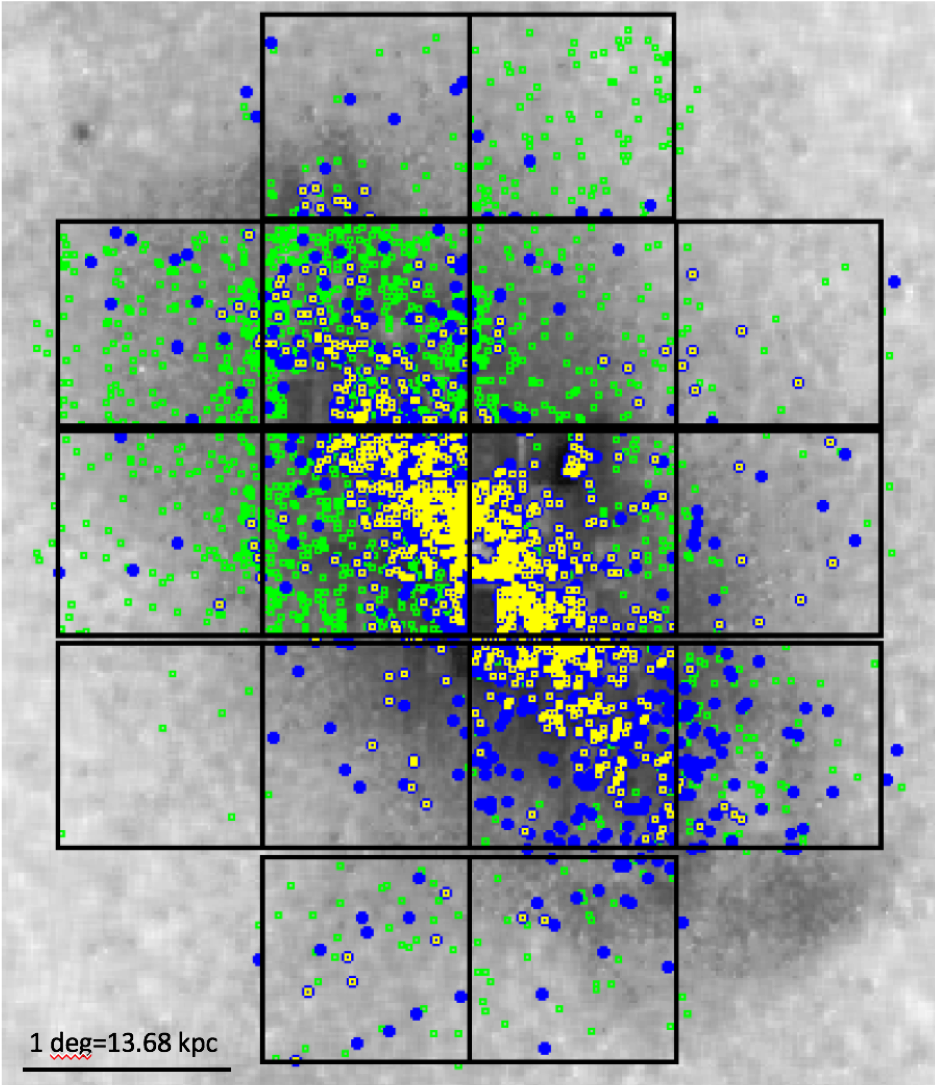}
	\caption{The PNe identifed by the survey (blue - PNe  brighter than m$\rm_{5007} = 25.64$ which is the 50\% completeness limit of the shallowest field (Field\# 33\_4), green - PNe with fainter than m$\rm_{5007} = 25.64$ with photometric depth varying with field) are overlaid on the map of RGB stars identified by the PAndAS survey. The M06 PNe re-identified by our survey are shown in yellow. North is up, east is left. }
	\label{fig:spatial}
\end{figure*}

\begin{figure}[t]
	\centering
	\includegraphics[width=\columnwidth,angle=0]{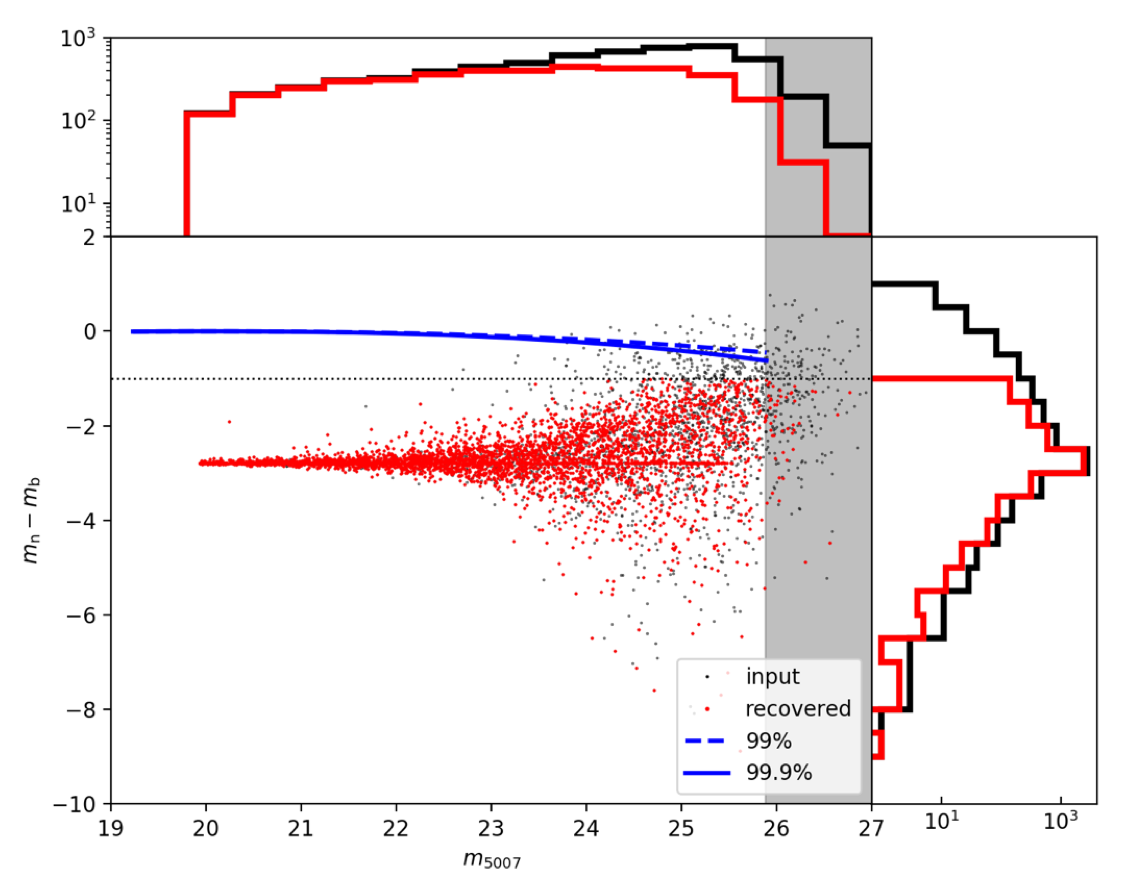}
	\caption{{The CMD for a single field showing all the simulated sources in black and those recovered as PNe in red (See Section~\ref{compcorr1}). The 99\% and 99.99\% limits for the continuum sources are shown in blue while the region beyond the limiting magnitude is shown in grey. The dotted black line shows the colour selection adopted to limit contamination from background galaxies. The histograms (in logarithmic scale) show the number of PNe recovered as a function of colour (right) and magnitude (top) with the simulated sources in black and those recovered as PNe in red.}}
	\label{fig:sel_comp}
\end{figure}

\subsection{Point-like selection}
\label{pts}
PNe are typically unresolved point-like objects at extragalactic distances and to differentiate them from extended ones (e.g. background galaxies or other extended objects with strong [\ion{O}{iii}] emission), we analyze the light distribution of the simulated sources, {as described in Section~\ref{sect:lim}}, on the on-band image. For each field, we use the half-light radius of the simulated sources, R$_h$, the radius within which half of the object’s total flux is contained, to determine its upper limit, R$_{hmax}$, corresponding to 95\%-percentile of the simulated population. We considered sources as point-like if they satisfied the following two criteria: these sources (i) have a half-light radius such that 1 < R$_h$ < R$_{hmax}$, and (ii) they fall in the region where the difference between m$\rm_n$ and m$\rm_{core}$, magnitude of the source for flux within an aperture of 5 pixels, is within the 95\%-limits of the simulated population. The point-like selection criteria are shown in Figure~\ref{fig:pt}, as applied to the colour-selected simulated population and to the real sources. 

{In order to estimate the number of continuum sources that may be misidentified as PNe in any field, we count the number of point-like continuum sources (excluded as PNe by the colour selection criteria) and multiply by 0.01\%. We estimate that in each field, our identified PNe sample may be contaminated by 2-4 continuum sources which lie in the faint magnitudes $> 25$.}

\begin{figure}[htb]
	\centering
	\includegraphics[width=\columnwidth,angle=0]{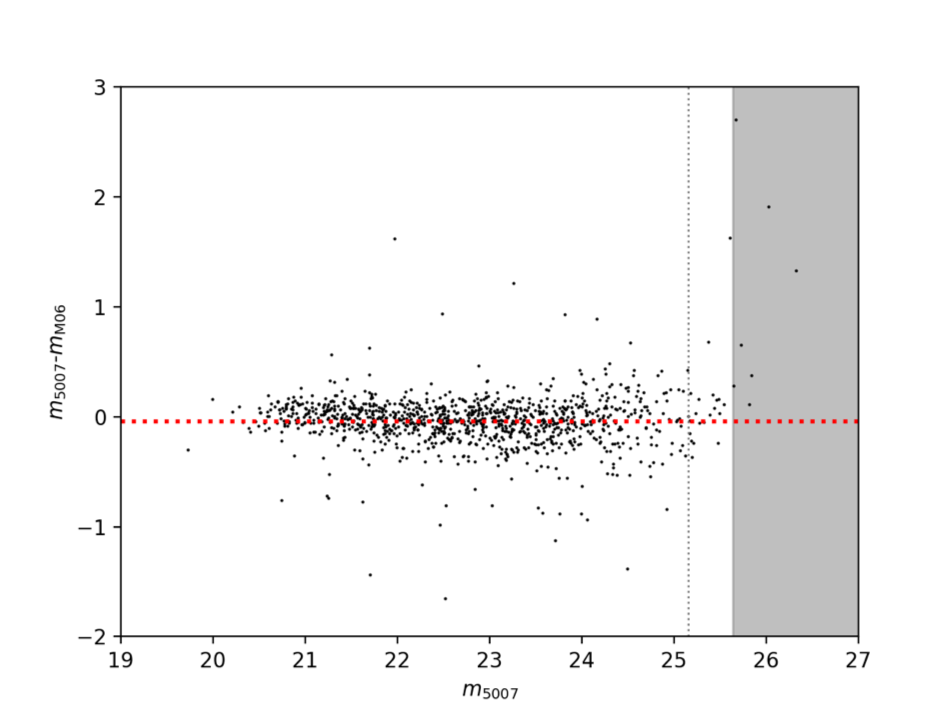}
	\caption{The difference between the narrow-band magnitudes of this work and M06 obtained for the matched sources, plotted against m$\rm_{5007}$. The red dashed line corresponds to the mean offset. The region beyond the limiting magnitude of the shallowest field (Field\# 33\_4) is shown in grey. The grey dotted line shows the 90\% completeness limit of the shallowest field.}
	\label{fig:mer}
\end{figure}

\subsection{PNe catalog}
{For each field, the [\ion{O}{iii}] sources that are brighter than the limiting magnitude and fulfill both the point-like and colour selection criteria are considered as PNe candidate.} They are shown in the CMD in Figure~\ref{fig:cmd} for a single field. Since the bandwidth of the broad-band filter also covers that of the narrow-band filter, the bright PNe are expected to show some remnant flux in the broad-band as well. This remnant broad-band flux is in the ratio of the filter widths and leads to nearly a constant colour excess, m$\rm_n - m\rm_b = 2.5 \rm log(\frac{\Delta\lambda_{[\ion{O}{iii}]}}{\Delta\lambda_g}) =-2.95 $, for the bright PNe. {The observed constant colour excess is slightly less negative due to the flux contributed to the broad-band from the [\ion{O}{iii}] 4959 $\AA$ line which is expected to be $\sim~1/3$ of the brightness of the [\ion{O}{iii}] 5007 $\AA$ line in PNe, as seen in Figure~\ref{fig:cmd}.} The final catalog of confirmed PNe is then checked for spurious sources by eye. Regions of spurious sources, typically caused by saturated stars, are masked and the final catalog of confirmed PNe is obtained for each field. {Counting the PNe identified in the overlapping regions of adjacent fields only once, we identify an unprecedented 4289 PNe in M31 in our survey.} Their spatial distribution, overlaid on a map of RGB stars identified by the PAndAS survey, is shown in Figure~\ref{fig:spatial}. Our survey is uniformly complete {(in the effective survey area)} down to m$\rm_{5007} = 25.64$ which is the 50\% completeness limit of the shallowest field (Field\# 33\_4). However photometric depth varies with fields and we find PNe down to m$\rm_{5007} = 26.4$ for the deepest field (Field\# 35\_4). 

\subsection{Completeness correction}
\label{compcorr1}
{Our color and point-like selection criteria would exclude a number of PNe which are affected by photometric errors especially in the fields covering the bright M31 disk. Thus, in order to determine the selection completeness of our extracted sample, we follow the procedure outlined in \citet{longobardi13} and \citet{hartke17}. We simulated a population of $10^4$ point-like sources (Sect~\ref{sect:lim}) on to the narrow-band image. On the broad-band image, we also simulate sources at the same positions with their fluxes scaled down by the ratio of the filter widths\footnote{We neglect the contribution from the [\ion{O}{iii}] 4959 $\AA$ line in the broad-band image.}.
We then use SExtractor in dual-mode to simultaneously extract m$\rm_{n}$ and m$\rm_b$ of these simulated sources. Figure~\ref{fig:sel_comp} shows the CMD of the simulated population for a single field. The colour excess remains nearly constant for the bright simulated sources, as expected from the ratio of their filter widths, but many are missed at fainter magnitudes. We can thus determine the selection completeness of the simulated population at different magnitude ranges for each field. To a single candidate extracted in a given field, we assign the value of the selection completeness at that magnitude as a probability of being detected after the selection effects. We also compute the detection completeness of the simulated population  at different magnitude ranges from the recovery fraction and similarly assign a detection probability to each PN candidate extracted in each field. The completeness correction is thus obtained from both the selection and detection probability. It is further detailed in Sect~\ref{compcorr}.}

\subsection{Comparison with Merrett et al. (2006) PNe catalog}
We identify those sources in our PNe catalog that have a counterpart in the catalog of PNe identified by M06 by matching them spatially within a 3$''$ aperture (Figure~\ref{fig:mer}). The astrometry of the M06 sample is reliable up to 3$''$ \citep{vey14}. We match 1099 such sources (in yellow in Figure~\ref{fig:spatial}). The narrow-band magnitudes of the matched PNe in our survey are 0.045 mag brighter than the corresponding value in M06, well-within the photometric uncertainty of the M06 sample. We thus validate the [\ion{O}{iii}] 5007 $\AA$  photometry of our PNe with the sample of M06. The photometry of the M06 PN sample is not very accurate for fainter sources. This is because the PN.S instrument used by M06 has not been optimized for photometry measurements but rather for the measurement of radial velocities. 

While we find most of the PNe found by M06 in the observed fields, we miss quite a few in the regions masked by us, mainly in the crowded bulge and the CCD edges. In our survey area, we recover 82.22\% of the M06 PNe. {Up to 25\% of the M06 PNe candidates were actually \ion{H}{ii} regions in the M31 disk \citep{vey14}. The spatial resolution of MegaCam bolstered by the favourable seeing allows us to improve upon the image quality and obtain an accurate PSF to identify PNe in M31. The \ion{H}{ii} regions would appear as extended objects and discarded as PNe by our point-like selection. We investigate the contamination of our PNe catalog by \ion{H}{ii} regions in the next section and further discuss it in Section~\ref{contamination}. Thus the 82.22\% recovery fraction of the M06 PNe is driven by the tighter morphological constraints on the light distribution of the detected [\ion{O}{iii}] sources in our survey. Since the limiting magnitude of our survey is 1.5-2 mag fainter than that of M06, we find a much larger number of new PN candidates in the overlapping area. }

\section{Counterparts in HST imaging}
\label{hst_phat}
To  test the image quality of our PNe survey, and to estimate the possible contamination from \ion{H}{ii} regions and other sources, we identified PNe counterparts in the HST data available from the Panchromatic Hubble Andromeda Treasury \citep{dal12} covering a portion of the M31 disk. 

\begin{figure}[t]
	\centering
	\includegraphics[width=\columnwidth,angle=0]{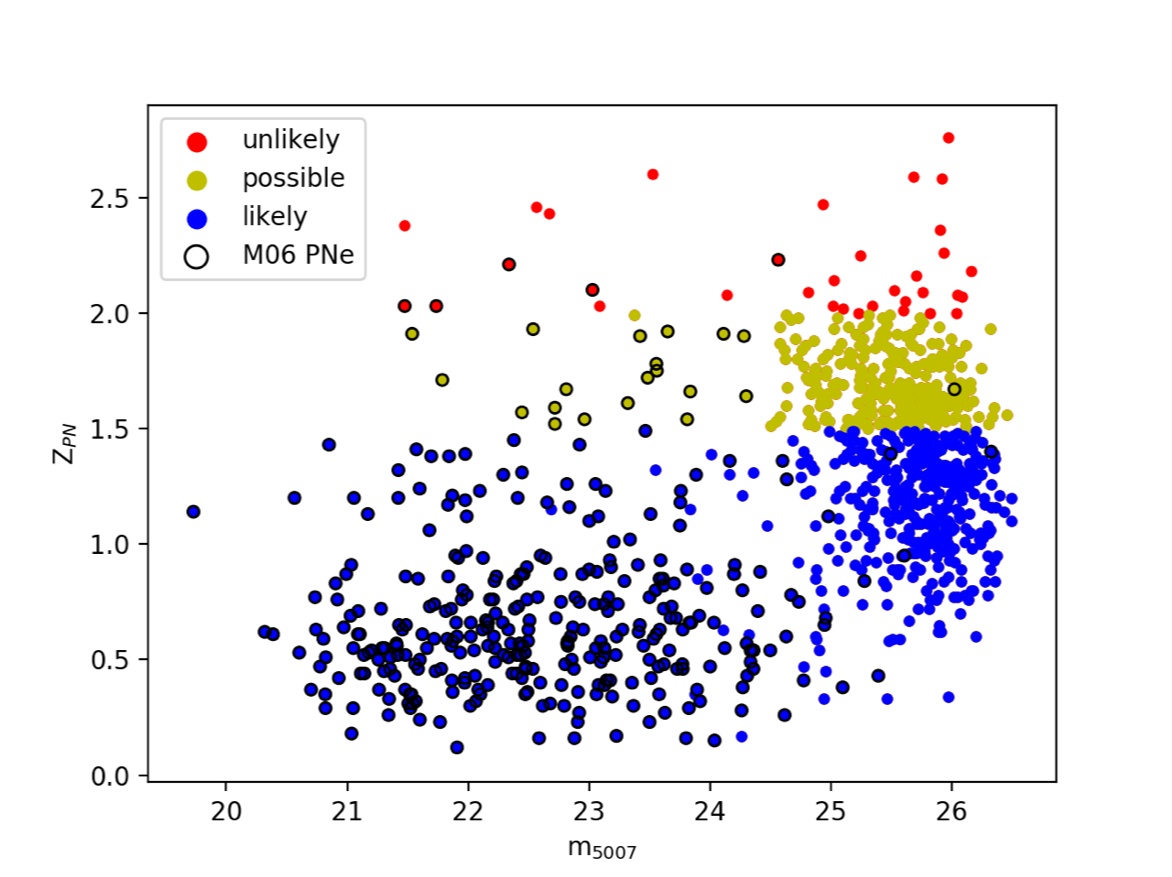}
	\caption{{The $Z\rm_{PN}$ of the PHAT-matched PNe plotted against m$\rm_{5007}$. The ``likely'', ``possible'' and ``unlikely'' PNe have been shown in blue, yellow and red respectively. The PNe previously found by M06 have been encircled in black.}}
	\label{fig:zpn}
\end{figure}

\begin{figure}[t]
	\centering
	\includegraphics[width=\columnwidth,angle=0]{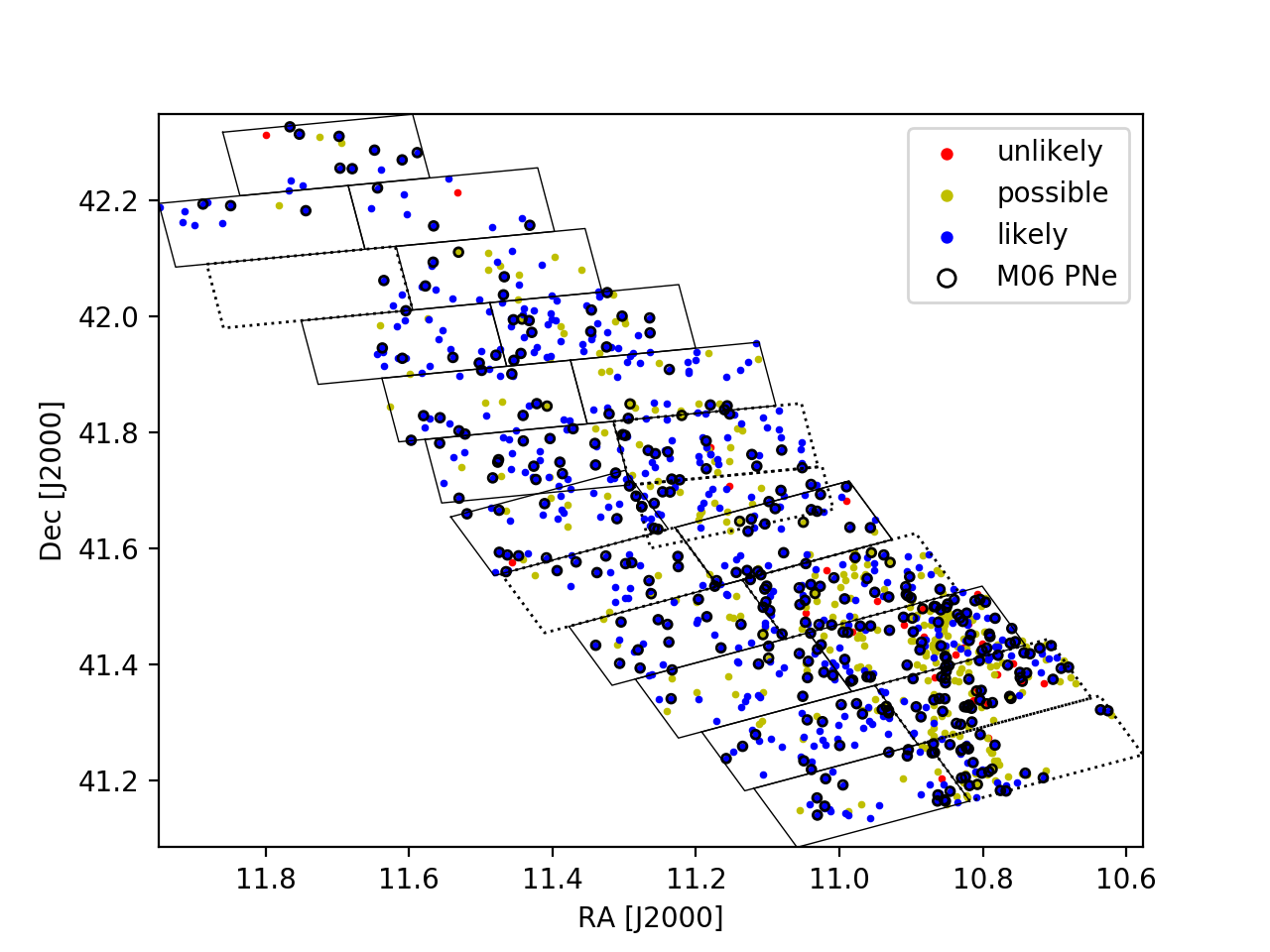}
	\caption{The spatial distribution of the PHAT-matched PNe in the PHAT footprint. The solid bricks are those previously analyzed by \citet{vey14} while the dashed ones have been analyzed here for the first time. The ``likely'', ``possible'' and ``unlikely'' PNe have been shown in blue, yellow and red respectively. The PNe previously found by M06 have been encircled in black.}
	\label{fig:phatspat}
\end{figure}

\begin{figure*}[t]
	\centering
	\includegraphics[width=0.65\columnwidth,angle=0]{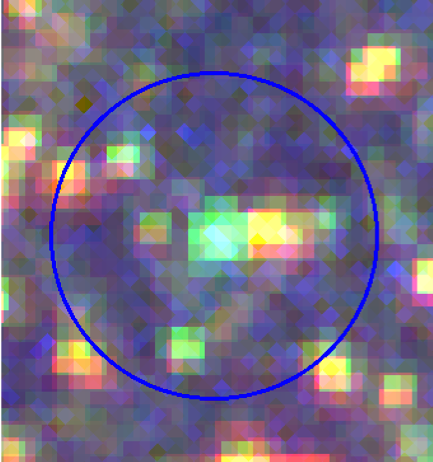}
	\includegraphics[width=0.66\columnwidth,angle=0]{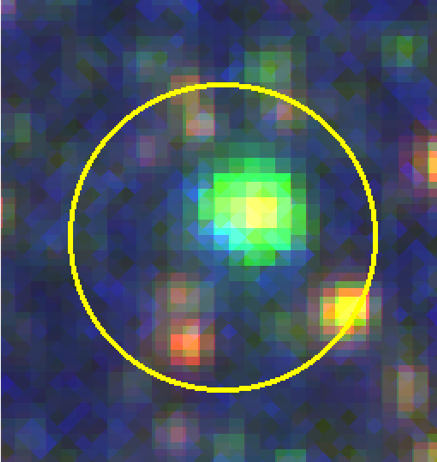}
	\includegraphics[width=0.65\columnwidth,angle=0]{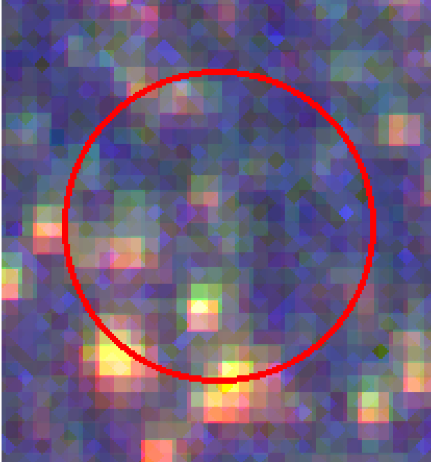}
	\caption{True colour images of PHAT-matched PNe. F814W, F475W, and F336W were used for the red, green, and blue images respectively. One ``likely'' (left), ``possible'' (center) and ``unlikely'' (right) PN has been shown. The circle denotes 0.5$''$ around the PN position.}
	\label{fig:phatpost}
\end{figure*}

\subsection{The Panchromatic Hubble Andromeda Treasury (PHAT)}
The Panchromatic Hubble Andromeda Treasury (PHAT\footnote{https://archive.stsci.edu/prepds/phat}) survey, covers $\sim$ 1/3 of the star-forming disk of M31 in six bands from the near-UV to the near-IR using HST’s imaging cameras (WFC3/IR, WFC3/UVIS, and ACS/WFC cameras). It combines the wide-field coverage typical of ground-based surveys with the precision of HST observations. The overall survey strategy, initial photometry and data quality (DQ) assessments were described in detail in \citet{dal12}. We utilize the second generation of photometric measurements of the resolved stars in the PHAT imaging \citep{wil14}, which takes advantage of all available information by carrying out photometry simultaneously in all six filters, resulting in a significant increase in the depth and accuracy the photometry over that presented in \cite{dal12}.

\begin{figure}[htb]
	\centering
	\includegraphics[width=\columnwidth,angle=0]{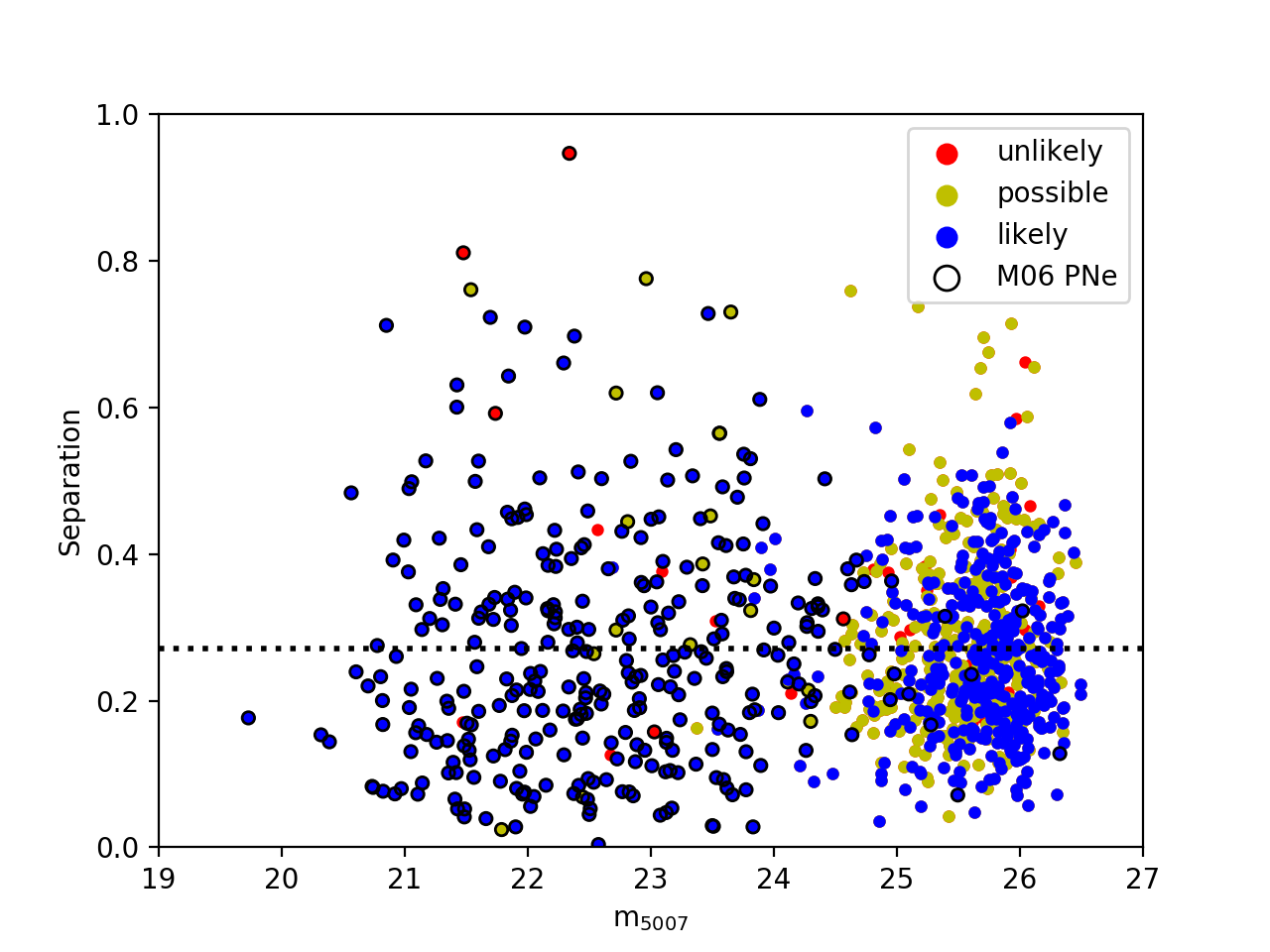}
	\caption{The positional separation between the PNe and their PHAT counterpart plotted against m$\rm_{5007}$. The dashed line shows the mean positional separation. The ``likely'', ``possible'' and ``unlikely'' PNe have been shown in blue, yellow and red respectively. The PNe previously found by M06 have been encircled in black.}
	\label{fig:phatsep}
\end{figure}

\begin{figure}[htb]
	\centering
	\includegraphics[width=\columnwidth,angle=0]{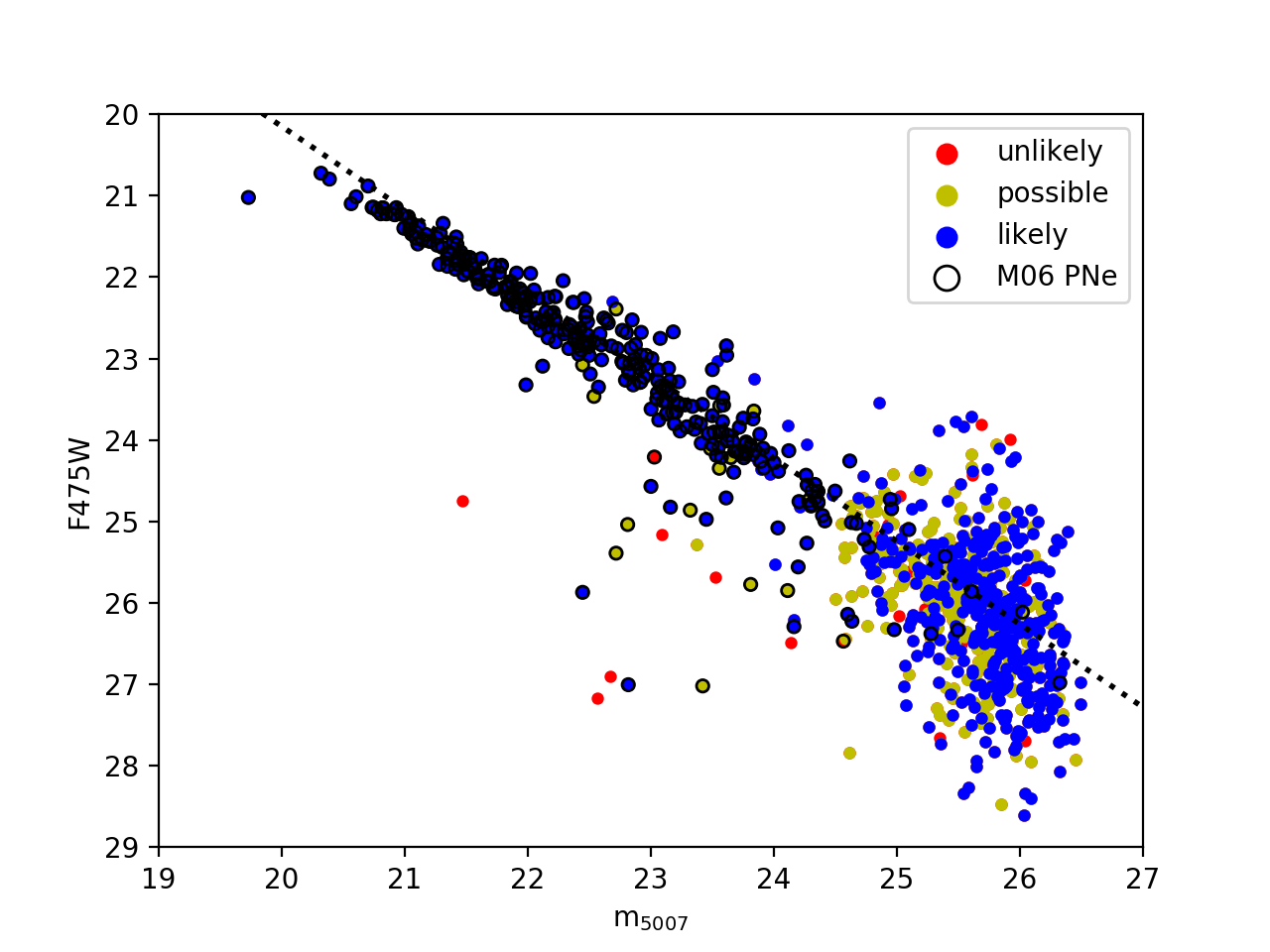}
	\caption{The F475W mag of the PHAT-matched PNe plotted against its m$\rm_{5007}$. The dashed line shows their relation described by \citet{vey14}. The ``likely'', ``possible'' and ``unlikely'' PNe have been shown in blue, yellow and red respectively. The PNe previously found by M06 have been encircled in black.}
	\label{fig:phatmag}
\end{figure}

\begin{figure}[htb]
	\centering
	\includegraphics[width=\columnwidth,angle=0]{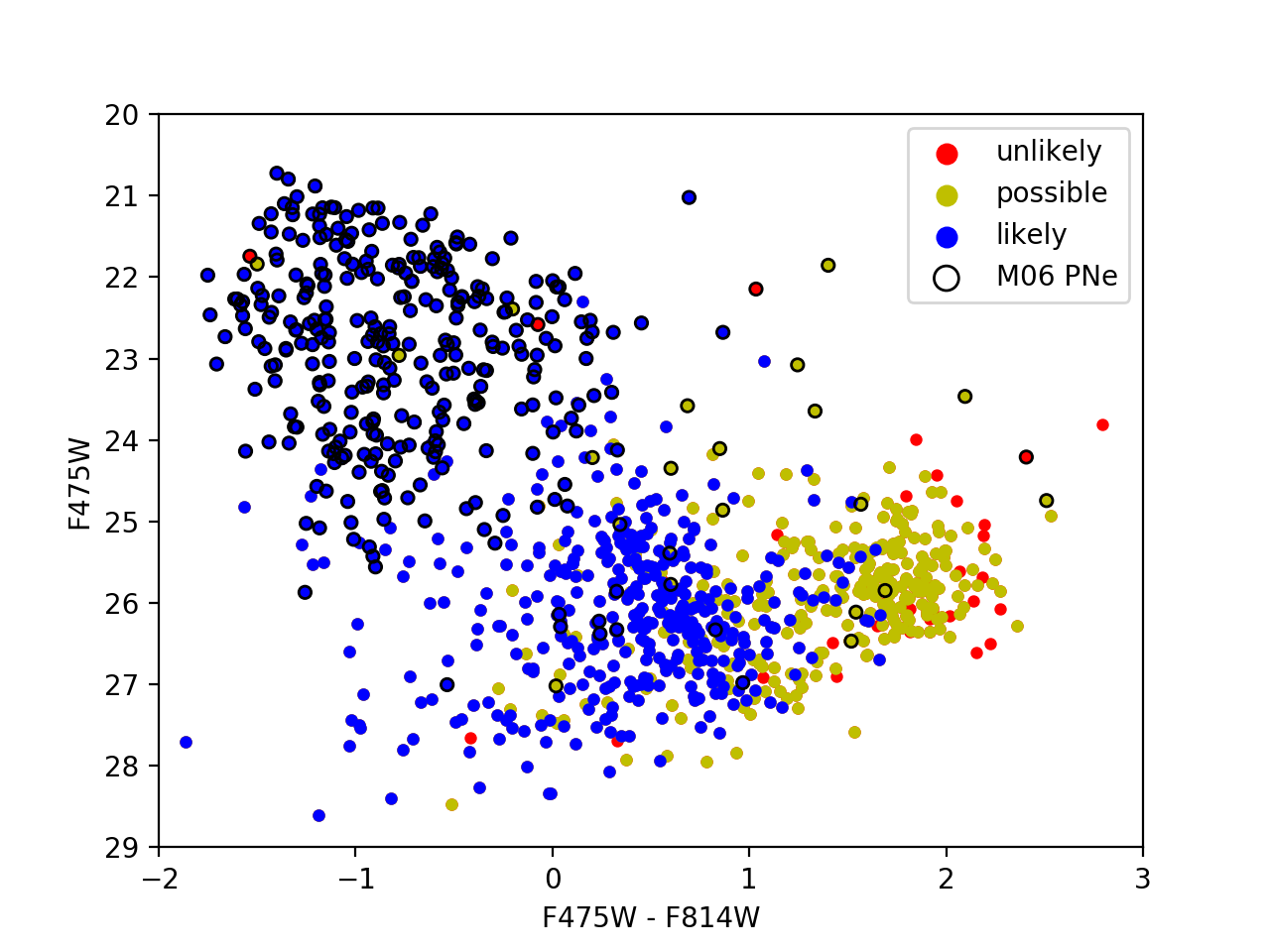}
	\caption{The F475W mag of the PHAT-matched PNe plotted against its F475W $-$ F814W colour. The ``likely'', ``possible'' and ``unlikely'' PNe have been shown in blue, yellow and red respectively. The PNe previously found by M06 have been encircled in black.}
	\label{fig:phatcol}
\end{figure}

\subsection{Finding PNe counterparts in PHAT}
\citet{vey14} conducted a search for M06 PNe counterparts in 16 of the 23 bricks in the PHAT footprint. {They find that $\sim 25\%$ of the PNe identified by M06 in the M31 disk are either \ion{H}{ii} regions resolved in HST or stellar contaminants.} They utilized the photometry presented in \citet{dal12}. They found a linear relation between the M06 m$\rm_{5007}$ magnitude and the PHAT F475W magnitude given by:
\begin{equation}
$$F475W = -0.2240 + 1.0187 \times \rm m_{5007}$$
\end{equation}
We adopt a similar method to identify PHAT counterparts to our PNe in the entire PHAT footprint. However, we search for our counterparts in the updated PHAT photometry from \citet{wil14}. We only consider those sources from the v2 star files whose square of the sharpness parameter in the F475W filter is below 0.2, to avoid cosmic rays and extended objects. The selection is done on the basis of the following parameters: 1. Difference between F475W mag and the F475W mag expected from the m$\rm_{5007}$ using the relation described by \citet{vey14}; 2. The F475W $-$ F814W colour; 3. The positional separation between the PNe and the PHAT counterpart; 4. The roundness of the PSF. To automatically select candidates on the basis of the above parameters, we construct an initial training set with the PNe in common with \citet{vey14}. We calculated the average value ($\bar{X}\rm_{PN}$) and 1$\sigma$ spread ($\sigma\rm_{PN}$) of each identification parameter. We found that the separation is within $0.7''$. For every PNe, the differences between the training set ($\bar{X}\rm_{PN}$) and the source ($X\rm_s$) parameter values were normalized by the $\sigma\rm_{PN}$ of the parameter values for each source in PHAT within $1''$ of our PNe location. These normalized parameters are of the form:
\begin{equation}
$$Z\rm_x = \frac{|X\rm_s-\bar{X}\rm_{PN}|}{\sigma\rm_{PN}}$$
\end{equation}
We obtain the sum of these normalized parameters, with half the weight given to roundness, as the merit function $Z\rm_{PN}$. 
\begin{equation}
$$Z\rm_{PN} = \frac{Z\rm_{F475W} + Z\rm_{colour} + Z\rm_{sep} + 0.5 \times Z\rm_{round}}{3.5}$$
\end{equation}
For each PNe, the PHAT source with the lowest value of the $Z\rm_{PN}$ is considered as the counterpart. {The distribution of the assigned $Z\rm_{PN}$ with the m$\rm_{5007}$ of the PHAT-matched PNe is shown in Figure~\ref{fig:zpn}.} Thresholds in $Z\rm_{PN}$ are used to classify the PNe based on the $Z\rm_{PN}$ assigned to those PHAT sources which were not PNe counterparts. PNe with $Z\rm_{PN}<1.5$ are classified as ``likely'', those with $1.5<Z\rm_{PN}<2$ are classified as ``possible'' and those with $Z\rm_{PN}>2$ are classified as ``unlikely''. The ``likely'' PHAT sources have a significantly lower $Z\rm_{PN}$ than that of other PHAT sources in the search region, but those which are ``possible'' do not stand out quite so much. Those classified as ``unlikely'' may not be PNe at all as stellar PHAT sources can have $Z\rm_{PN} \sim 2$ even though most of them have $Z\rm_{PN}>4$. The spatial distribution of PHAT PNe counterparts is presented in Figure~\ref{fig:phatspat}.

\subsection{Characteristics of the PHAT-matched PNe}
Of the 1023 PNe in the PHAT footprint, 700 are classified as ``likely'', 292 as ``possible'' and 31 as ``unlikely''. True colour images of three PNe, one of each classification, is shown in Figure~\ref{fig:phatpost}. None of the PHAT PNe are resolved as \ion{H}{ii} regions. The 31 ``unlikely'' PNe ($\sim$3\% of the PHAT PNe) maybe stellar contaminants. Of the \ion{H}{ii} regions spectroscopically identified by \citet{san12}, 81 are present in the PHAT footprint. Many of these were misidentified by M06 as PNe. We misidentify only 3 of these \ion{H}{ii} regions, probably compact \ion{H}{ii} regions, as PNe which is a testament to the photometric quality and improved spatial resolution of our survey. 

The mean positional separation of our PNe and their PHAT counterparts is 0.27$''$ which is testament to the accurate astrometry of the image and the Elixir pipeline greatly benefiting our survey. Figure~\ref{fig:phatsep} shows the variation in separation with m$_{5007}$. The separation remains largely uniform throughout the range of m$_{5007}$. The correlation between m$\rm_{5007}$ and the PHAT F475W mag is also seen for our PNe (Figure~\ref{fig:phatmag}) although there is larger dispersion at the faint end. Similarly the variation of the the PHAT F475W mag with the F475W $-$ F814W colour (Figure~\ref{fig:phatcol}) shows that most of the faint sources that are classified as ``possible'' and ``unlikely'' have a larger F475W $-$ F814W colour. The dispersion in F475W mag and the larger colour for some of the faint sources may be due to the [\ion{O}{iii}] 5007 $\AA$ line not being as prominent compared to the continuum flux in F475W for the fainter PNe.




\section{Planetary Nebula luminosity-specific frequency}
The PNe population follows the sampled (bolometric) luminosity of its parent stellar population. The PN luminosity-specific frequency \citep[$\alpha$-parameter;][]{jacoby80} is the ratio of the total number of PNe, $N_{\rm PN}$, to the total bolometric luminosity of the parent stellar population, $L_{\rm bol}$, given by: 
\begin{equation}
$$\alpha$ =  $\frac {N_{\rm PN}}{L_{\rm bol}} = B\tau_{\rm PN}$$
\end{equation}
where B is the specific evolutionary flux ($\rm stars~yrs^{-1}~L_{\odot}^{-1}$) and $\tau_{\rm PN}$ is the PN visibility lifetime \citep{buz06}. The $\alpha$-parameter value provides the number of PNe expected per unit bolometric light. We determine the radial PN number density profile and by comparison with the stellar surface brightness profile of M31, derived from broad-band photometric studies, we compute the $\alpha$-parameter value, similar to the procedure followed for M87 \citep{longobardi13} and M49 \citep{hartke17}. 

\begin{figure}[t]
	\centering
	\includegraphics[width=\columnwidth,angle=0]{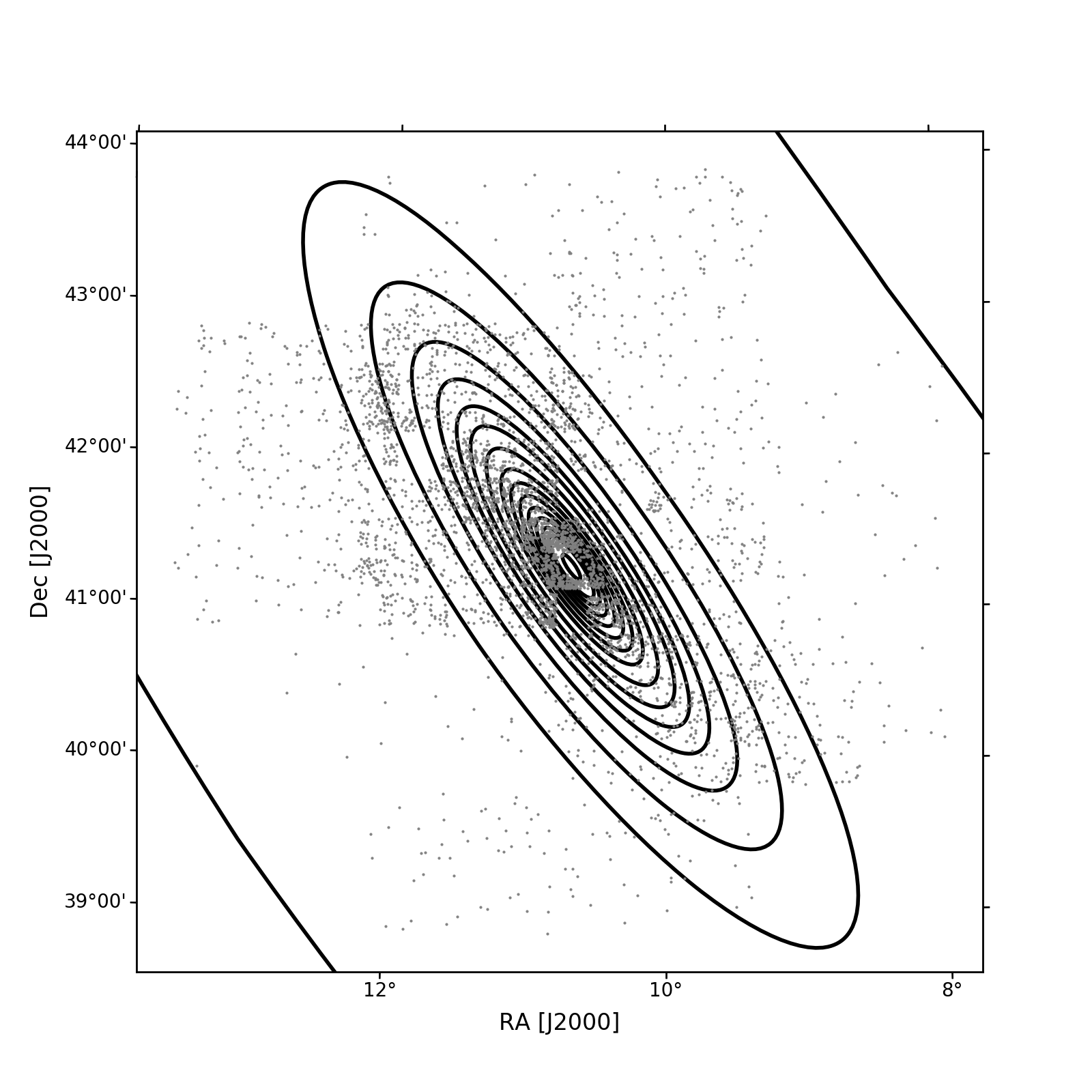}
	\caption{The PNe are shown spatially divided into elliptical bins containing equal number of PNe brighter than the 50\% completeness limit of the shallowest field (Field\# 33\_4).}
	\label{fig:alphaell}
\end{figure}

\begin{figure}[t]
	\centering
	\includegraphics[width=\columnwidth,angle=0]{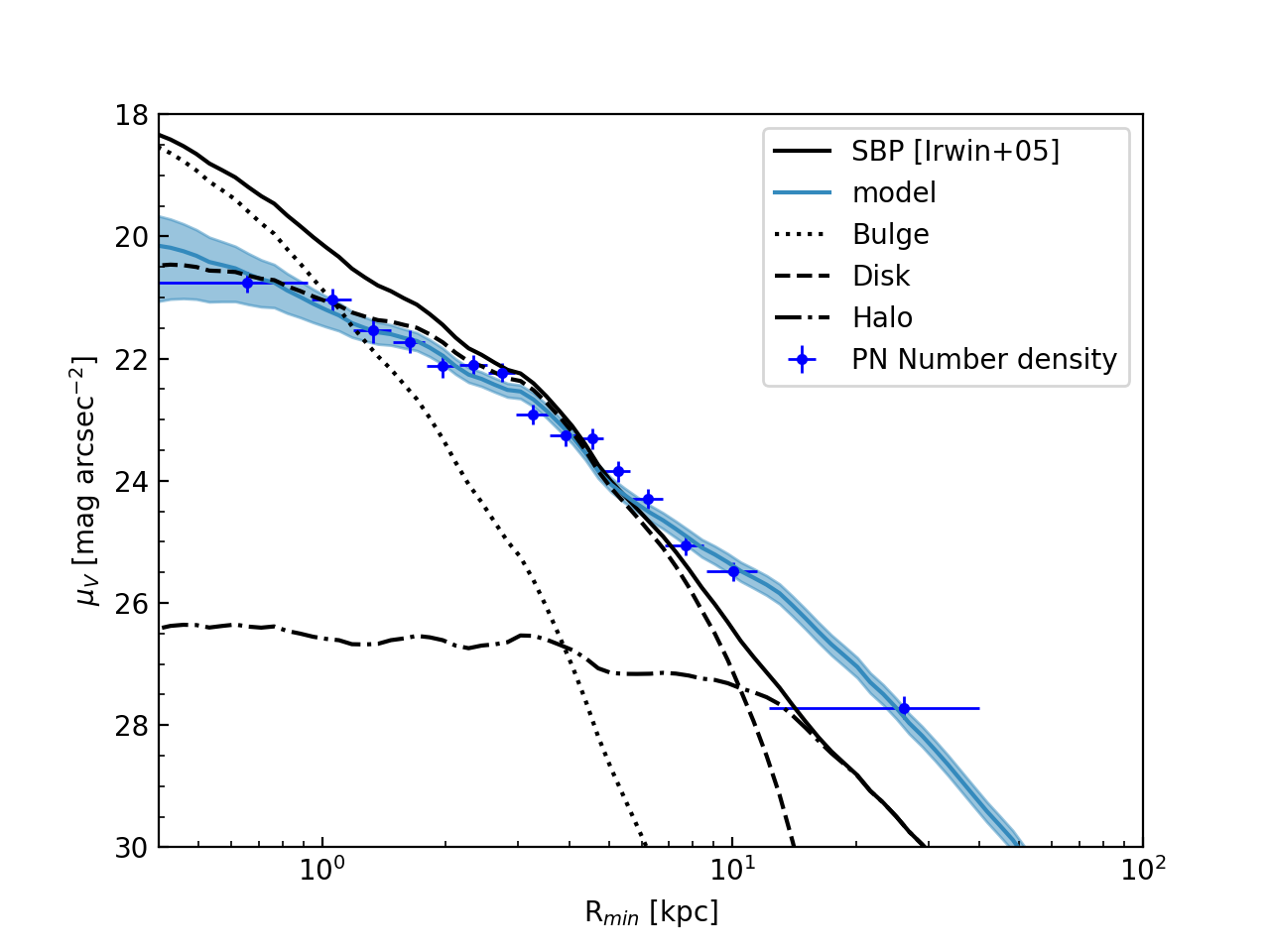}
	\caption{The radial surface brightness profile of M31 \citep{irwin05} is shown in black while the PN density, {using only the  PNe within 2.5 mag of the bright cut-off}, obtained at different elliptical bins are shown in blue. The \citet{cou11} decomposition of the M31 luminosity profile into the bulge, disk and halo components, scaled to the V-band surface brightness profile are also shown. The three-component photometric model for the predicted PN surface density is shown in blue.}
	\label{fig:alphaden}
\end{figure}

\subsection{The radial PN number density profile}
Figure~\ref{fig:alphaell} shows the distribution of PNe on the sky together with 15 elliptical bins aligned using the known position angle, PA = 38$^\circ$ of M31 and its ellipticity = 0.73 \citep{wk88}. The binning has been chosen such that each bin contains the same number of PNe  brighter than m$\rm_{5007} = 25.64$, which is the 50\% completeness limit of the shallowest field (Field\# 33\_4). These bins correspond to 0.65-26.13 kpc radial distances projected on the minor axis of M31, R$_{min}$. The PN logarithmic number density profile is defined as:
\begin{equation}
$$\mu_{\rm PN}(r)=-2.5 {\rm log_{10}}\Sigma_{\rm PN}(r)+\mu_{\rm0}$$
\end{equation}
where $\mu_{\rm0}$ is a constant in order to match the PN number density
profile with the stellar surface brightness profile and $\Sigma_{\rm PN}$ is the completeness-corrected PN number density which is in-turn given by:
\begin{equation}
$$\Sigma_{\rm PN}(r)=\frac{N_{\rm PN,corr}(r)}{A(r)}$$
\end{equation}
$N_{\rm PN,corr}$ is the completeness-corrected PN number in any elliptical bin. The completeness correction accounts for both detection and colour incompleteness, detailed in Sect~\ref{compcorr}. $A(r)$ is the observed area in any elliptical bin. {We obtain the density profile using only the  PNe within 2.5 mag of the bright cut-off (m$\rm_{5007} < 22.56$).} The resulting PN density profile in each elliptical bin is matched to the V-band surface brightness profile, obtained by \citet{irwin05} using photometry and number counts along the minor-axis of M31. It is shown in Figure~\ref{fig:alphaden}. We compute $\mu_{\rm0}=11.085 \pm 0.004$.  \citet{cou11} decomposed the M31 luminosity profile into the bulge, disk and halo components finding that the halo component becomes significant from  R$_{min} \sim8$ kpc and it dominates the luminosity profile over the disk from R$_{min} \sim11$ kpc. The PN density profile follows the surface brightness profile from R$_{min} \sim2-8$ kpc, where the disk dominates the luminosity profile. The M31 B-V colour, available out to R$_{min} \sim10$ kpc, shows a gradient towards bluer colour at larger radii \citep{wk88,tem10}. We find a flattening of $\mu_{\rm PN}$ with respect to the observed surface brightness profile at larger radial distances. This correlates with the B-V colour gradient and the increasing dominance of the halo component in the light-profile. Additionally, the light from the inner-halo substructures is underestimated by the surface brightness profile obtained along the minor-axis of M31 at larger radii. This may also lead to a flattening of $\mu_{\rm PN}$.

\begin{figure}[t]
	\centering
	\includegraphics[width=\columnwidth,angle=0]{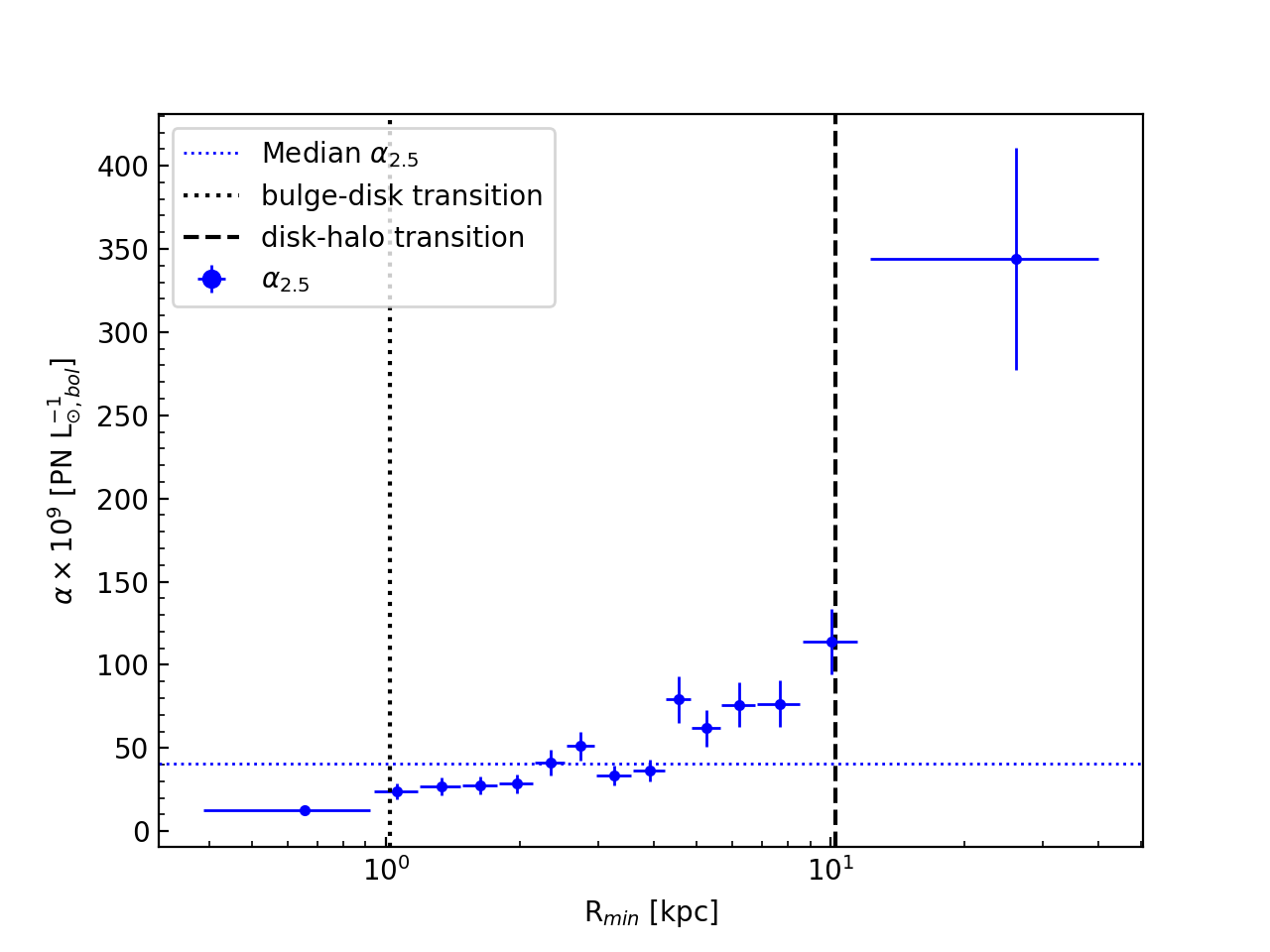}
	\caption{The $\alpha_{2.5}$ values obtained at different elliptical bins are shown in blue. {The dotted black line and the dashed black line show the bulge-disk transition and the disk-halo transition respectively from the \citet{cou11} decomposition of the M31 luminosity profile.}}
	\label{fig:alpharad}
\end{figure}

\subsection{The PN luminosity-specific number – the $\alpha$-parameter}
We can compute the value of $\alpha$ for the observed M31 PNe from $\mu_{\rm0}$ using the following relation:
\begin{equation}
$$\alpha$ =  $\frac{1}{s^2}10^{0.4 (\mu_{\rm0}-K-(BC_{\odot}-BC_V))}$$
\end{equation}
where the $K=26.4$ mag arcsec$^{-2}$ is the V-band conversion factor to physical
units L$_{\odot}$pc$^{-2}$ and $BC_{\odot}= -0.07$ is the solar bolometric correction. $s = D/206265$ is a scale factor related to the galaxy distance $D$. A fixed value of $BC_V = 0.85$ can be assumed with 10\% accuracy based on the study of stellar population
models for different galaxy types \citep[irregular to elliptical, see ][]{buz06}.

For the PNe within 2.5 mag of the bright cut-off, we obtain $\alpha_{2.5} = (40.55 \pm 3.74) \times 10^{-9}$ PN $L_{\rm \odot, bol}^{-1}$. For a PN population following the PNLF described by \citet{ciardullo89}, this corresponds to a $\rm log\alpha = \rm log(\alpha_{2.5}/0.1) = -6.39 \pm 0.04$ and $\tau_{\rm PN}=(22527.78 \pm 207.78)$ yr \citep[based on relations described in][]{buz06}. M06 obtained $\alpha_{2.5} = (15 \pm 2) \times 10^{-9}$ PN $L_{\rm \odot, bol}^{-1}$ but only within 1.8 kpc of the minor-axis radius. Scaling our PN number density profile to the surface brightness profile only within this interval, we find $\mu_{\rm0}=10.33 \pm 0.01$ corresponding to $\alpha_{2.5} = (20.16 \pm 1.87) \times 10^{-9}$ PN $L_{\rm \odot, bol}^{-1}$ which is closer to the value obtained by M06. This may be due to the bulge, disk and halo components of M31 having differing $\alpha$-parameter values. 

\subsection{A three-component photometric model for M31}
\citet{longobardi13} and \citet{hartke17} also saw differences in the $\alpha$-parameter values associated with different components of the surface brightness profile, although they looked at halo and Intra-group light components. Similar to their procedure, we describe a photometric model for the predicted PN surface density as:
\begin{equation}
$$\widetilde{\Sigma}_{\rm PN}(r) =  (\alpha_{2.5,\rm bulge} I_{\rm bulge}(r) + \alpha_{2.5,\rm disk} I_{\rm disk}(r) + \alpha_{2.5,\rm halo} I_{\rm halo}(r)) s^2$$
\end{equation}
where $\alpha_{2.5,\rm bulge}$, $\alpha_{2.5,\rm disk}$ and $\alpha_{2.5,\rm halo}$ are the $\alpha_{2.5}$ values associated with the bulge, disk and halo components of M31 respectively. $I_{\rm bulge}$, $I_{\rm disk}$ and $I_{\rm halo}$ are the surface brightness profiles of the bulge, disk and halo components of M31 respectively as per the \citet{cou11} decomposition of the M31 luminosity profile into the bulge, disk and halo components, scaled to the V-band surface brightness profile observations by \citet{irwin05}. 

We simultaneously fit the different $\alpha$-parameter values of the photometric model to the observed PN surface density to obtain $\alpha_{2.5,\rm bulge} = (5.28 \pm 6.25) \times 10^{-9}$ PN $L_{\rm \odot, bol}^{-1}$, $\alpha_{2.5,\rm disk} = (39.16 \pm 3.33) \times 10^{-9}$ PN $L_{\rm \odot, bol}^{-1}$ and $\alpha_{2.5,\rm halo} = (273.89 \pm 41.31) \times 10^{-9}$ PN $L_{\rm \odot, bol}^{-1}$. The model is also shown in Figure~\ref{fig:alphaden}. While the $\alpha_{2.5,\rm bulge}$ is not very well constrained, we find the $\alpha_{2.5,\rm disk}$ is quite close to the $\alpha_{2.5}$ obtained for the whole PN sample which is expected since the surface brightness profile of M31 is dominated by the disk component in our survey. We also find that the $\alpha_{2.5,\rm halo} \sim 7 \alpha_{2.5,\rm disk}$. 
 {The observed variation in $\alpha$-parameter values in different galaxies that spans almost 2 orders of magnitude is studied by \citet{buz06} who show that late-type spiral and irregular galaxies, with bluer B-V colours, are expected to have larger $\alpha$-parameter values than the redder early-type galaxies. Their analysis is based on population synthesis models of galaxies with different ages, metallicities and morphological types.} The different stellar populations in these galaxies thus exhibit very different $\alpha$-parameter values. \citet{hartke17} also found that $\alpha$-parameter values for the intra-group light of M49 is 3 times larger than that of its halo. \citet{hartke18} later confirmed that the halo PNe and intra-group light PNe were kinematically different corresponding to two separate parent stellar populations. We can hence infer that the larger $\alpha$-parameter value of the bluer halo of M31 may indicate that the stellar population of the inner halo is different from that the disk. {While the bluer halo is thus expected to have a higher $\alpha_{2.5}$ than the disk, its absolute value measured for the M31 halo is the highest observed in any galaxy.}


\begin{figure}[t]
	\centering
	\includegraphics[width=\columnwidth,angle=0]{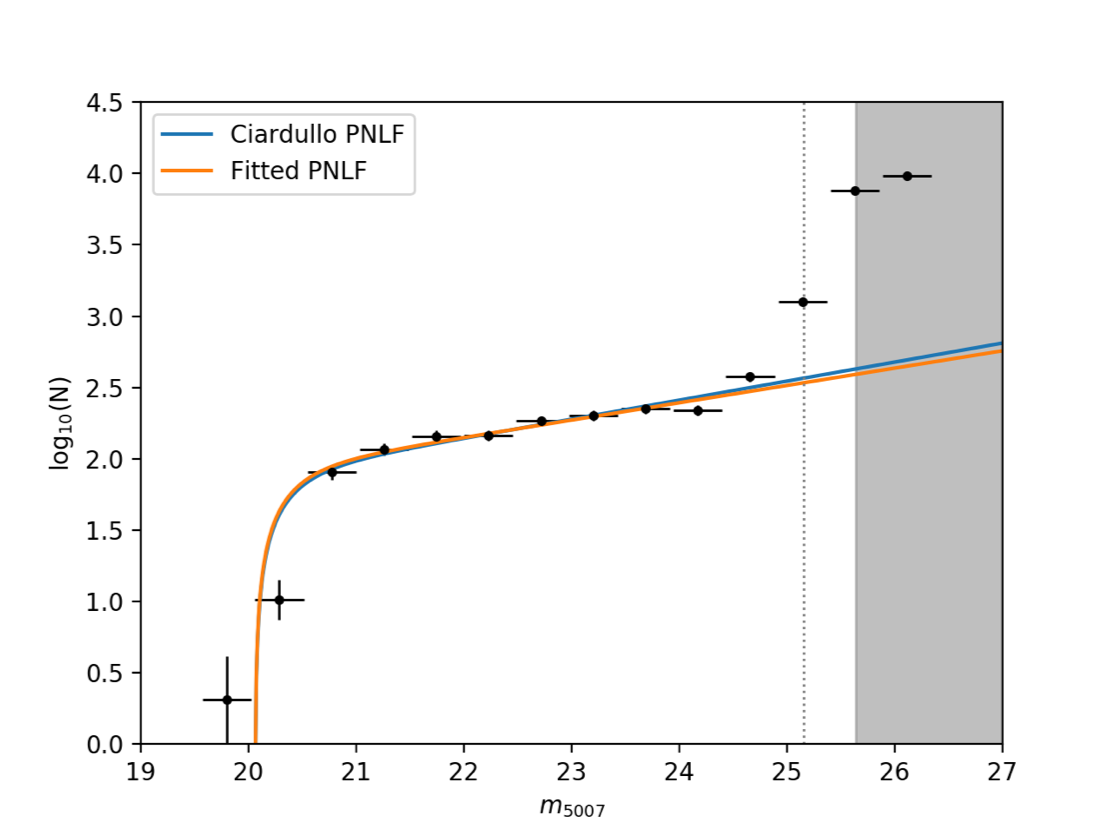}
	\caption{The completeness-corrected PNLF for the whole catalog of M31 PNe is shown fitted by both the generalised analytical formula (in orange) and the \citet{ciardullo89} analytical LF (in blue).The region beyond the limiting magnitude of the shallowest field (Field\# 33\_4) is shown in grey. The grey dotted line shows the 90\% completeness limit of the shallowest field.}
	\label{fig:pnlf}
\end{figure}

We also obtain the radial variation of the $\alpha$-parameter values (Figure~\ref{fig:alpharad}) by scaling the PN number density individually in each elliptical bin, which shows an increase at larger radii. The increasing trend in the $\alpha$-parameter values is a direct consequence of the flattening of $\mu_{\rm PN}$, given the B-V colour gradient of M31. The radial variation of the $\alpha$-parameter values found by M06 also showed an slight increasing trend within the errors at larger radii.  However, for the furthest radial bin in Figure~\ref{fig:alpharad}, the minor axis surface brightness profile may be underestimating the light from the known inner halo substructures of M31 primarily found along the disk major-axis. This may also be contributing to the large $\alpha_{2.5,\rm halo}$ value. More accurate surface brightness profile measurements need to be performed in order to quantify the contribution of substructures to the minor axis surface brightness profile of M31.

\section{Planetary Nebula Luminosity Function}
\label{sec:pnlf}
For different galaxies, the generalised analytical formula for the PNLF introduced by \citet{longobardi13} is:
\begin{equation}
$$N(M)=c_1e^{c_2M}(1-e^{3(M^{*}-M)})$$
\end{equation}
where $c_1$ is a normalisation constant, $c_2$ is the slope at the faint end, and $M^{*}$ is the absolute magnitude of the LF’s bright cut-off. The \citet{ciardullo89} analytical LF is then a specific case of the generalised analytical formula with $c_2$ = 0.307 that reproduces their best fit to the PNLF of M31. Observations suggest that
the slope described by the parameter $c_2$ is correlated with the star formation history of the parent stellar population \citep{ciardullo04, ciardullo10,longobardi13,hartke17}. 
In order to ascertain the robustness of the morphology of the PNLF at magnitude ranges not reached before, we investigate the M31 PNLF from our survey with different independent methods.

\begin{figure}[t]
	\centering
	\includegraphics[width=\columnwidth,angle=0]{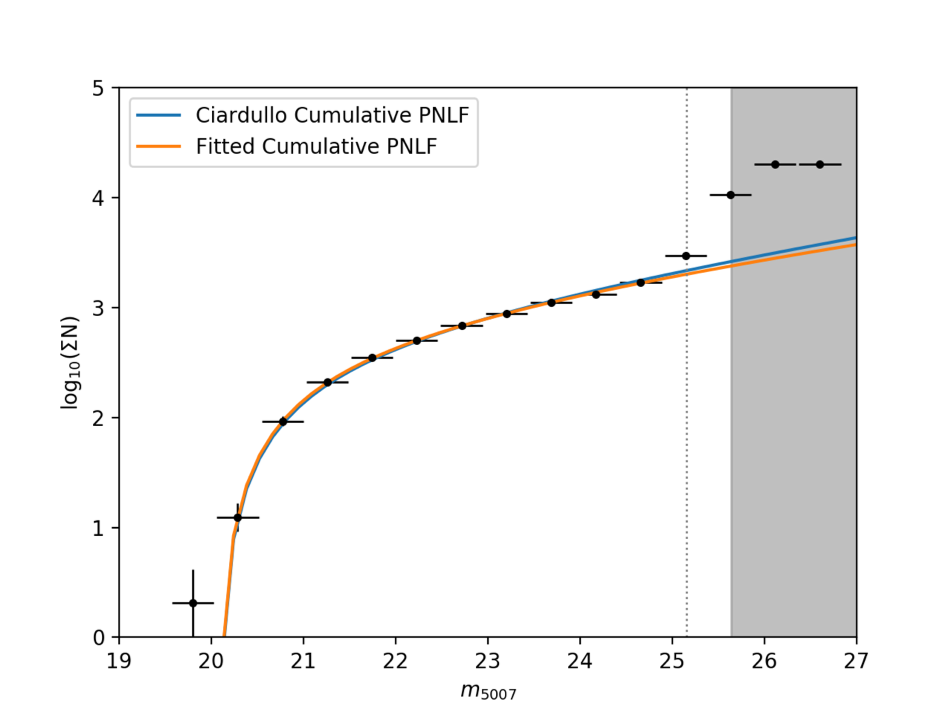}
	\caption{The completeness-corrected cumulative PNLF (binned for visual clarity) for the whole catalog of M31 PNe is shown fitted by both the generalised analytical formula for the cumulative PNLF (in orange) and the  cumulative PNLF corresponding to the \citet{ciardullo89} analytical formula (in blue). The region beyond the limiting magnitude of the shallowest field (Field\# 33\_4) is shown in grey. The grey dotted line shows the 90\% completeness limit of the shallowest field.}
	\label{fig:cum_pnlf}
\end{figure}

\subsection{PNLF of M31}
{The PNLF is corrected for detection completeness and also for selection completeness due to the colour and point-like selection, detailed in Sect~\ref{compcorr1}.} Figure~\ref{fig:pnlf} shows the PNLF of M31 for all the PNe identified by our survey, fitted by the generalised analytical formula with $c_2 = 0.279 \pm 0.024$, which agrees well with that previously found by \citet{ciardullo89}. Only the data corresponding to m$\rm_{5007} < 24$ are considered for the fit. The bright cut-off remains consistent with the known value of M*. The faint-end of PNLF shows a rise as compared to the fitted function. Such a rise is seen at m$\rm_{5007} > 24.5$ in all the fields of the survey. {This rise was not seen by M06 whose sample was photometrically complete to  brighter magnitude (m$\rm_{5007} = 23.5$). }

\begin{figure}[t]
	\centering
	\includegraphics[width=\columnwidth,angle=0]{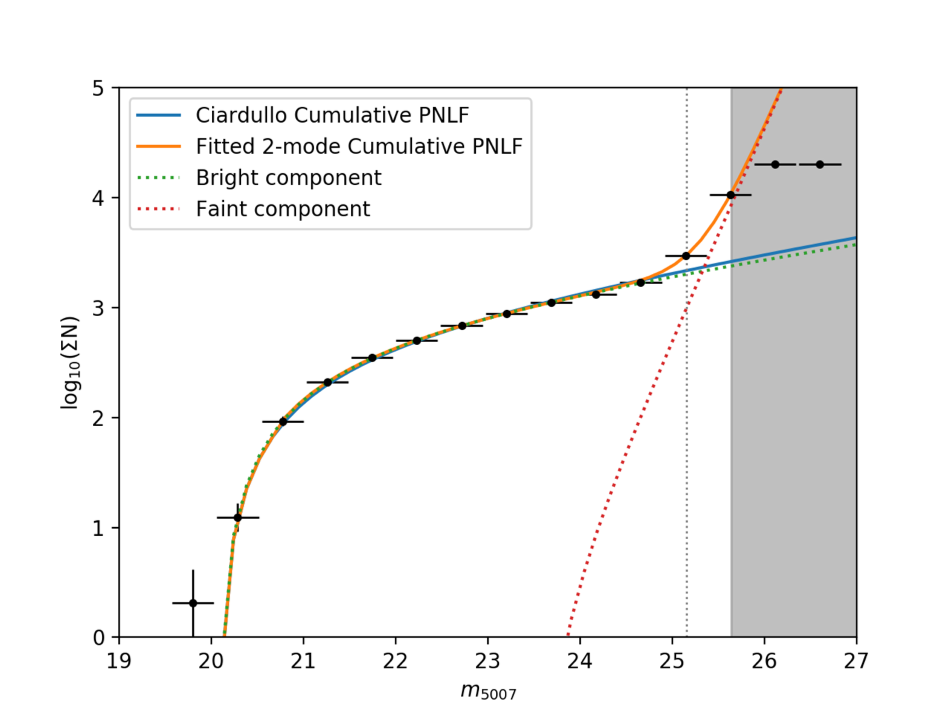}
	\caption{{The completeness-corrected cumulative PNLF (binned for visual clarity) for the whole catalog of M31 PNe is shown. It is fitted with two-modes of the generalised analytical formula for the cumulative PNLF (in orange) with one component dominating the brighter-end (in green) and another dominating the faint-end (in red). The cumulative PNLF corresponding to the \citet{ciardullo89} analytical formula (in blue) is also shown. The region beyond the limiting magnitude of the shallowest field (Field\# 33\_4) is shown in grey. The grey dotted line shows the 90\% completeness limit of the shallowest field.}}
	\label{fig:2m_pnlf}
\end{figure}

\begin{figure}[t]
	\centering
	\includegraphics[width=\columnwidth,angle=0]{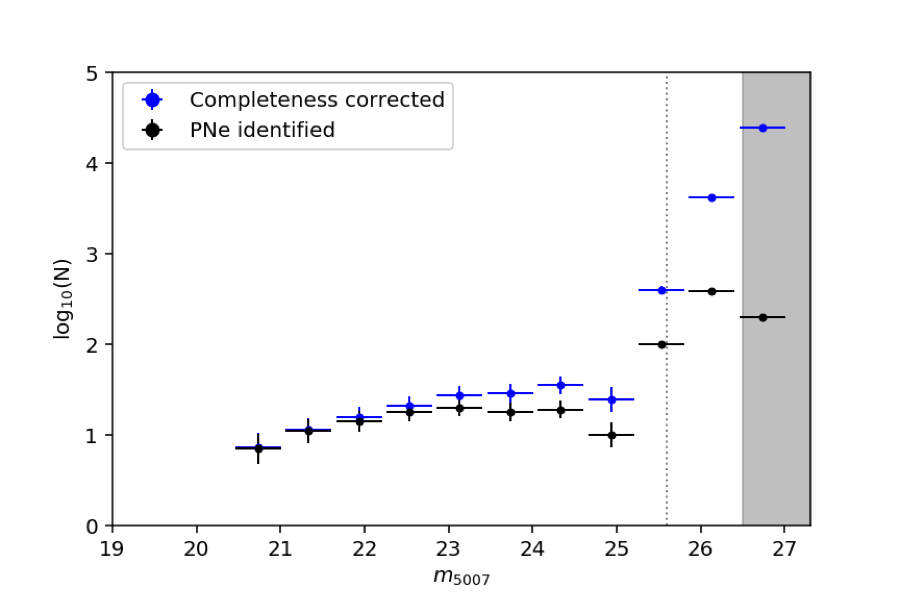}
	\caption{{For the deepest field (Field\# 35\_4), the observed number of PNe in each magnitude bin is shown in black, while the completeness corrected number is shown in blue. The region beyond the limiting magnitude of this field is shown in grey. The grey dotted line indicates the 90\% completeness limit of this field.}}
	\label{fig:pnlf_35_4}
\end{figure}

\begin{figure}[t]
	\centering
	\includegraphics[width=\columnwidth,angle=0]{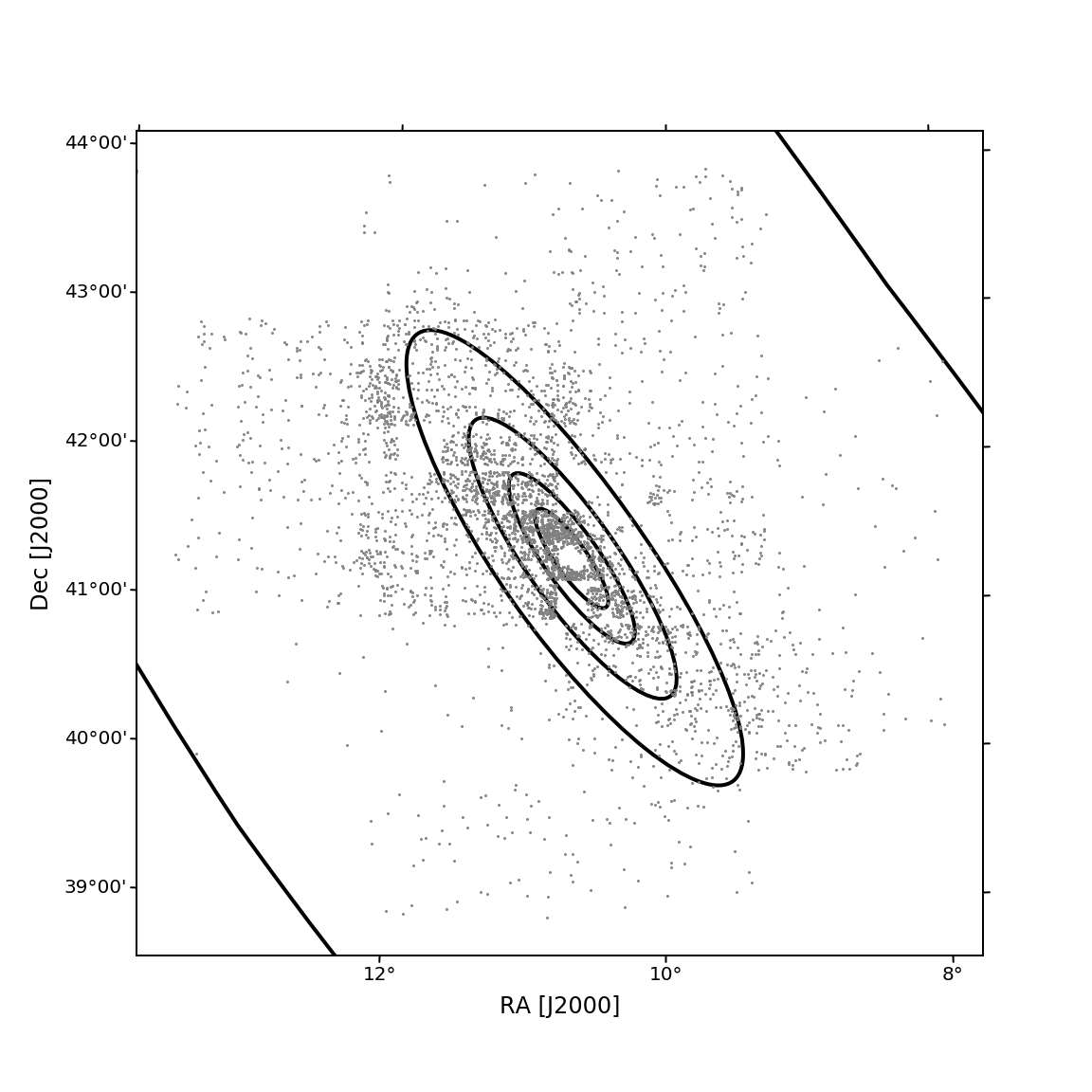}
	\caption{Same as Figure~\ref{fig:alphaell} but with 5 elliptical bins.}
	\label{fig:ellbins}
\end{figure}

\begin{figure*}[t]
	\centering
	\includegraphics[width=0.65\columnwidth,angle=0]{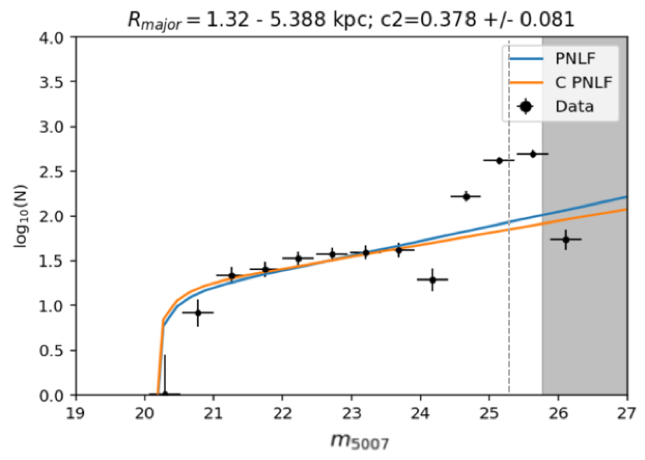}
	\includegraphics[width=0.65\columnwidth,angle=0]{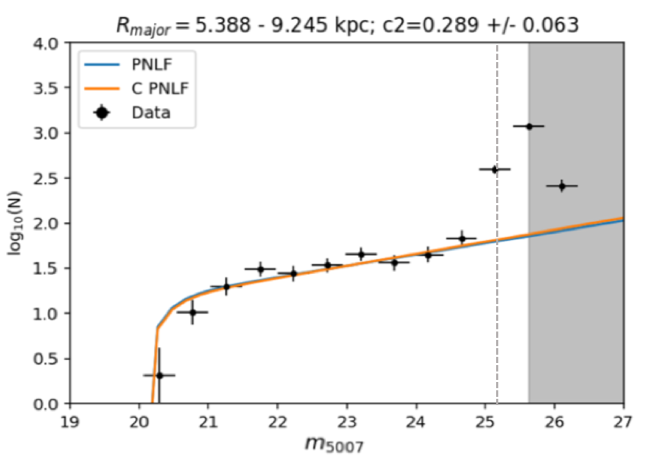}
	\includegraphics[width=0.65\columnwidth,angle=0]{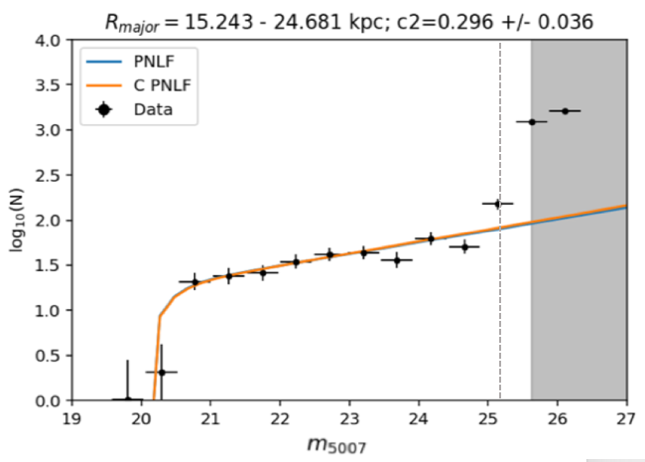}
	\includegraphics[width=0.65\columnwidth,angle=0]{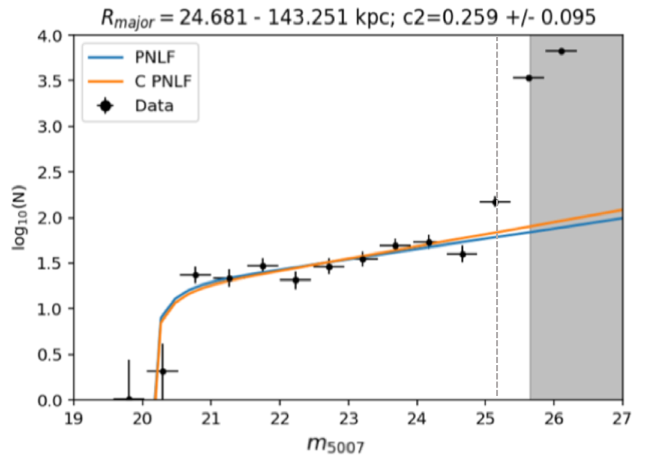}
	\includegraphics[width=0.65\columnwidth,angle=0]{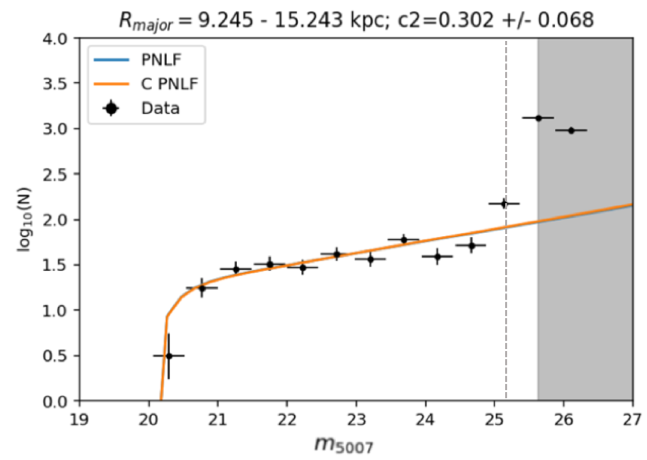}
	\caption{The completeness-corrected PNLF for each elliptical bin is shown fitted by both the generalised analytical formula (in blue) and the \citet{ciardullo89} analytical LF (in orange). The projected major axis elliptical radii covered by each bin is noted along with the fitted slope of the PNLF. The region beyond the limiting magnitude of the shallowest field (Field\# 33\_4) is shown in grey. The grey dotted line shows the 90\% completeness limit of the shallowest field.}
	\label{fig:ellpnlf}
\end{figure*}

\subsection{Cumulative PNLF}
The M31 PNe can be fitted to a cumulative luminosity function  \citep[e.g.][for NGC 3109]{pena07} to avoid potential histogram binning considerations like the bin size, limits, or the position of the first bin. However, some important features of the canonical PNLF (like the dip) could be lost. \citet{rod15} presented the Cumulative PNLF corresponding to the analytical PNLF described by \citet{ciardullo89}. The Cumulative PNLF for the generalised PNLF is presented as follows:
\begin{equation}
$$I(M)=c_1e^{c_2M}[\frac{1}{c_2}e^{c_2M}+\frac{1}{3-c_2}e^{3(M^{*}+\mu)-(3-c_2)M}-(\frac{1}{c_2}+\frac{1}{3-c_2})e^{c_2(M^{*}+\mu)}]$$
\end{equation}
Figure~\ref{fig:cum_pnlf} shows the cumulative PNLF of M31 for all the PNe identified by our survey, fitted by the generalised analytical formula with $c_2 = 0.257 \pm 0.011$, which agrees well with that found from the canonical PNLF thereby corroborating that the effect histogram binning considerations on the canonical PNLF is negligible. Only the data corresponding to m$\rm_{5007} < 24$ are considered for the fit. The faint-end of cumulative PNLF also shows the rise as compared to the fitted function.

\subsection{Two-mode PNLF}
\label{2mode}
{The rise in the faint-end of the PNLF may be an indication that there are two PN populations, one dominating the brighter end and a second one at the fainter end. Thus we fit the observed cumulative PNLF with two modes similar to \citet{rod15}. The two-mode PNLF is defined as $I_{2m}(M) =I_{b}(M)+I_{f}(M)$, where $I_{b}(M)$ is the cumulative function for the generalized PNLF set to parameters found in the previous section, which accurately represents the observed PNLF before the rise in the faint end, and $I_{f}(M)$ is a cumulative function for the generalized PNLF considered with free parameters, $c_{f1}$, $c_{f2}$ and $M^{*}_{f}$ which are the normalisation constant, slope and bright cut-off of a possible second faint PNe population. \citet{rod15} had included an additional magnitude cut-off in their definition of the two-mode PNLF where the contribution of the brighter mode was set to zero at this magnitude and only the faint-end contribution was present. We do not include such a magnitude cut-off because any PN population, such as the one dominating in the brighter-end, should have a contribution down to $\sim8$ mag below the bright cut-off \citep{buz06}. The two-mode fit of the observed PNLF is shown in Figure~\ref{fig:2m_pnlf}. For the second possible PNe population, we find $c_{f2}= 4.4 \pm 0.1$ and $M^{*}_{f}= -1 \pm 0.3$. Thus, the observed cumulative PNLF may have a second PN population with a much steeper slope and very different bright cut-off. }

\subsection{PNLF of the deepest field}
\label{deepest}
{The deepest field of our survey is Field\# 35\_4 which has a limiting magnitude at m$\rm_{5007} = 26.4$ and is 90\% complete at m$\rm_{5007} = 25.6$. The PNLF for the PNe observed in this field is shown in Figure~\ref{fig:pnlf_35_4} both before and after completeness correction. We note that the rise in the PNLF at m$\rm_{5007} > 25$ is present in the observed PNLF within the 90\% completeness limit. Thus the rise of the PNLF is a physical property of the observed PN sample.}

\subsection{Radial variation of PNLF}
\label{rad}
We divide the PNe spatially into 5 elliptical bins (Figure~\ref{fig:ellbins}) similar to Figure~\ref{fig:alphaell}. The PNLF corresponding to each bin is shown in Figure~\ref{fig:ellpnlf}. Only the data corresponding to m$\rm_{5007} < 24$ are considered to fit each PNLF. We observe that the rise in the PNLF remains invariant as we go radially outwards indicating that the rise is ubiquitous throughout the surveyed area and not a function of photometry in crowded areas. \citet{pas13} identified PNe in the high metallicity nuclear region within 80 pc of the center of M31 with HST and SAURON data. They found a PNLF with a paucity of bright PNe within $\sim1$ mag of the bright cut-off. We find that the bright cut-off remains invariant as we go radially outwards but this is expected because the median metallicity of the disk and inner halo stars is largely uniform as the metal-poor inner halo is colligated with stars associated with the more metal-rich substructures \citep{ibata14}. We note that the PNLF corresponding to the innermost bin, covering part of the bulge of M31 which is mostly saturated in our survey, has a steeper fitted slope as expected from PNe corresponding to an older parent stellar population \citep{longobardi13}.

\begin{figure}[t]
	\centering
	\includegraphics[width=\columnwidth,angle=0]{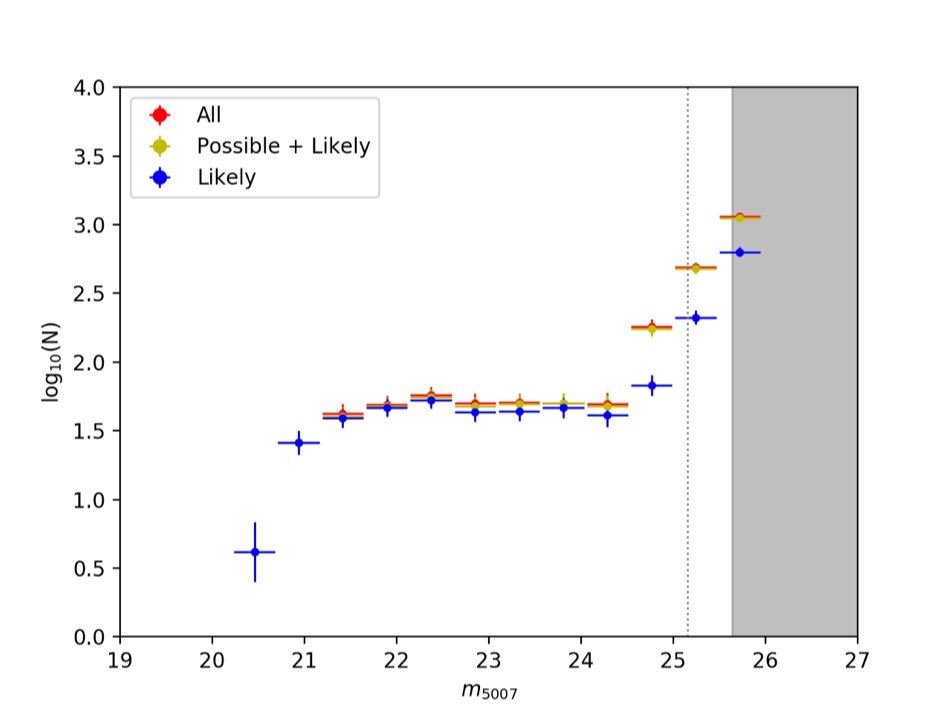}
	\caption{The completeness-corrected PNLF for PHAT-matched PNe is shown for the ``likely'' (blue), ``possible'' + ``likely'' (yellow) and ``possible'' + ``likely'' +``unlikely'' (red). The region beyond the limiting magnitude of the shallowest field (Field\# 33\_4) is shown in grey. The grey dotted line shows the 90\% completeness limit of the shallowest field.}
	\label{fig:phatpnlf}
\end{figure}

\subsection{PNLF of PHAT-matched PNe}
\label{phat_pnlf}
Figure~\ref{fig:phatpnlf} shows the PNLF of the PHAT-matched PNe. The faint-end of the PNLF still shows a rise even for the conservatively-selected ``likely'' PNe. This indicates that the faint PNe have colour and magnitude consistent with those of the bright PNe. 

\begin{figure}[t]
	\centering
	\includegraphics[width=\columnwidth,angle=0]{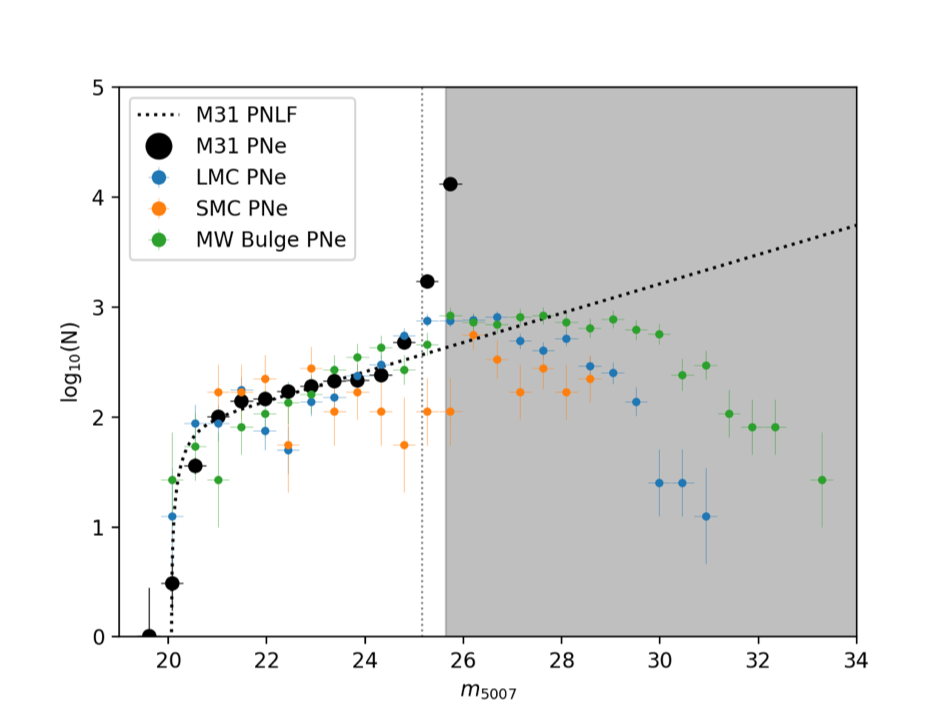}
	\caption{The completeness-corrected PNLF for the whole catalog of M31 PNe (black points) is shown fitted by the \citet{ciardullo89} analytical LF (black dotted line). The LMC \citep{rp10}, SMC \citep{jd02} and MW Bulge PNe \citep{kov11} shifted to the distance of M31 are shown in blue, orange, and green respectively. The region beyond the limiting magnitude of the shallowest field (Field\# 33\_4) is shown in grey. The grey dotted line shows the 90\% completeness limit of the shallowest field.}
	\label{fig:pnlfmc}
\end{figure}

\subsection{Comparison with other deep PN surveys in Local Group galaxies}

We compare the shape of the M31 PNLF with that of other surveys in different galaxies that sample similar magnitude intervals from the bright cut-off. A catalog of PNe with m$\rm_{5007}$ down to $\sim$ 10 mag below the bright cut-off of the PNLF is available for the Large Magellanic Cloud (LMC) from \citet{rp10}. This catalog is not completeness corrected as the m$\rm_{5007}$ magnitudes are estimated from spectroscopy but is expected to be largely complete to $\sim6$ mag below the bright cut-off, i.e., the magnitude interval covered by our survey of PNe in M31. Another catalog of PNe is available for the Small Magellanic Cloud (SMC) from \citet{jd02} with m$\rm_{5007}$ down to $\sim$ 8 mag below the bright cut-off. It is also not corrected for completeness but has significantly fewer PNe than the LMC and may suffer from significant completeness issues $\sim6$ mag below the bright cut-off. A catalog of PNe identified in the MW bulge is also available \citep{kov11} and assuming that all their PNe are at a constant 8 kpc distance \citep{majaess10}, the average distance to the MW bulge, a PNLF maybe constructed. The distance approximation maybe inaccurate and completeness information is also unavailable.

Considering the LMC, SMC and MW bulge PNe at the distance of M31 normalized for the number of PNe in M31, we can compare the shape of their PNLF with that of M31 (Figure~\ref{fig:pnlfmc}). We note that the difference in the bright cut-off is expected from the difference in metallicity between the three galaxies. The dip in the PNLF seen for both the LMC and the SMC, albeit at different magnitudes, is not seen for M31. The MW bulge PNe seem to show a different slope but no rise is evident. However, we can not disregard the possibility that a dip or a rise in the faint end of the PNLF may be seen pending accurate distance determination. The rise in the faint end of the PNLF of M31 is much steeper than any of the others. 

\section{Discussion}
\subsection{Possible sources of contamination}
\label{contamination}
{As discussed in Section~\ref{pts}}, up to 50 continuum sources may contaminate our PNe catalog, most of which should be fainter than m$\rm_{5007} = 25$. {This is also corroborated for the PHAT counterparts where $\sim$3\% of the matched PNe may be infact stellar contaminants.} The contamination from Milky Way (MW) halo and disk PNe is negligible as there are no MW halo PNe at m$\rm_{5007}$ > 20.5 \citep{kov11}. Background galaxies at redshift z = 0.345 which are [OII]3727 $\AA$ emitters can be another source of contamination but they are not known to have EW$_{obs}$ > 95 $\AA$ \citep{Colless90,Hammer97,Hogg98}. Our colour selection thus renders their contamination negligible. 

\ion{H}{ii} regions are also bright in the [\ion{O}{iii}] 5007 $\AA$ line and are present in the same region of the CMD as PNe. {However, \ion{H}{ii} regions appear extended at the distance of M31 and with the photometric quality of our survey, we are able to significantly limit their contamination as seen in Section~\ref{hst_phat}}. Some compact \ion{H}{ii} regions, especially in the disk of M31, may still contaminate our survey. M06 had classified as PNe 101 of the 253 \ion{H}{ii} regions identified by \citet{san12} later from spectroscopy. We find only 15 of these in our survey as PNe, thereby corroborating the excellent photometric quality of our survey. Other [\ion{O}{iii}] 5007 $\AA$ sources like Symbiotic Stars may also mimic PNe and contaminate our survey. {We investigate in the following whether the contribution from Symbiotic Stars is responsible for the two-mode PNLF investigated in Section~\ref{2mode}.}

\begin{figure}[t]
	\centering
	\includegraphics[width=\columnwidth,angle=0]{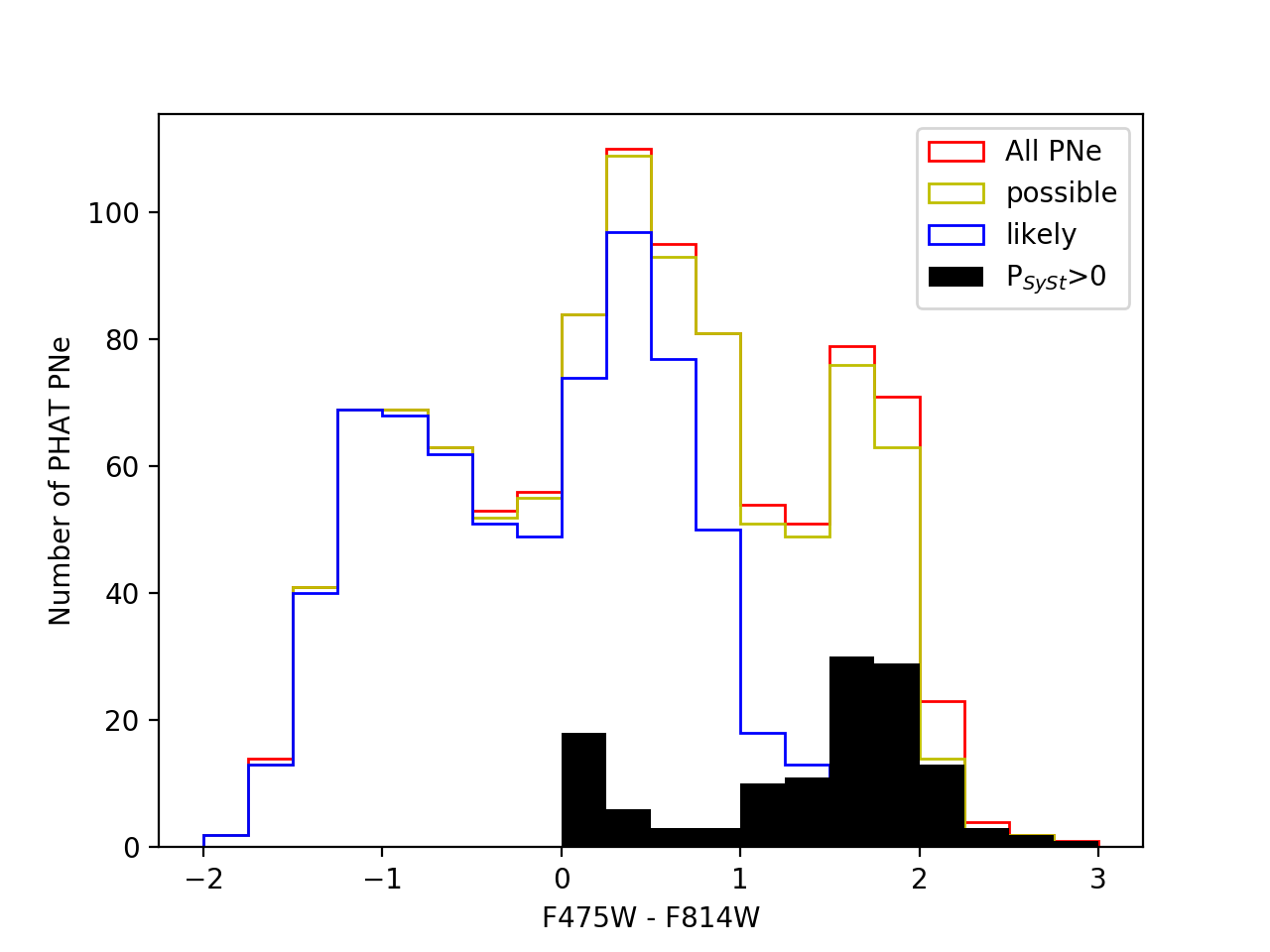}
	\caption{The histogram for PHAT-matched PNe is shown for the ``likely'' (blue), ``possible'' + ``likely'' (yellow) and ``possible'' + ``likely'' +``unlikely'' (red). The histogram of possible SySt are shown in black.}
	\label{fig:phatsyst}
\end{figure}

\begin{figure}[t]
	\centering
	\includegraphics[width=\columnwidth,angle=0]{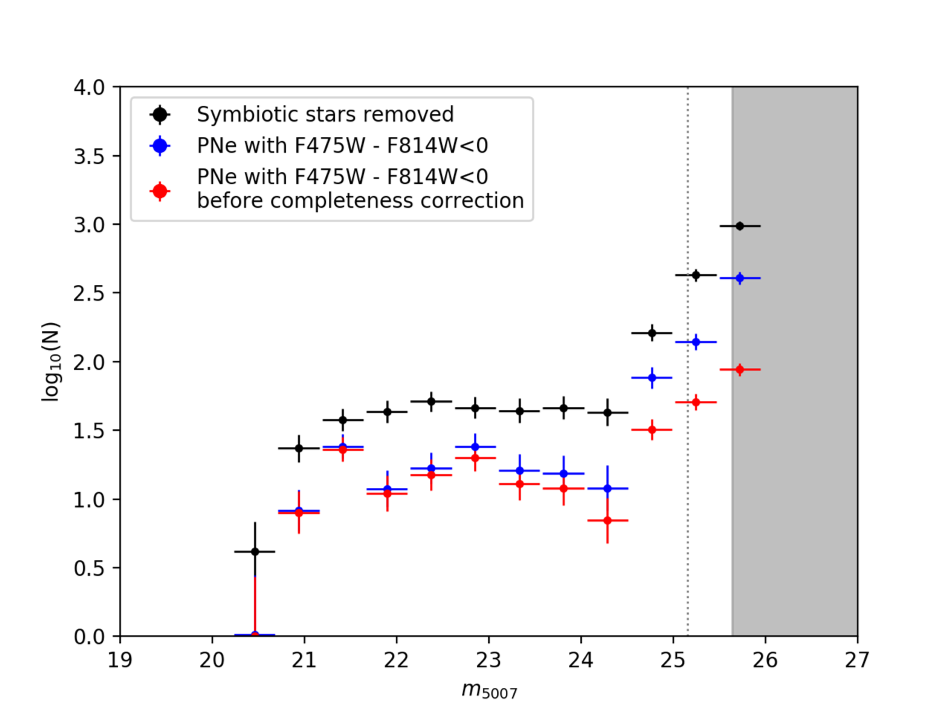}
	\caption{The completeness-corrected PNLF for PHAT-matched PNe with the contribution from SySt removed is shown in black. The completeness-corrected PNLF for only those PNe with $F475W - F814W < 0$ is shown in blue. {The PNLF for the observed PHAT-matched PNe with $F475W - F814W < 0$ is shown in red. The region beyond the limiting magnitude of the shallowest field (Field\# 33\_4) is shown in grey. The grey dotted line shows the 90\% completeness limit of the shallowest field. }}
	\label{fig:pnlfsyst}
\end{figure}

\subsection{Symbiotic Star contribution to PNLF}
Symbiotic Stars (SySt) are among the longest orbital period interacting binaries, consisting of an evolved cool giant and an accreting, hot, luminous companion (usually a White Dwarf) surrounded by a dense ionized nebula. Depending on the nature of the cool giant, there are two main classes of SySt: the S-types (stellar), which are normal M giants with orbital periods of the order of a few years, and the D-types (dusty), which contain Mira variable primaries surrounded by warm dust with orbital periods of a decade or longer.
\citet{mik14} found 31 confirmed symbiotic stars in M31, 10 of which had unambiguous PHAT counterparts. Of these, 9 were S-type symbiotic stars and their spectra show that the continuum emission is high in the F814W filter. Indeed \citet{mik14} utilized this to identify the PHAT counterpart of their SySt as that PHAT source which had the brightest F814W mag in their 0.75$''$ search radius. It is also seen that the F475W $-$ F814W colour of these SySt generally have a high positive value which can be used to distinguish them from PNe. 

We estimate the probability of each of our PHAT-matched PN being a SySt by checking how close it was in F814W mag ($F814W\rm_{PN}$) to the brightest F814W mag source ($F814W\rm_{high}$) in our search radius, and also if it had a colour excess in F814W as compared to F475W. The probability, $P\rm_{SySt}$ is given by the following formula:
\begin{equation}
$$P\rm_{SySt} = \frac{F814W\rm_{PN}-\overline{F814W}}{F814W\rm_{high}-\overline{F814W}} \times P\rm_{col}$$
\end{equation}
Here $\overline{F814W}$ is the mean F814W mag of all sources in the search radius. $P\rm_{col}$ is the colour excess check which equals 1 if $F475W - F814W > 1$ (most of the \citealt{mik14} SySt have this), 1/2 if $0 < F475W - F814W < 1$ (some of the confirmed PNe have this), and 0 otherwise. PHAT-matched PN with F814W below the $\bar{F814W}$ are set to have a $P\rm_{SySt} = 0$. We find that there is a systematic over-estimation of $P\rm_{SySt}$ since PHAT-matched PNe just above their $\overline{F814W}$ would still be assigned a small value of $P\rm_{SySt}$. We thus update the $P\rm_{SySt}$ by subtracting the mean of the probability, $\overline{P\rm_{SySt}}$ from each source and setting those with negative values to zero. We find that most PHAT-matched PNe with a probability of being a SySt are clustered around a high $F475W - F814W$ colour (Figure~\ref{fig:phatsyst}). Removing the contribution of SySt from the PNLF (Figure~\ref{fig:pnlfsyst}), it is evident that the rise in the faint end of the PNLF can not be explained by SySt. Figure~\ref{fig:pnlfsyst} also shows the PNLF for only those PHAT-matched PNe with $F475W - F814W < 0$ to show that the rise in the PNLF is not dependent on the $F475W - F814W$ colour. {Continuum stars are unlikely to have $F475W - F814W < 0$ and considering that we limit contamination from \ion{H}{ii} regions, the sources with $F475W - F814W < 0$ are most likely genuine PNe. The rise in the observed PNLF is indeed visible, even prior to completeness correction.}

D-type SySt do not show an excess in the F814W filter. Thus, in our analysis a D-type SySt can not be distinguished from a PNe. Yet, their numbers are expected to be far lower than that of a S-type SySt and so there should not be any significant number of D-type SySt plaguing our data. \citet{mik14} had one D-type SySt in the PHAT footprint which we indeed misidentify as a PNe in our study.

\begin{figure}[t]
	\centering
	\includegraphics[width=\columnwidth,angle=0]{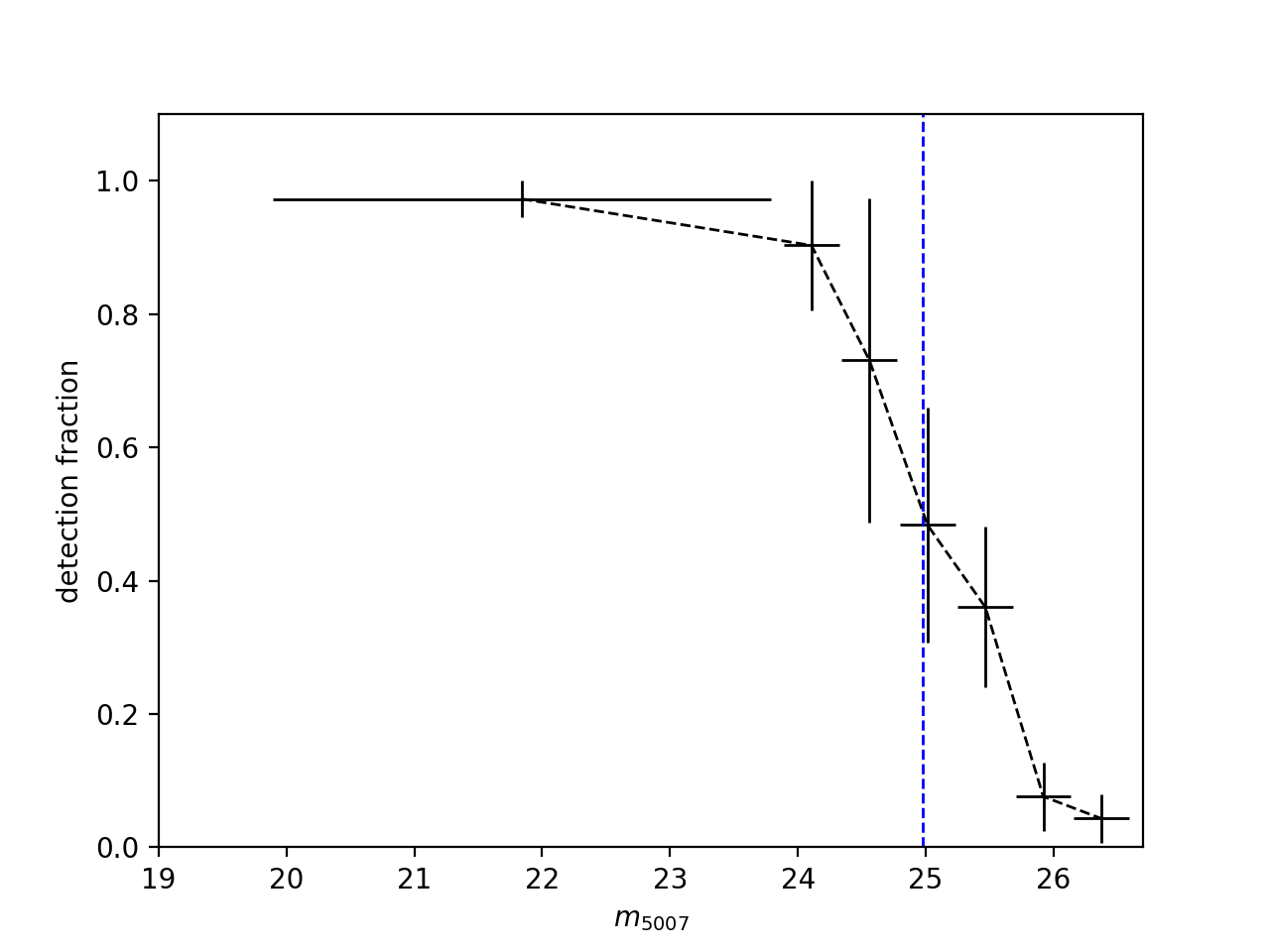}
	\caption{The fraction of PNe targeted with spectroscopic observations where a narrow [\ion{O}{iii}] 5007 $\AA$ emission-line was detected. {The uncertainty in detection fraction is the binomial proportion confidence-interval of observed PNe in any magnitude bin obtained using the Wilson score interval method \citep{Wilson27}. The blue dashed line shows the 50\% detection limit of the spectroscopic follow-up.}}
	\label{fig:detspec}
\end{figure}

\begin{figure}[t]
	\centering
	\includegraphics[width=\columnwidth,angle=0]{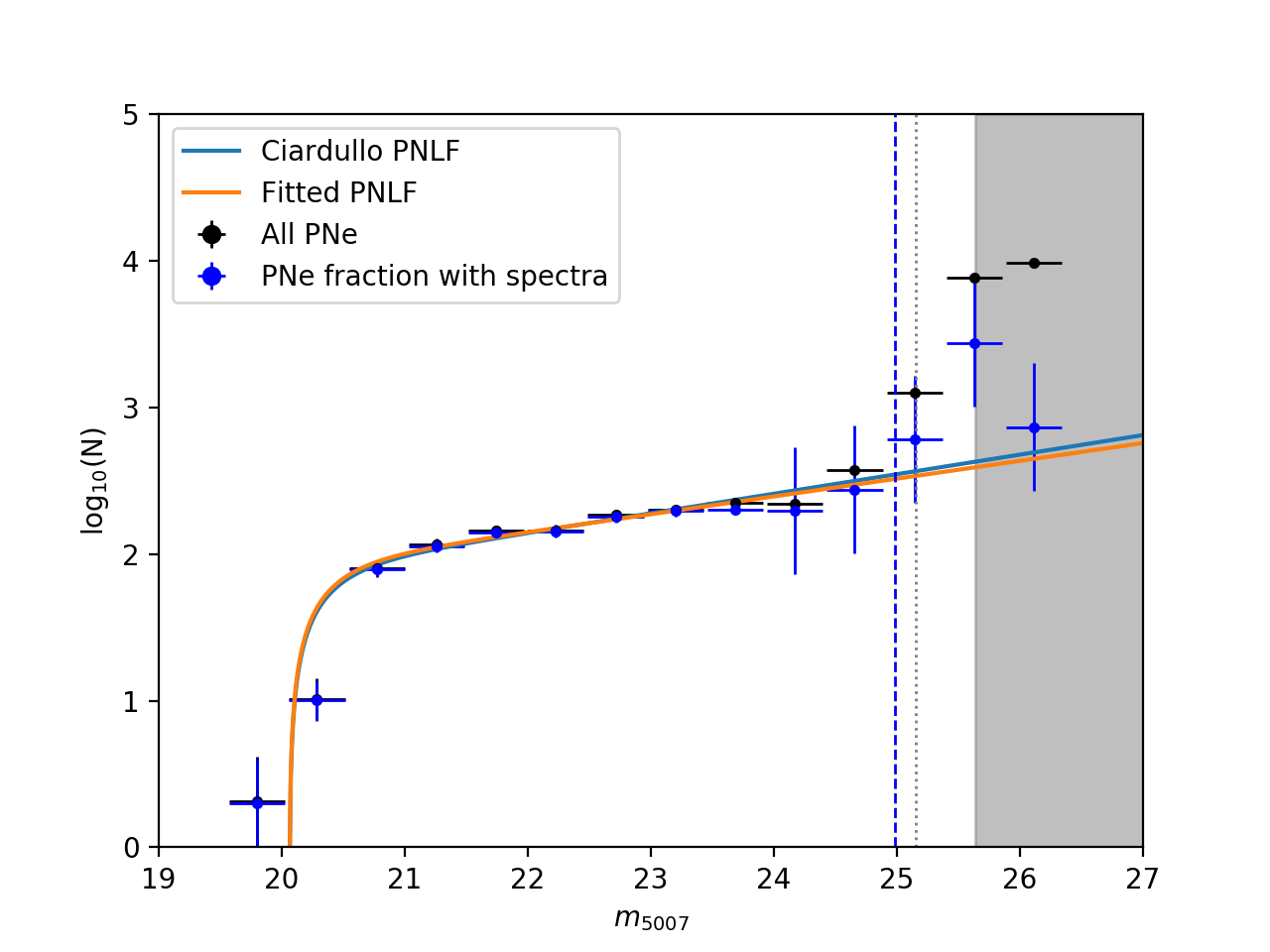}
	\caption{The completeness-corrected PNLF for all PNe is shown in black fitted by both the generalised analytical formula for the cumulative PNLF (in orange) and the  cumulative PNLF corresponding to the \citet{ciardullo89} analytical formula (in blue). The completeness-corrected PNLF accounting for the detection fraction from spectroscopy is shown in blue. The region beyond the limiting magnitude of the shallowest field (Field\# 33\_4) is shown in grey. The grey dotted line shows the 90\% completeness limit of the shallowest field. {The blue dashed line shows the 50\% detection limit of the spectroscopic follow-up.}}
	\label{fig:specpnlf}
\end{figure}

\subsection{Spectroscopically confirmed PNe and the faint-end of the PNLF}

{A spectroscopic follow-up of a complete subsample of the M31 PNe was carried out with the Hectospec multifiber positioner and spectrograph on the MMT \citep{fab05}. Observations were made on September 15, 2018 and October 10, 2018 with an exposure time of 9000 seconds each and also on December 4, 2018 with an exposure time of 3600 seconds.} The Hectospec 270 gpm grating was used and provided spectral coverage from 3650 to 9200 $\AA$ at a resolution of $\sim5~\AA$. Each Hectospec fiber subtends 1.5$\arcsec$ on the sky. {The fibres were placed on PNe candidates to maximise the observation of bright PNe, obtain their LOSV and determine their chemical abundances. Of the 343 PNe candidates observed in M31, 129 had confirmed detection of the [\ion{O}{iii}] 4959/5007 $\AA$  emission-lines. The [\ion{O}{iii}] 5007 $\AA$ emission-line was detected in all cases with a signal-to-noise ratio greater than 8. All of them also had the H$\alpha$ line present. Since our original PNe sample is largely devoid of \ion{H}{ii} regions, these observed sources are expected to be PNe. Details of the spectroscopy confirming them as PNe, along with their LOSV distribution and chemical abundances would be presented in a forthcoming paper (Bhattacharya et al. in preparation).} The fraction of PNe detected as a function of magnitude is shown in Figure~\ref{fig:detspec}. 

{The spectroscopic follow-up shows that all the PNe candidates observed were confirmed as PNe, from the presence of [\ion{O}{iii}] 4959/5007 $\AA$ and H$\alpha$ emission-lines, down to 24.5 magnitude but only a fraction of the targeted PNe candidates at fainter magnitudes could be confirmed. While the required emission lines may not have been detected due to low signal-to-noise in the spectra for the faint targeted PNe candidates, it is also possible that those faint sources are continuum contaminants or background galaxies instead.} Making the conservative assumption that all such sources are not PNe, we can modify the PNLF to account for this behavior. Since the rise in the PNLF is seen at different radii at about the same extent, we can compute the expected PNLF for the fraction of PNe that are conservatively confirmed spectroscopically as shown in Figure~\ref{fig:specpnlf}. {The PNLF is still consistent with the rise in the faint-end of the PNLF although the uncertainty in the fraction with spectral confirmation is large}.

\subsection{Summary of the observational evidence for the rise in the faint-end of the PNLF}
\label{summary}
{Our investigation of the morphology of the PNLF is carried out at different positions and radial distances, and considering the effects of possible contaminants. The rise in the faint-end of the PNLF is observed not only in the PNLF for the whole survey but also in individual fields, including the deepest field discussed in Section~\ref{deepest}. In this field, it is already visible for the observed PNe even before completeness correction.} 

{The rise in the faint-end of the PNLF occurs from m$\rm_{5007} \sim 24.5-25$ and is present at varying radial distances from the center of M31 (Section~\ref{rad}). If the rise in the faint-end of the PNLF was due to compact \ion{H}{ii} regions, it would be largely present only in the disk of M31 and not in all radial elliptical bins. In the survey region in common with \textit{HST} data from PHAT, we also see the rise in the PNLF for those PNe with reliable PHAT counterparts (Section~\ref{phat_pnlf}). The rise is especially evident when looking at the PNLF from PHAT-matched PNe with $F475W - F814W < 0$, a region in colour occupied almost exclusively by PNe, even prior to completeness correction (Figure~\ref{fig:pnlfsyst}). The spectroscopic follow-up also shows that the rise in the faint-end of the PNLF is seen in confirmed PNe, albeit with large errors. }

{For the aforementioned observational evidence, it is reasonable to believe that the rise of the PNLF is indeed physical and associated with the PN population and their parent stars.  \citet{rod15} fitted the two-mode PNLF for NGC 6822 (an irregular galaxy in the local group) and from its star formation history they show that the two-modes of the PNLF may correspond to PNe associated with the two episodes of star formation, with the older parent stellar population corresponding to the fainter PNe. It is possible that the second PN population in M31 is associated with an older stellar population.} With data from the PHAT survey, \citet{wil17} found that there was a burst of star formation $\sim2-4$ Gyr ago throughout the M31 stellar disc. \citet{bernard15} studied individual HST fields associated with the disk and inner halo substructures of M31 to find that all the fields show a burst of star formation $\sim 2$ Gyr ago even though most of the stars in the M31 outer disk formed $> 8$ Gyr ago. Since the slope of the PNLF associated with PNe belonging to an older stellar population is seen to be steeper {(Section~\ref{2mode})}, it is possible that the steep rise in the faint-end of the PNLF is caused by PNe associated with the older population while the PNe associated with the $\sim 2$ Gyr old burst of star formation populate the brighter magnitudes of the PNLF. 

Another possible reason for the rise in the faint-end of the PNLF could be a change in opacity of the nebula of the PNe. The PNe occupying the bright-end of the PNLF could be more opaque, with the nebula more efficiently reprocessing the ionizing flux of the central star into [\ion{O}{iii}] 5007 $\AA$ while the faint-end maybe populated by the more transparent PNe. The significant dependence of the PNLF on the considerations of the nebular transparency is also evident in the models by \citet{ges18}. Such an scenario has previously been invoked to describe the dip in the PNLF seen in the SMC \citep{jd02}.





\section{Conclusions}
We present a 16 sq. deg. survey  in the disk and inner halo of M31 with the MegaCam at CFHT using the narrow-band [\ion{O}{iii}] 5007 $\AA$ filter and the broad-band \textit{g} filter. We identify point-lke PNe from the colour excess between the narrow-band and broad-band images down to a limiting magnitude between m$\rm_{5007} = 25.64$ for the shallowest observed field to m$\rm_{5007} = 26.4$ for the deepest observed field. We obtain 4289 PNe, the largest sample outside the MW, of which 1099 were previously observed by M06. Using a simulated PN population, we are able to account for incompleteness. We find counterparts in PHAT for the 1023 of our identified PNe in the PHAT footprint. We find no resolved H II regions and a very small fraction ($\sim$3\%) of PNe which could be stellar contamination. The $\alpha$-parameter value obtained shows an increasing trend as we go further away from the center of M31 but matches reasonably with the value obtained by M06 for the central regions. The high value of the obtained $\alpha_{2.5,\rm halo}$ may be attributed to the bluer halo of M31 indicating that the stellar population in the halo, at radii larger than $\sim 10$ kpc,  may be different from that of the inner disk.

The PNLF of the whole sample is complete down to $\sim$5.5 mag fainter than the bright cut-off and shows a significant rise at the faint-end. The generalised analytical formula fitted to the completeness corrected PNLF, for magnitudes brighter than m$\rm_{5007} < 24$ , returns parameters that are in agreement with that previously determined by \citet{ciardullo89}. The rise in the faint end is seen at different elliptical radial distances from the center and also in the PHAT-matched PNe implying that such rise is not associated with crowding. The rise is steeper than that seen in the LMC and the SMC. It is also not caused by contaminating sources like symbiotic stars and seems to be a property of the parent population of the PNe, {evident from the comparison with PHAT}. Early findings from the spectroscopic follow-up also show that the rise in the number of PNe at faint magnitudes is physical,{ albeit with large uncertainty}. It is possible that the PNe in M31 associated with the population created from the burst of star formation $\sim 2$ Gyr ago populate the brighter magnitudes of the PNLF while those at the faint-end are associated with the older population. It is also possible that the faint-end is populated by PNe with more transparent nebulae while those with opaque nebulae populate the bright-end.

\begin{acknowledgements}
      SB and JH acknowledge support from the IMPRS on Astrophysics at the LMU Munich. We are grateful to the anonymous referee for constructive comments that improved the manuscript. This research made use of IRAF, distributed by the National Optical Astronomy Observatory, which is operated by the Association of Universities for Research in Astronomy (AURA) under a cooperative agreement with the National Science Foundation. This research made use of Astropy-- a community-developed core Python package for Astronomy \citep{Rob13}, Numpy \citep{numpy} and Matplotlib \citep{matplotlib}. This research also made use of NASA’s Astrophysics Data System (ADS\footnote{https://ui.adsabs.harvard.edu}).
\end{acknowledgements}


\bibliographystyle{aa} 
\bibliography{ref_pne.bib}

\begin{thebibliography}{76}
\expandafter\ifx\csname natexlab\endcsname\relax\def\natexlab#1{#1}\fi

\bibitem[{{Arnaboldi} {et~al.}(2002){Arnaboldi}, {Aguerri}, {Napolitano},
  {Gerhard}, {Freeman}, {Feldmeier}, {Capaccioli}, {Kudritzki}, \&
  {M{\'e}ndez}}]{arnaboldi02}
{Arnaboldi}, M., {Aguerri}, J. A.~L., {Napolitano}, N.~R., {et~al.} 2002, \aj,
  123, 760

\bibitem[{{Arnaboldi} {et~al.}(1998){Arnaboldi}, {Freeman}, {Gerhard},
  {Matthias}, {Kudritzki}, {M{\'e}ndez}, {Capaccioli}, \& {Ford}}]{arnaboldi98}
{Arnaboldi}, M., {Freeman}, K.~C., {Gerhard}, O., {et~al.} 1998, \apj, 507, 759

\bibitem[{{Arnaboldi} {et~al.}(1996){Arnaboldi}, {Freeman}, {Mendez},
  {Capaccioli}, {Ciardullo}, {Ford}, {Gerhard}, {Hui}, {Jacoby}, {Kudritzki},
  \& {Quinn}}]{arnaboldi96}
{Arnaboldi}, M., {Freeman}, K.~C., {Mendez}, R.~H., {et~al.} 1996, \apj, 472,
  145

\bibitem[{{Arnaboldi} {et~al.}(2003){Arnaboldi}, {Freeman}, {Okamura},
  {Yasuda}, {Gerhard}, {Napolitano}, {Pannella}, {Ando}, {Doi}, {Furusawa},
  {Hamabe}, {Kimura}, {Kajino}, {Komiyama}, {Miyazaki}, {Nakata}, {Ouchi},
  {Sekiguchi}, {Shimasaku}, \& {Yagi}}]{arnaboldi03}
{Arnaboldi}, M., {Freeman}, K.~C., {Okamura}, S., {et~al.} 2003, \aj, 125, 514

\bibitem[{{Astropy Collaboration} {et~al.}(2013){Astropy Collaboration},
  {Robitaille}, {Tollerud}, {Greenfield}, {Droettboom}, {Bray}, {Aldcroft},
  {Davis}, {Ginsburg}, {Price-Whelan}, {Kerzendorf}, {Conley}, {Crighton},
  {Barbary}, {Muna}, {Ferguson}, {Grollier}, {Parikh}, {Nair}, {Unther},
  {Deil}, {Woillez}, {Conseil}, {Kramer}, {Turner}, {Singer}, {Fox}, {Weaver},
  {Zabalza}, {Edwards}, {Azalee Bostroem}, {Burke}, {Casey}, {Crawford},
  {Dencheva}, {Ely}, {Jenness}, {Labrie}, {Lim}, {Pierfederici}, {Pontzen},
  {Ptak}, {Refsdal}, {Servillat}, \& {Streicher}}]{Rob13}
{Astropy Collaboration}, {Robitaille}, T.~P., {Tollerud}, E.~J., {et~al.} 2013,
  \aap, 558, A33

\bibitem[{{Barmby} {et~al.}(2006){Barmby}, {Ashby}, {Bianchi}, {Engelbracht},
  {Gehrz}, {Gordon}, {Hinz}, {Huchra}, {Humphreys}, {Pahre},
  {P{\'e}rez-Gonz{\'a}lez}, {Polomski}, {Rieke}, {Thilker}, {Willner}, \&
  {Woodward}}]{barmby06}
{Barmby}, P., {Ashby}, M.~L.~N., {Bianchi}, L., {et~al.} 2006, \apj, 650, L45

\bibitem[{{Bernard} {et~al.}(2015){Bernard}, {Ferguson}, {Chapman}, {Ibata},
  {Irwin}, {Lewis}, \& {McConnachie}}]{bernard15}
{Bernard}, E.~J., {Ferguson}, A. M.~N., {Chapman}, S.~C., {et~al.} 2015,
  \mnras, 453, L113

\bibitem[{{Bertin} \& {Arnouts}(1996)}]{bertin96}
{Bertin}, E. \& {Arnouts}, S. 1996, Astronomy and Astrophysics Supplement
  Series, 117, 393

\bibitem[{{Bertin} {et~al.}(2002){Bertin}, {Mellier}, {Radovich}, {Missonnier},
  {Didelon}, \& {Morin}}]{bertin02}
{Bertin}, E., {Mellier}, Y., {Radovich}, M., {et~al.} 2002, in Astronomical
  Data Analysis Software and Systems XI, ed. D.~A. {Bohlender}, D.~{Durand}, \&
  T.~H. {Handley}, Vol. 281, 228

\bibitem[{{Bonnarel} {et~al.}(2000){Bonnarel}, {Fernique}, {Bienaym{\'e}},
  {Egret}, {Genova}, {Louys}, {Ochsenbein}, {Wenger}, \& {Bartlett}}]{bon00}
{Bonnarel}, F., {Fernique}, P., {Bienaym{\'e}}, O., {et~al.} 2000, Astronomy
  and Astrophysics Supplement Series, 143, 33

\bibitem[{{Boulade} {et~al.}(2003){Boulade}, {Charlot}, {Abbon}, {Aune},
  {Borgeaud}, {Carton}, {Carty}, {Da Costa}, {Deschamps}, {Desforge},
  {Eppell{\'e}}, {Gallais}, {Gosset}, {Granelli}, {Gros}, {de Kat}, {Loiseau},
  {Ritou}, {Rouss{\'e}}, {Starzynski}, {Vignal}, \& {Vigroux}}]{boulade03}
{Boulade}, O., {Charlot}, X., {Abbon}, P., {et~al.} 2003, in Instrument Design
  and Performance for Optical/Infrared Ground-based Telescopes, ed. M.~{Iye} \&
  A.~F.~M. {Moorwood}, Vol. 4841, 72--81

\bibitem[{{Buzzoni} {et~al.}(2006){Buzzoni}, {Arnaboldi}, \& {Corradi}}]{buz06}
{Buzzoni}, A., {Arnaboldi}, M., \& {Corradi}, R. L.~M. 2006, \mnras, 368, 877

\bibitem[{{Ciardullo}(2010)}]{ciardullo10}
{Ciardullo}, R. 2010, Publications of the Astronomical Society of Australia,
  27, 149

\bibitem[{{Ciardullo} {et~al.}(2004){Ciardullo}, {Durrell}, {Laychak},
  {Herrmann}, {Moody}, {Jacoby}, \& {Feldmeier}}]{ciardullo04}
{Ciardullo}, R., {Durrell}, P.~R., {Laychak}, M.~B., {et~al.} 2004, \apj, 614,
  167

\bibitem[{{Ciardullo} {et~al.}(2002){Ciardullo}, {Feldmeier}, {Jacoby}, {Kuzio
  de Naray}, {Laychak}, \& {Durrell}}]{ciardullo02}
{Ciardullo}, R., {Feldmeier}, J.~J., {Jacoby}, G.~H., {et~al.} 2002, \apj, 577,
  31

\bibitem[{{Ciardullo} {et~al.}(2013){Ciardullo}, {Gronwall}, {Adams}, {Blanc},
  {Gebhardt}, {Finkelstein}, {Jogee}, {Hill}, {Drory}, {Hopp}, {Schneider},
  {Zeimann}, \& {Dalton}}]{ciardullo13}
{Ciardullo}, R., {Gronwall}, C., {Adams}, J.~J., {et~al.} 2013, \apj, 769, 83

\bibitem[{{Ciardullo} \& {Jacoby}(1992)}]{ciardullo92}
{Ciardullo}, R. \& {Jacoby}, G.~H. 1992, \apj, 388, 268

\bibitem[{{Ciardullo} {et~al.}(1989){Ciardullo}, {Jacoby}, {Ford}, \&
  {Neill}}]{ciardullo89}
{Ciardullo}, R., {Jacoby}, G.~H., {Ford}, H.~C., \& {Neill}, J.~D. 1989, \apj,
  339, 53

\bibitem[{{Coccato} {et~al.}(2009){Coccato}, {Gerhard}, {Arnaboldi}, {Das},
  {Douglas}, {Kuijken}, {Merrifield}, {Napolitano}, {Noordermeer},
  {Romanowsky}, {Capaccioli}, {Cortesi}, {De Lorenzi}, \& {Freeman}}]{coc09}
{Coccato}, L., {Gerhard}, O., {Arnaboldi}, M., {et~al.} 2009, \mnras, 394, 1249

\bibitem[{{Colless} {et~al.}(1990){Colless}, {Ellis}, {Taylor}, \&
  {Hook}}]{Colless90}
{Colless}, M., {Ellis}, R.~S., {Taylor}, K., \& {Hook}, R.~N. 1990, \mnras,
  244, 408

\bibitem[{{Cortesi} {et~al.}(2013){Cortesi}, {Arnaboldi}, {Coccato},
  {Merrifield}, {Gerhard}, {Bamford}, {Romanowsky}, {Napolitano}, {Douglas},
  {Kuijken}, {Capaccioli}, {Freeman}, {Chies-Santos}, \& {Pota}}]{cor13}
{Cortesi}, A., {Arnaboldi}, M., {Coccato}, L., {et~al.} 2013, \aap, 549, A115

\bibitem[{{Courteau} {et~al.}(2011){Courteau}, {Widrow}, {McDonald},
  {Guhathakurta}, {Gilbert}, {Zhu}, {Beaton}, \& {Majewski}}]{cou11}
{Courteau}, S., {Widrow}, L.~M., {McDonald}, M., {et~al.} 2011, \apj, 739, 20

\bibitem[{{Dalcanton} {et~al.}(2012){Dalcanton}, {Williams}, {Lang}, {Lauer},
  {Kalirai}, {Seth}, {Dolphin}, {Rosenfield}, {Weisz}, {Bell}, {Bianchi},
  {Boyer}, {Caldwell}, {Dong}, {Dorman}, {Gilbert}, {Girardi}, {Gogarten},
  {Gordon}, {Guhathakurta}, {Hodge}, {Holtzman}, {Johnson}, {Larsen}, {Lewis},
  {Melbourne}, {Olsen}, {Rix}, {Rosema}, {Saha}, {Sarajedini}, {Skillman}, \&
  {Stanek}}]{dal12}
{Dalcanton}, J.~J., {Williams}, B.~F., {Lang}, D., {et~al.} 2012, The
  Astrophysical Journal Supplement Series, 200, 18

\bibitem[{{Davis} {et~al.}(2018){Davis}, {Ciardullo}, {Jacoby}, {Feldmeier}, \&
  {Indahl}}]{davis18}
{Davis}, B.~D., {Ciardullo}, R., {Jacoby}, G.~H., {Feldmeier}, J.~J., \&
  {Indahl}, B.~L. 2018, \apj, 863, 189

\bibitem[{{Douglas} {et~al.}(2002){Douglas}, {Arnaboldi}, {Freeman}, {Kuijken},
  {Merrifield}, {Romanowsky}, {Taylor}, {Capaccioli}, {Axelrod}, {Gilmozzi},
  {Hart}, {Bloxham}, \& {Jones}}]{dou02}
{Douglas}, N.~G., {Arnaboldi}, M., {Freeman}, K.~C., {et~al.} 2002,
  Publications of the Astronomical Society of the Pacific, 114, 1234

\bibitem[{{D'Souza} \& {Bell}(2018)}]{dsouza18}
{D'Souza}, R. \& {Bell}, E.~F. 2018, Nature Astronomy, 2, 737

\bibitem[{{Fabricant} {et~al.}(2005){Fabricant}, {Fata}, {Roll}, {Hertz},
  {Caldwell}, {Gauron}, {Geary}, {McLeod}, {Szentgyorgyi}, {Zajac}, {Kurtz},
  {Barberis}, {Bergner}, {Brown}, {Conroy}, {Eng}, {Geller}, {Goddard},
  {Honsa}, {Mueller}, {Mink}, {Ordway}, {Tokarz}, {Woods}, {Wyatt}, {Epps}, \&
  {Dell'Antonio}}]{fab05}
{Fabricant}, D., {Fata}, R., {Roll}, J., {et~al.} 2005, Publications of the
  Astronomical Society of the Pacific, 117, 1411

\bibitem[{{Fang} {et~al.}(2015){Fang}, {Garc{\'\i}a-Benito}, {Guerrero}, {Liu},
  {Yuan}, {Zhang}, \& {Zhang}}]{fang15}
{Fang}, X., {Garc{\'\i}a-Benito}, R., {Guerrero}, M.~A., {et~al.} 2015, \apj,
  815, 69

\bibitem[{{Fang} {et~al.}(2018){Fang}, {Garc{\'\i}a-Benito}, {Guerrero},
  {Zhang}, {Liu}, {Morisset}, {Karakas}, {Miller Bertolami}, {Yuan}, \&
  {Cabrera-Lavers}}]{fang18}
{Fang}, X., {Garc{\'\i}a-Benito}, R., {Guerrero}, M.~A., {et~al.} 2018, \apj,
  853, 50

\bibitem[{{Ferguson} \& {Mackey}(2016)}]{fm16}
{Ferguson}, A. M.~N. \& {Mackey}, A.~D. 2016, in Tidal Streams in the Local
  Group and Beyond, ed. H.~J. {Newberg} \& J.~L. {Carlin}, Vol. 420, 191

\bibitem[{{Gesicki} {et~al.}(2018){Gesicki}, {Zijlstra}, \& {Miller
  Bertolami}}]{ges18}
{Gesicki}, K., {Zijlstra}, A.~A., \& {Miller Bertolami}, M.~M. 2018, Nature
  Astronomy, 2, 580

\bibitem[{{Hammer} {et~al.}(1997){Hammer}, {Flores}, {Lilly}, {Crampton}, {Le
  F{\`e}vre}, {Rola}, {Mallen-Ornelas}, {Schade}, \& {Tresse}}]{Hammer97}
{Hammer}, F., {Flores}, H., {Lilly}, S.~J., {et~al.} 1997, \apj, 481, 49

\bibitem[{{Hammer} {et~al.}(2018){Hammer}, {Yang}, {Wang}, {Ibata}, {Flores},
  \& {Puech}}]{ham18}
{Hammer}, F., {Yang}, Y.~B., {Wang}, J.~L., {et~al.} 2018, \mnras, 475, 2754

\bibitem[{{Hartke} {et~al.}(2018){Hartke}, {Arnaboldi}, {Gerhard}, {Agnello},
  {Longobardi}, {Coccato}, {Pulsoni}, {Freeman}, \& {Merrifield}}]{hartke18}
{Hartke}, J., {Arnaboldi}, M., {Gerhard}, O., {et~al.} 2018, \aap, 616, A123

\bibitem[{{Hartke} {et~al.}(2017){Hartke}, {Arnaboldi}, {Longobardi},
  {Gerhard}, {Freeman}, {Okamura}, \& {Nakata}}]{hartke17}
{Hartke}, J., {Arnaboldi}, M., {Longobardi}, A., {et~al.} 2017, \aap, 603, A104

\bibitem[{{Henize} \& {Westerlund}(1963)}]{hw63}
{Henize}, K.~G. \& {Westerlund}, B.~E. 1963, \apj, 137, 747

\bibitem[{{Hern{\'a}ndez-Mart{\'\i}nez} \& {Pe{\~n}a}(2009)}]{hmp09}
{Hern{\'a}ndez-Mart{\'\i}nez}, L. \& {Pe{\~n}a}, M. 2009, \aap, 495, 447

\bibitem[{{Hogg} {et~al.}(1998){Hogg}, {Cohen}, {Blandford}, \&
  {Pahre}}]{Hogg98}
{Hogg}, D.~W., {Cohen}, J.~G., {Blandford}, R., \& {Pahre}, M.~A. 1998, \apj,
  504, 622

\bibitem[{{Hui} {et~al.}(1995){Hui}, {Ford}, {Freeman}, \& {Dopita}}]{hui95}
{Hui}, X., {Ford}, H.~C., {Freeman}, K.~C., \& {Dopita}, M.~A. 1995, \apj, 449,
  592

\bibitem[{Hunter(2007)}]{matplotlib}
Hunter, J.~D. 2007, Computing In Science \& Engineering, 9, 90

\bibitem[{{Ibata} {et~al.}(2014){Ibata}, {Lewis}, {McConnachie}, {Martin},
  {Irwin}, {Ferguson}, {Babul}, {Bernard}, {Chapman}, {Collins}, {Fardal},
  {Mackey}, {Navarro}, {Pe{\~n}arrubia}, {Rich}, {Tanvir}, \&
  {Widrow}}]{ibata14}
{Ibata}, R.~A., {Lewis}, G.~F., {McConnachie}, A.~W., {et~al.} 2014, \apj, 780,
  128

\bibitem[{{Irwin} {et~al.}(2005){Irwin}, {Ferguson}, {Ibata}, {Lewis}, \&
  {Tanvir}}]{irwin05}
{Irwin}, M.~J., {Ferguson}, A. M.~N., {Ibata}, R.~A., {Lewis}, G.~F., \&
  {Tanvir}, N.~R. 2005, \apj, 628, L105

\bibitem[{{Jacoby}(1980)}]{jacoby80}
{Jacoby}, G.~H. 1980, The Astrophysical Journal Supplement Series, 42, 1

\bibitem[{{Jacoby}(1989)}]{jacoby89}
{Jacoby}, G.~H. 1989, \apj, 339, 39

\bibitem[{{Jacoby} \& {De Marco}(2002)}]{jd02}
{Jacoby}, G.~H. \& {De Marco}, O. 2002, \aj, 123, 269

\bibitem[{{Jones} \& {Boffin}(2017)}]{jb17}
{Jones}, D. \& {Boffin}, H. M.~J. 2017, Nature Astronomy, 1, 0117

\bibitem[{{Kovacevic} {et~al.}(2011){Kovacevic}, {Parker}, {Jacoby}, {Sharp},
  {Miszalski}, \& {Frew}}]{kov11}
{Kovacevic}, A.~V., {Parker}, Q.~A., {Jacoby}, G.~H., {et~al.} 2011, \mnras,
  414, 860

\bibitem[{{Li} {et~al.}(2018){Li}, {Li}, {Dong}, {Fang}, \& {Xu}}]{Li18}
{Li}, A., {Li}, Z., {Dong}, H., {Fang}, X., \& {Xu}, X. 2018, ArXiv e-prints,
  arXiv:1805.12092

\bibitem[{{Longobardi} {et~al.}(2013){Longobardi}, {Arnaboldi}, {Gerhard},
  {Coccato}, {Okamura}, \& {Freeman}}]{longobardi13}
{Longobardi}, A., {Arnaboldi}, M., {Gerhard}, O., {et~al.} 2013, \aap, 558, A42

\bibitem[{{Longobardi} {et~al.}(2015){Longobardi}, {Arnaboldi}, {Gerhard}, \&
  {Hanuschik}}]{longobardi15}
{Longobardi}, A., {Arnaboldi}, M., {Gerhard}, O., \& {Hanuschik}, R. 2015,
  \aap, 579, A135

\bibitem[{{Mackey} {et~al.}(2010){Mackey}, {Huxor}, {Ferguson}, {Irwin},
  {Tanvir}, {McConnachie}, {Ibata}, {Chapman}, \& {Lewis}}]{mackey10}
{Mackey}, A.~D., {Huxor}, A.~P., {Ferguson}, A.~M.~N., {et~al.} 2010, \apj,
  717, L11

\bibitem[{{Magnier} \& {Cuillandre}(2004)}]{magnier04}
{Magnier}, E.~A. \& {Cuillandre}, J.~C. 2004, Publications of the Astronomical
  Society of the Pacific, 116, 449

\bibitem[{{Majaess}(2010)}]{majaess10}
{Majaess}, D. 2010, \actaa, 60, 55

\bibitem[{{Marmo} \& {Bertin}(2008)}]{marmo08}
{Marmo}, C. \& {Bertin}, E. 2008, in Astronomical Data Analysis Software and
  Systems XVII, ed. R.~W. {Argyle}, P.~S. {Bunclark}, \& J.~R. {Lewis}, Vol.
  394, 619

\bibitem[{{Martin} {et~al.}(2018){Martin}, {Drissen}, \& {Melchior}}]{martin18}
{Martin}, T.~B., {Drissen}, L., \& {Melchior}, A.-L. 2018, \mnras, 473, 4130

\bibitem[{{McConnachie} {et~al.}(2018){McConnachie}, {Ibata}, {Martin},
  {Ferguson}, {Collins}, {Gwyn}, {Irwin}, {Lewis}, {Mackey}, {Davidge},
  {Arias}, {Conn}, {Cote}, {Crnojevic}, {Huxor}, {Penarrubia}, {Spengler},
  {Tanvir}, {Valls-Gabaud}, {Babul}, {Barmby}, {Bate}, {Bernard}, {Chapman},
  {Dotter}, {Harris}, {McMonigal}, {Navarro}, {Puzia}, {Rich}, \&
  {Thomas}}]{mcc18}
{McConnachie}, A.~W., {Ibata}, R., {Martin}, N., {et~al.} 2018, ArXiv e-prints,
  arXiv:1810.08234

\bibitem[{{McConnachie} {et~al.}(2009){McConnachie}, {Irwin}, {Ibata},
  {Dubinski}, {Widrow}, {Martin}, {C{\^o}t{\'e}}, {Dotter}, {Navarro},
  {Ferguson}, {Puzia}, {Lewis}, {Babul}, {Barmby}, {Bienaym{\'e}}, {Chapman},
  {Cockcroft}, {Collins}, {Fardal}, {Harris}, {Huxor}, {Mackey},
  {Pe{\~n}arrubia}, {Rich}, {Richer}, {Siebert}, {Tanvir}, {Valls-Gabaud}, \&
  {Venn}}]{mcc09}
{McConnachie}, A.~W., {Irwin}, M.~J., {Ibata}, R.~A., {et~al.} 2009, \nat, 461,
  66

\bibitem[{{M{\'e}ndez} {et~al.}(2001){M{\'e}ndez}, {Riffeser}, {Kudritzki},
  {Matthias}, {Freeman}, {Arnaboldi}, {Capaccioli}, \& {Gerhard}}]{men01}
{M{\'e}ndez}, R.~H., {Riffeser}, A., {Kudritzki}, R.~P., {et~al.} 2001, \apj,
  563, 135

\bibitem[{{Merrett} {et~al.}(2006){Merrett}, {Merrifield}, {Douglas},
  {Kuijken}, {Romanowsky}, {Napolitano}, {Arnaboldi}, {Capaccioli}, {Freeman},
  {Gerhard}, {Coccato}, {Carter}, {Evans}, {Wilkinson}, {Halliday}, \&
  {Bridges}}]{merrett06}
{Merrett}, H.~R., {Merrifield}, M.~R., {Douglas}, N.~G., {et~al.} 2006, \mnras,
  369, 120

\bibitem[{{Miko{\l}ajewska} {et~al.}(2014){Miko{\l}ajewska}, {Caldwell}, \&
  {Shara}}]{mik14}
{Miko{\l}ajewska}, J., {Caldwell}, N., \& {Shara}, M.~M. 2014, \mnras, 444, 586

\bibitem[{{Oke} \& {Gunn}(1983)}]{Oke83}
{Oke}, J.~B. \& {Gunn}, J.~E. 1983, \apj, 266, 713

\bibitem[{Oliphant(2015)}]{numpy}
Oliphant, T.~E. 2015, Guide to NumPy, 2nd edn. (USA: CreateSpace Independent
  Publishing Platform)

\bibitem[{{Pastorello} {et~al.}(2013){Pastorello}, {Sarzi}, {Cappellari},
  {Emsellem}, {Mamon}, {Bacon}, {Davies}, \& {de Zeeuw}}]{pas13}
{Pastorello}, N., {Sarzi}, M., {Cappellari}, M., {et~al.} 2013, \mnras, 430,
  1219

\bibitem[{{Pe{\~n}a} {et~al.}(2007){Pe{\~n}a}, {Richer}, \&
  {Stasi{\'n}ska}}]{pena07}
{Pe{\~n}a}, M., {Richer}, M.~G., \& {Stasi{\'n}ska}, G. 2007, \aap, 466, 75

\bibitem[{{Pulsoni} {et~al.}(2018){Pulsoni}, {Gerhard}, {Arnaboldi}, {Coccato},
  {Longobardi}, {Napolitano}, {Moylan}, {Narayan}, {Gupta}, {Burkert},
  {Capaccioli}, {Chies-Santos}, {Cortesi}, {Freeman}, {Kuijken}, {Merrifield},
  {Romanowsky}, \& {Tortora}}]{pul18}
{Pulsoni}, C., {Gerhard}, O., {Arnaboldi}, M., {et~al.} 2018, \aap, 618, A94

\bibitem[{{Reid} \& {Parker}(2010)}]{rp10}
{Reid}, W.~A. \& {Parker}, Q.~A. 2010, \mnras, 405, 1349

\bibitem[{{Rodr{\'\i}guez-Gonz{\'a}lez}
  {et~al.}(2015){Rodr{\'\i}guez-Gonz{\'a}lez}, {Hern{\'a}ndez-Mart{\'\i}nez},
  {Esquivel}, {Raga}, {Stasi{\'n}ska}, {Pe{\~n}a}, \& {Mayya}}]{rod15}
{Rodr{\'\i}guez-Gonz{\'a}lez}, A., {Hern{\'a}ndez-Mart{\'\i}nez}, L.,
  {Esquivel}, A., {et~al.} 2015, \aap, 575, A1

\bibitem[{{Sanders} {et~al.}(2012){Sanders}, {Caldwell}, {McDowell}, \&
  {Harding}}]{san12}
{Sanders}, N.~E., {Caldwell}, N., {McDowell}, J., \& {Harding}, P. 2012, \apj,
  758, 133

\bibitem[{{Tempel} {et~al.}(2010){Tempel}, {Tamm}, \& {Tenjes}}]{tem10}
{Tempel}, E., {Tamm}, A., \& {Tenjes}, P. 2010, \aap, 509, A91

\bibitem[{{Teplitz} {et~al.}(2000){Teplitz}, {Malkan}, {Steidel}, {McLean},
  {Becklin}, {Figer}, {Gilbert}, {Graham}, {Larkin}, {Levenson}, \&
  {Wilcox}}]{teplitz00}
{Teplitz}, H.~I., {Malkan}, M.~A., {Steidel}, C.~C., {et~al.} 2000, \apj, 542,
  18

\bibitem[{{Veljanoski} {et~al.}(2014){Veljanoski}, {Mackey}, {Ferguson},
  {Huxor}, {C{\^o}t{\'e}}, {Irwin}, {Tanvir}, {Pe{\~n}arrubia}, {Bernard},
  {Fardal}, {Martin}, {McConnachie}, {Lewis}, {Chapman}, {Ibata}, \&
  {Babul}}]{vel14}
{Veljanoski}, J., {Mackey}, A.~D., {Ferguson}, A.~M.~N., {et~al.} 2014, \mnras,
  442, 2929

\bibitem[{{Veyette} {et~al.}(2014){Veyette}, {Williams}, {Dalcanton}, {Balick},
  {Caldwell}, {Fouesneau}, {Girardi}, {Gordon}, {Kalirai}, {Rosenfield}, \&
  {Seth}}]{vey14}
{Veyette}, M.~J., {Williams}, B.~F., {Dalcanton}, J.~J., {et~al.} 2014, \apj,
  792, 121

\bibitem[{{Walterbos} \& {Kennicutt}(1988)}]{wk88}
{Walterbos}, R.~A.~M. \& {Kennicutt}, R.~C., J. 1988, \aap, 198, 61

\bibitem[{{Williams} {et~al.}(2017){Williams}, {Dolphin}, {Dalcanton}, {Weisz},
  {Bell}, {Lewis}, {Rosenfield}, {Choi}, {Skillman}, \& {Monachesi}}]{wil17}
{Williams}, B.~F., {Dolphin}, A.~E., {Dalcanton}, J.~J., {et~al.} 2017, \apj,
  846, 145

\bibitem[{{Williams} {et~al.}(2014){Williams}, {Lang}, {Dalcanton}, {Dolphin},
  {Weisz}, {Bell}, {Bianchi}, {Byler}, {Gilbert}, {Girardi}, {Gordon},
  {Gregersen}, {Johnson}, {Kalirai}, {Lauer}, {Monachesi}, {Rosenfield},
  {Seth}, \& {Skillman}}]{wil14}
{Williams}, B.~F., {Lang}, D., {Dalcanton}, J.~J., {et~al.} 2014, The
  Astrophysical Journal Supplement Series, 215, 9

\bibitem[{Wilson(1927)}]{Wilson27}
Wilson, E.~B. 1927, Journal of the American Statistical Association, 22, 209

\end{thebibliography}

\begin{appendix} 
\section{Catalog Extraction}

\subsection{Flux calibration for m$_{5007}$}
\label{flux}
The integrated flux,  $F_{5007}$,  from the [\ion{O}{iii}] 5007 $\AA$ line is related to the m$\rm_{5007}$ magnitude \citep{jacoby89} as follows:
\begin{equation}
$$m_{5007}=-2.5 {\rm log} F_{5007} -13.74$$
\end{equation}
where the flux is in units of erg cm$^2$s$^-1$. The \textit{AB} magnitude \citep{Oke83} relates to $F_{5007}$ through the MegaCam narrow-band [\ion{O}{iii}] filter characteristics as:
\begin{equation}
$$m\rm_{n}=-2.5 {\rm log} F_{5007}- 2.5{\rm log}\frac{\lambda\rm_c^2}{\Delta\lambda_{\rm eff}c}-48.59$$
\end{equation}
where $\lambda\rm_c = 5007 \AA$ and $\Delta\lambda_{\rm eff}= \Delta\lambda \times T$ for the on-band with $\Delta\lambda= 102 \AA $ and effective transmission, $T=0.91$. The relation between the \textit{AB} and m$\rm_{5007}$ magnitudes, for this narrow-band filter, is thus determined \citep[following][]{arnaboldi03} to be:
\begin{equation}
$$m_{5007}=m\rm_{n} + 2.27$$
\end{equation}

\begin{table}[htb]
\caption{Parameters for the best fitting PSF}
\centering
\begin{tabular}{cccc}
\hline
Radius (px) & Ellipticity & $\beta$ & Moffat FWHM (px)\\
\hline
15.2 & 0 & 5.02 & 5.15 \\
\hline
\end{tabular}

\label{table : psfparam}

\end{table}

\begin{figure}[htb]
	\centering
	\includegraphics[width=\columnwidth,angle=0]{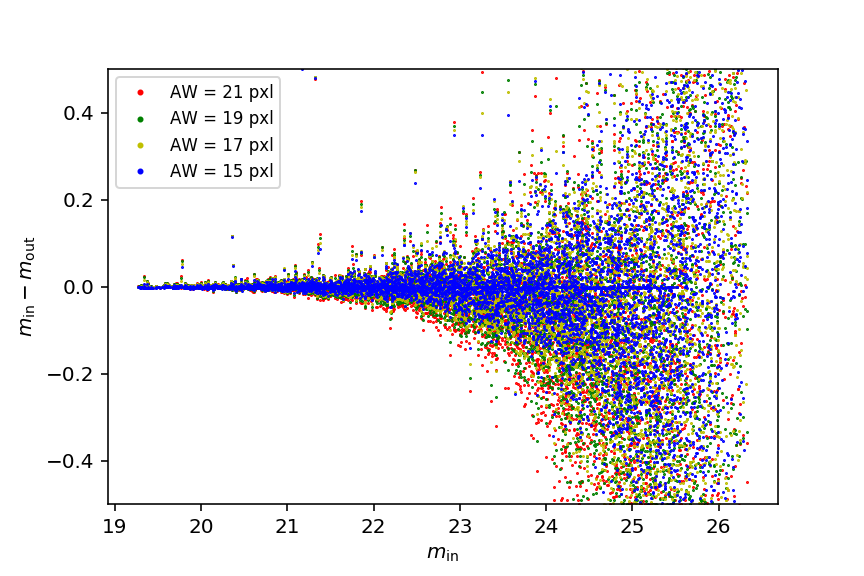}
	\caption{The recovery of the input magnitude of simulated sources using different SExtractor aperture widths, AW, for a single field.}
	\label{fig:aper}
\end{figure}

\begin{figure}[t]
	\centering
	\includegraphics[width=\columnwidth,angle=0]{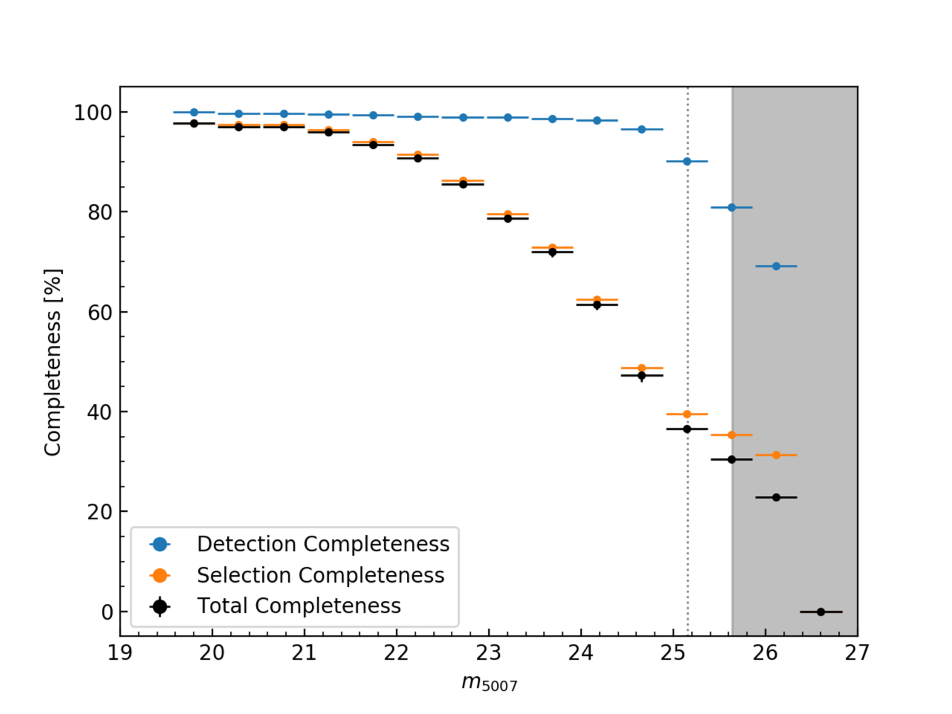}
	\caption{{For the entire survey, the number of PNe in each magnitude bin is shown in black, while the completeness corrected number is shown in blue for a single field. The region beyond the limiting magnitude of the shallowest field (Field\# 33\_4) is shown in grey. The grey dotted line shows the 90\% completeness limit of the shallowest field.}}
    \label{fig:comp}
\end{figure}

\subsection{The point-spread function}
\label{psf}
The point-spread function (PSF) is the instrumental response to a point-like object. The degree of spreading of the  PSF represents the quality of the instrument imaging. To determine the PSF, We first used the IRAF tasks {\it daofind}, {\it daophot} and {\it pstsel} to respectively extract all the positions of detected sources, determine their magnitude and select the brighter stars. We then used the IRAF task {\it psf}, which allowed us to select the best stars to fit the PSF, and reject bad objects (e.g. saturated or located near an another source). The confirmed stars are used to calculate the PSF. The PSF is modeled with different luminosity profiles (e.g. Gaussian, Moffat function) and the best fit is selected using $\chi^2$-minimization (goodness-of-fit). The best fit is a Moffat function (parameters in Table~\ref{table : psfparam}) of the form :
\begin{equation}
$$f(r,\alpha,\beta)=\frac{\beta-1}{\pi\alpha}[1+(\frac{r}{\alpha})^2]^{-\beta}$$
\end{equation}

\subsection{Masking of noisy regions and CCD edges}
\label{mask}
After identifying sources on the images, the image regions which were affected by dithering or saturation are masked on the on-band and off-band images. This is carried out for each field. Dithering leads to different exposure depths at the image edges. Due to the combination of the 42 MegaCam CCDs, the same effect also affects columns at the borders of the individual CCDs. Additionally, in order to mask those regions with a high background value (e.g. due to charge transfer or saturated stars) we used the rms-background map created by SExtractor and create a pixel-mask with all values higher than 3 times the median background.

\subsection{Choice of aperture width}
\label{aper}
After simulating sources on to the on-band, we determine the magnitude aperture most suited to recovering them. Figure~\ref{fig:aper} shows that different SExtractor aperture widths, AW, of 15, 17, 19 and 21 pixels all recover the  magnitudes well but an AW of 15 pixels can recover sources most accurately even for fainter sources.

\begin{figure*}[t]
	\centering
	\includegraphics[width=\columnwidth,angle=0]{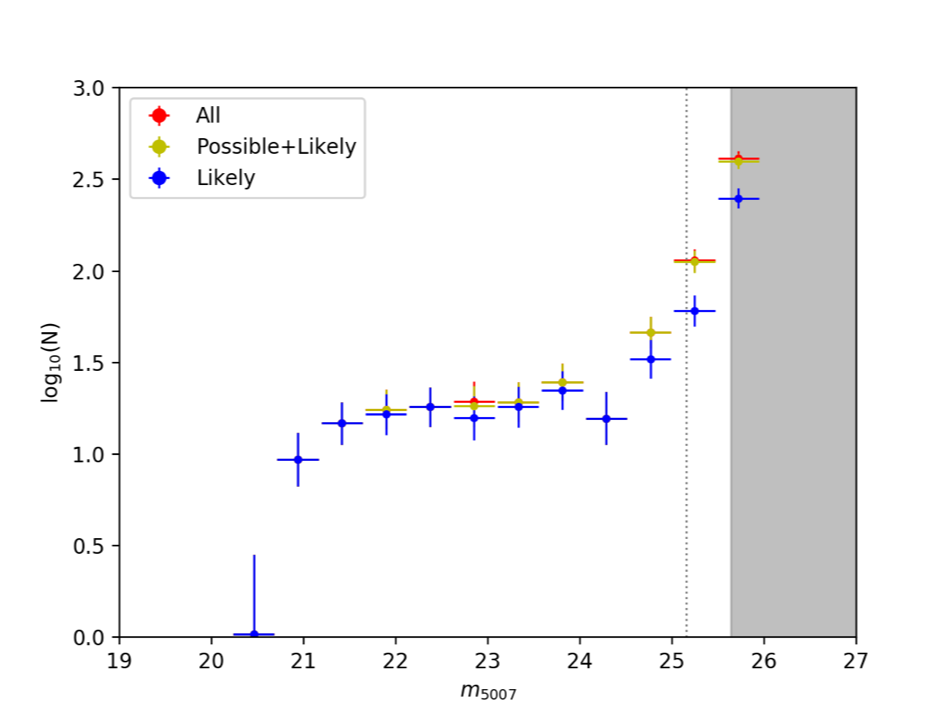}
	\includegraphics[width=\columnwidth,angle=0]{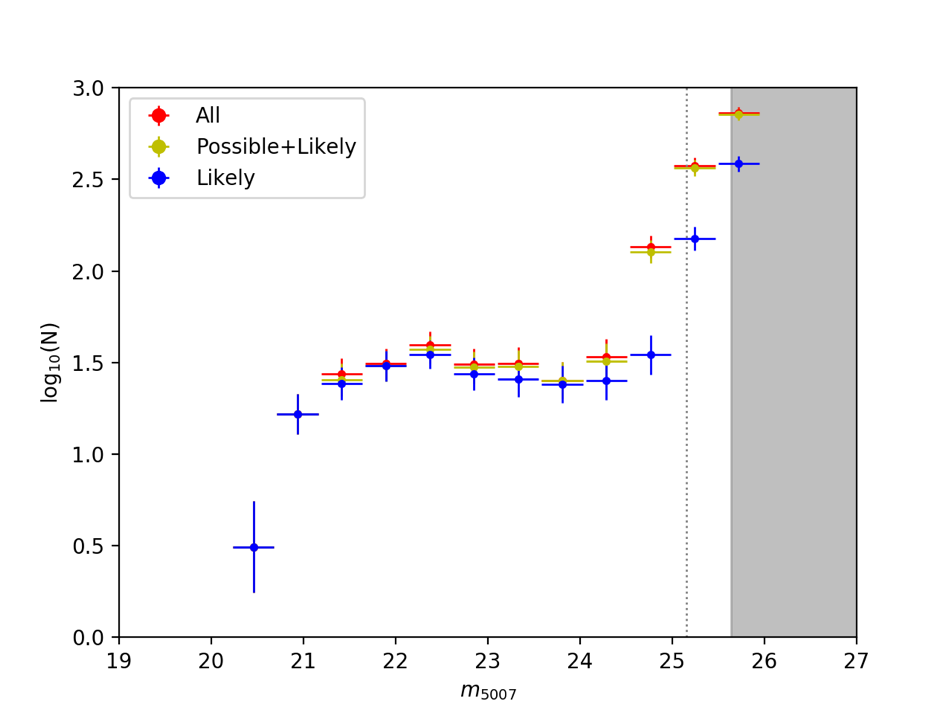}
	\caption{The completeness-corrected PNLF for PHAT-matched PNe within (left) and outside (right) 0.3$\deg$ of the center of M31 is shown for the ``likely'' (blue), ``possible'' + ``likely'' (yellow) and ``possible'' + ``likely'' +``unlikely'' (red). The region beyond the limiting magnitude of the shallowest field (Field\# 33\_4) is shown in grey. The grey dotted line shows the 90\% completeness limit of the shallowest field.}
	\label{fig:phatnearfar}
\end{figure*}

\begin{figure}[htb]
	\centering
	\includegraphics[width=\columnwidth,angle=0]{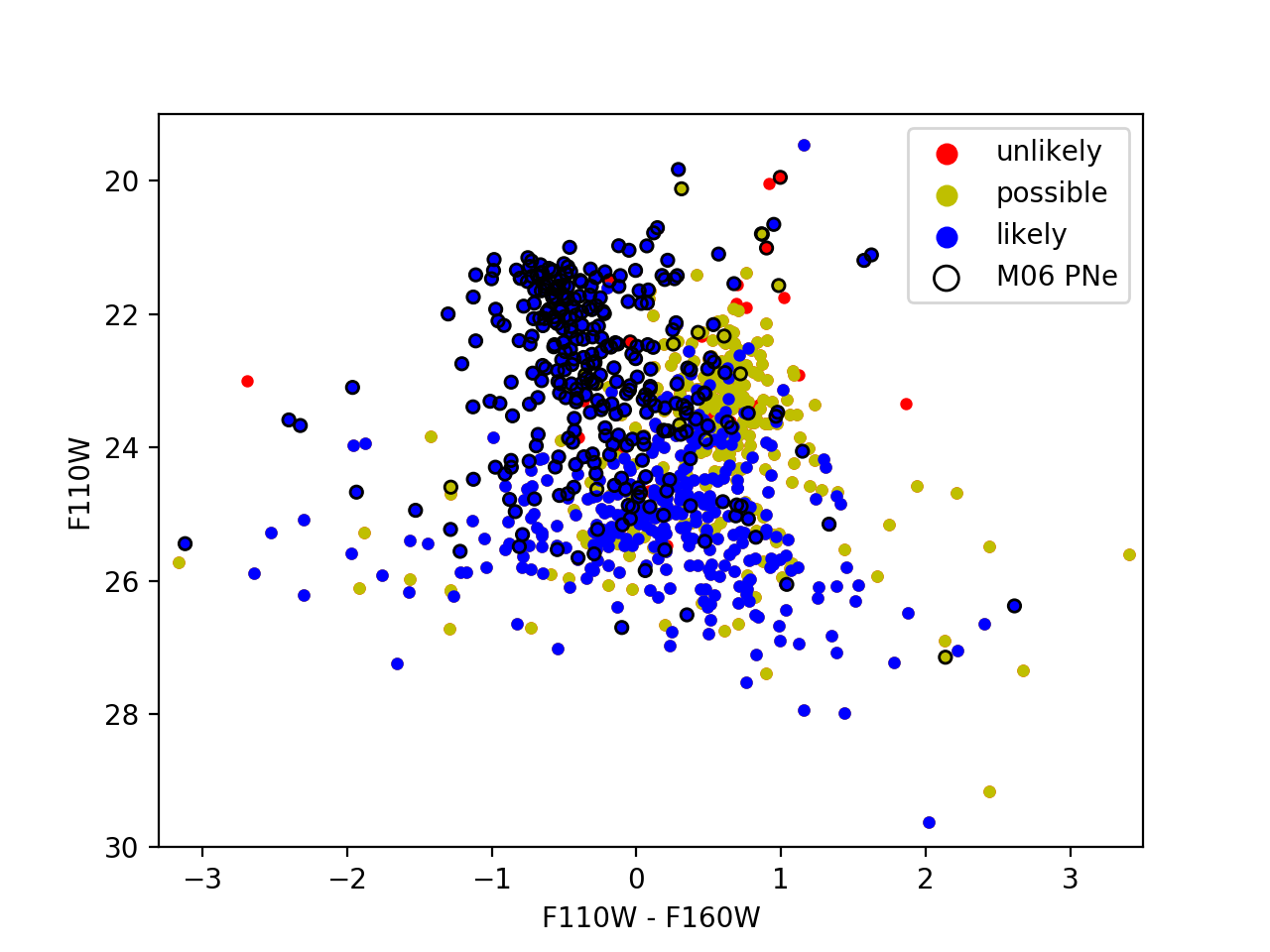}
	\caption{The F110W mag of the PHAT-matched PNe plotted against its F110W $-$ F160W colour. The ``likely'', ``possible'' and ``unlikely'' PNe have been shown in blue, yellow and red respectively. The PNe previously found by M06 have been encircled in black.}
	\label{fig:phatnir}
\end{figure}

\begin{figure}[htb]
	\centering
	\includegraphics[width=\columnwidth,angle=0]{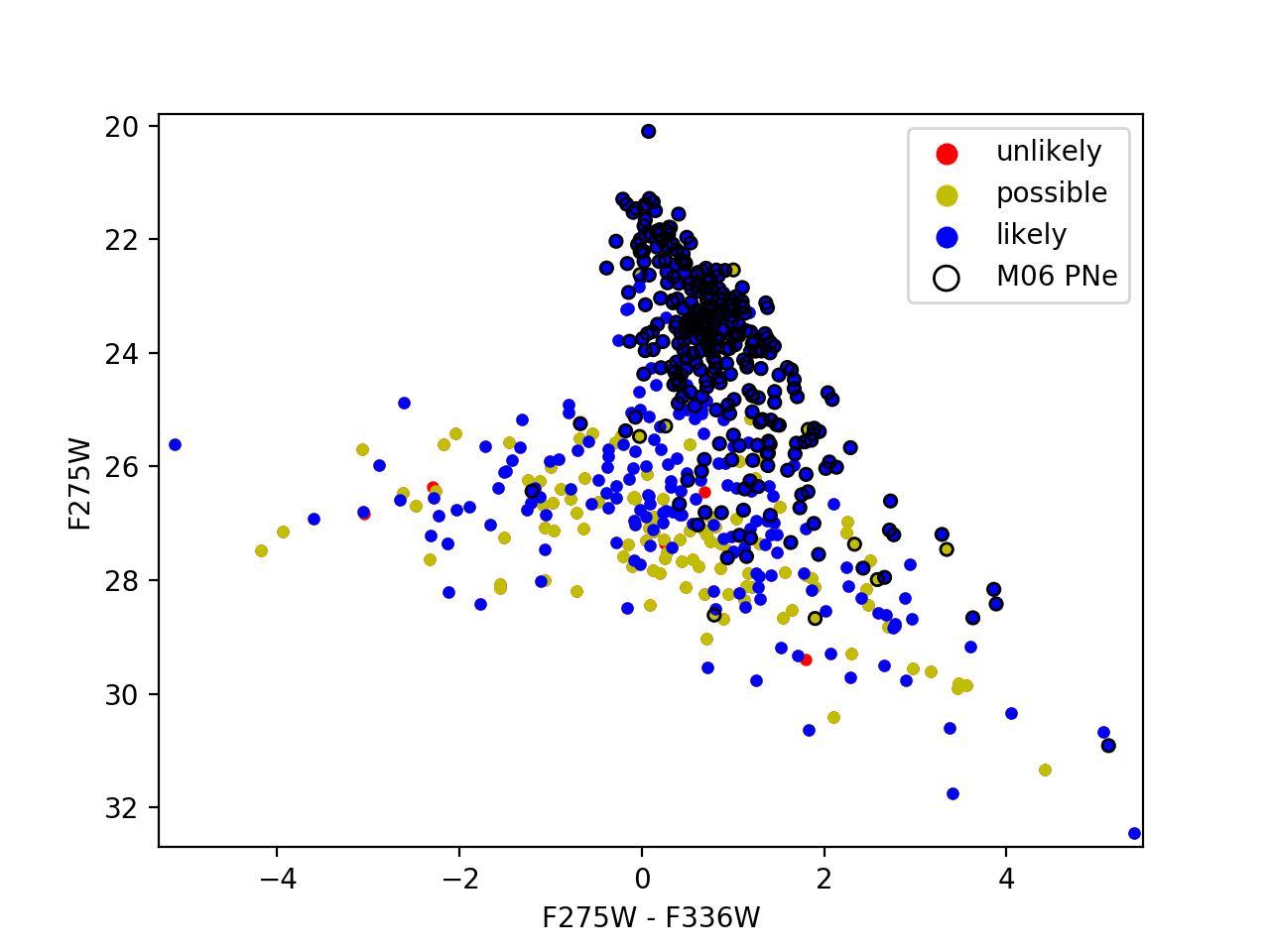}
	\caption{The F275W mag of the PHAT-matched PNe plotted against its F275W $-$ F336W colour. The ``likely'', ``possible'' and ``unlikely'' PNe have been shown in blue, yellow and red respectively. The PNe previously found by M06 have been encircled in black.}
	\label{fig:phatuv}
\end{figure}

\subsection{Completeness correction in the survey}
\label{compcorr}
{For the entire survey, the mean percentage total completeness is shown as a function of the m$\rm_{5007}$ magnitude in Figure~\ref{fig:comp} in black while the mean percentage detection completeness and percentage selection completeness are shown in blue and orange respectively. It is clear that the completeness correction applied is dominated by the selection completeness, correcting for the number of genuine PNe missed due to our colour and point-like selection. The percentage completeness shown is only representative in Figure~\ref{fig:comp}, as the completeness varies in each field and the completeness correction is applied to PNe in each field separately. }

\subsection{Detection check on background}
{For each image, as a by-product of the source extraction, we obtain a background image. In order to check for spurious sources in any image, we run SExtractor on this background image with the parameters described in Section~\ref{sext}. The sources detected on the background image and their distribution as a function of magnitude would give us an indication of the number of spurious sources present in our PNe catalog. However, subjecting these sources to the PNe selection criteria as described in Sections~\ref{csel} and ~\ref{pts}, we find that even if these detected sources survive the colour selection, they would be rejected as PNe in the point-like selection as their half-light radius is much smaller than that expected from the simulated PNe population. }


\section{PNe counterparts in PHAT}

\subsection{The M06 PNe counterparts in PHAT}
For each M06 PNe, \citet{vey14} searched for counterparts within $3''$. They expected a strong relation between the M06 m$\rm_{5007}$ magnitude and the PHAT F475W magnitude, since the majority of F475W flux is due to the [\ion{O}{iii}] 5007 $\AA$ line, and a colour excess between the PHAT F475W and F814W filters, since PNe are not expected to show a strong continuum excess. Additionally they utilized the separation between the M06 PNe and the PHAT counterpart, and the roundness and sharpness of its PSF. Of the 711 PNe in the M06 catalog that overlap the PHAT images, they found that only 467 had counterparts consistent with being PNe. The rest were either \ion{H}{ii} regions or possible stellar sources with no PHAT counterpart consistent with being a PNe.

\subsection{Further characteristics of the PHAT-matched PNe}

We also check the PNLF for consistency both near the crowded center (within 0.3 $\deg$) and in the less crowded outer disk (outside 0.3 $\deg$)as shown in Figure~\ref{fig:phatnearfar}. The PNLF is similar in both cases, indicating again that the rise in the PNLF does not depend on crowding, but the one in the less crowded disk shows a more pronounced dip in the PNLF. This maybe due to the presence of the 10kpc star forming ring in M31 \citep{barmby06} which may be populating the brightest 2.5 mag of the PNLF with PNe evolved from young massive stars. 954 PNe have detection in the PHAT NIR filters. Figure~\ref{fig:phatnir} shows the F110W mag of the PHAT-matched PNe plotted against its F110W $-$ F160W colour. 644 PNe have detection in the PHAT UV filters. Figure~\ref{fig:phatuv} shows the F275W mag of the PHAT-matched PNe plotted against its F275W $-$ F336W colour. The trends in both the UV and the NIR are similar to those found by \citet{vey14} for the M06 PNe. 

\end{appendix}

\end{document}